%% file: ms.tex
\documentclass[12pt]{article}
	
	\usepackage{amsmath,amssymb,amsthm}
	\usepackage{array}
	\usepackage[margin=1in]{geometry}
	\usepackage{graphicx}
	\usepackage{setspace}
	\usepackage{natbib}
	\usepackage{tikz,pgfplots}
	\usepackage{xstring}
	\usepackage{subcaption}
	\usepackage[hidelinks]{hyperref} 
	
	\usepgfplotslibrary{fillbetween}
	\pgfplotsset{compat=1.14}

	\definecolor{carolina blue}{RGB}{075,156,211}
	\definecolor{carolina dark blue}{RGB}{000,127,174}
	\definecolor{oxford primary blue}{RGB}{000,033,071}
	\definecolor{gopuff primary blue}{RGB}{000,164,255}
	\definecolor{hhu blue}{RGB}{0,106,179}
	
	\usetikzlibrary{patterns}
	\usetikzlibrary{decorations.pathreplacing}

	\date{29 April 2023}
	\def\bkemail{\href{mailto:kasberger@dice.hhu.de}{\color{hhu blue}kasberger@dice.hhu.de}}
	\def\kwemail{\href{mailto:kyle.woodward@gopuff.com}{\color{gopuff primary blue}kyle.woodward@gopuff.com}}

	\newtheorem{comparison}{Comparison}
	
	\newtheorem{corollary}{Corollary}
	\newtheorem{definition}{Definition}
	\newtheorem{lemma}{Lemma}
	\newtheorem{observation}{Observation}
	\newtheorem{proposition}{Proposition}
	
	\newtheorem{theorem}{Theorem}
	
	\theoremstyle{remark}
	\newtheorem{example}{Example}

	\def\arginf{\operatornamewithlimits{arg inf}}
	
	\def\argmin{\operatornamewithlimits{arg min}}

	\def\PAB{{\text{PAB}}}
	\def\UPA{{\text{UPA}}}
	\def\FRB{{\text{FRB}}}
	\def\LAB{{\text{LAB}}}

\author{Bernhard Kasberger\thanks{D\"{u}sseldorf Institute for Competition Economics (DICE), Heinrich Heine University D\"{u}sseldorf; \bkemail}\ \ and Kyle Woodward\thanks{Gopuff/University of North Carolina at Chapel Hill; \kwemail
\newline 
	We thank Justin Burkett, Piotr Dworczak, Marek Pycia, Orly Sade, Karl Schlag, Eran Shmaya, Alex Teytelboym, and Andy Zapechelnyuk, as well as a seminar audience at INFORMS, for valuable feedback and comments.
}}
\title{Bidding in Multi-Unit Auctions under Limited Information}

\begin{document}

\maketitle

\begin{abstract}
We study multi-unit auctions in which bidders have limited knowledge of opponent strategies and values. We characterize optimal prior-free bids; 
these bids minimize the maximal loss in expected utility resulting from uncertainty surrounding opponent behavior. Optimal bids are readily computable 
despite bidders having multi-dimensional private information, and in certain cases admit 
closed-form solutions. In the pay-as-bid auction the minimax-loss bid is unique; in the uniform-price auction the minimax-loss bid is unique if the bidder is 
allowed to determine the quantities for which they bid, as in many practical applications. We compare minimax-loss bids and auction outcomes across auction formats, and derive testable predictions.
\end{abstract}



\newpage

\input{./section-introduction.tex}

\input{./section-model.tex}

\input{./section-example.tex}

\input{./section-loss.tex}

\input{./section-minimax-loss-bids.tex}

\input{./section-conclusion.tex}

\bibliographystyle{plainnat}
\bibliography{kasberger-woodward}

\appendix

\input{./appendix-frb.tex}

\input{./appendix-divisible-goods.tex}

\input{./appendix-loss-minimizing-bids.tex}

\input{./appendix-applications.tex}

\end{document}

%% file: section-introduction.tex
\section{Introduction}
\label{section: introduction}

Multi-unit auctions play a critical role in many markets. For example, they are used to allocate generation capacity across power plants in electricity markets, and determine the interest rates at which governments can issue new debt.\footnote{For government securities, see \citet{Brenner+Galai+Sade-2009} and \citet{oecd-2021}. For electricity generation, see \citet{Maurer+Barroso-2011} and \citet{del-rio-ESD-2017}.} Traditional equilibrium analysis of these auctions relies on the common prior assumption, which may not be satisfied in practice; and, even when it is satisfied, computation of equilibrium strategies is typically intractable due to the multi-dimensionality of bidders' information 
{\setcitestyle{semicolon}\citep{Swinkels-Econometrica-2001A,Hortasu+Kastl+ECMA+2012}}. In this paper we relax the common prior and equilibrium assumptions in multi-unit auctions, and analyze worst-case loss minimizing bidders facing maximal uncertainty. We characterize optimal bids in this framework, show when optimal bids are unique (and when not), derive bounds on optimal bids in terms of novel \emph{iso-loss curves}, and provide comparative statics across common auction formats.

Real-world bidders often face uncertainty about the distribution of their opponents' bids, and this uncertainty is 
inconsistent with a Bayesian equilibrium. In a Bayesian equilibrium, bidders can compute for each bid the probability that its respective unit is won. 
However, surveying academic and professional auction consultants, \cite{kasberger-schlag} find that most real-world bidders cannot assign winning probabilities to their bids, which suggests that many real-world bidders are uncertain about the bid distribution. This uncertainty is particularly pronounced following policy shifts. 
In electricity auctions temporally close to deregulation, \cite{Doraszelski-Lewis-Pakes-AER-2018} show that bidder behavior is hard to anticipate, while in later auctions behavior can be explained by learning and eventual convergence to equilibrium. That is, economists---and bidders themselves---have limited understanding of initial play.
\footnote{Level-$k$ reasoning provides an alternative non-equilibrium approach that has been applied to initial play in multi-unit auctions \citep{10.1257/aer.20172015}. However, \cite{Rasooly-level-k} does not find support for the level-$k$ model in an experiment designed to disentangle level-$k$ from equilibrium behavior in single-unit auctions.}

To increase our understanding of 
bidder behavior in the presence of such uncertainty, we study how bidders bid in two major multi-unit auction formats when is hard to anticipate rivals' behavior and compare out-of-equilibrium auction outcomes. We study the pay-as-bid and uniform-price auction formats, both of which are frequently used to allocate homogeneous goods.\footnote{Members of the family of uniform-price auctions are defined by the selection of a market-clearing price. In the main text we analyze the last-accepted-bid auction, which is common in practice; in an appendix we analyze the first-rejected-bid auction. \citet{burkett+woodward-2020A} show that price selection may have a dramatic impact on equilibrium behavior, though this effect vanishes as the auction becomes competitive \citep{Swinkels-Econometrica-2001A}.} 
In these auctions bidders submit demand curves to the auctioneer. The auctioneer uses submitted demand curves to compute market-clearing prices and quantities. Each bidder receives their market-clearing quantity; in the pay-as-bid auction they pay their bid for each unit received, while in the uniform-price auction they pay the constant market-clearing price for each unit received. 
Little is known about equilibrium behavior in these auctions when bidders have general, multi-dimensional private values.\footnote{Bayesian equilibrium constructions in these auctions do exist in parameterized contexts. For example, \citet{Engelbrecht-Wiggans+Kahn-Southern-Economic-Journal-2002A} describe equilibrium when demand barely exceeds supply; \citet{Back+Zender-Review-of-Financial-Studies-1993A} and \citet{Wang+Zender-Economic-Theory-2002A} when the good is divisible and bidders have common values; \citet{Ausubel+Cramton+Pycia+Rostek+Weretka-The-Review-of-Economic-Studies-2014A} when bidders demand two units; \citet{burkett+woodward-2020A} when bidders' values are defined by order statistics; and \citet{pycia+woodward-2020A} when bidders have common, decreasing marginal values.}

Our prior-free non-equilibrium approach allows us to characterize the optimal bid functions for arbitrary multi-dimensional private marginal values. As the optimal bid functions depend non-linearly on all marginal values, closed-form solutions are available only when the number of parameters are relatively low (such as in the case of two-unit demand or under flat marginal values); however, we give a simple recursive construction showing that numerical solutions can always be computed straightforwardly. We characterize the optimal bids in three settings that appear in the literature and the real-world. The first setting is the standard discrete multi-unit auction. Our second empirically-relevant setting presumes that a large number of goods is available but that bidders are constrained to submit a relatively small number of bid points; bidders are free to choose the quantities at which bids are submitted. The implied bid function is a step function
, and the location and height of the steps are the bidders' choice variables.\footnote{Although step functions are mathematically simple they are economically complex: when bids are constant over wide intervals bidders are almost always rationed. When rationing occurs with positive probability Bayesian equilibrium bids must take bidding incentives for non-local units into account, and the equilibrium first-order conditions imply a complicated non-local differential system {\setcitestyle{semicolon}\citep{Kastl-Journal-of-Mathematical-Economics-2012A, woodward-2014}}. Our prior-free approach is computationally more tractable. We provide analytic solutions in the case of constant marginal values.} Finally, in the appendix we also characterize the solution for the continuous divisible-good case, which features prominently in theoretical analyses of auctions for homogeneous goods.

A key concept in our characterization is \emph{conditional regret}. Conditional on winning a certain number of units, the bid can be ex-post ``too high'' or ``too low.'' The distinct payment rules in the pay-as-bid and uniform-price auctions imply distinct approaches to loss minimization: in the pay-as-bid auction a bid for a given quantity is too high whenever a higher quantity is received, while in the uniform-price auction a bid is too high only when it sets the market-clearing price. Intuitively, the optimal bid trades off the loss in utility (regret) from bidding too high 
and the loss in utility from bidding too low---that is, from winning too few units due to shading bids below the bidder's true value.

We summarize our findings through a set of testable predictions. In the discrete multi-unit setting, we expect more variation in the bids in the uniform-price auction than in the pay-as-bid for a given multi-dimensional value. We reach this conclusion because the optimal bid is unique in the pay-as-bid auction but not in the uniform-price auction.\footnote{The multiplicity of optimal bids in the uniform-price auction is reminiscent of the multiplicity of Bayesian Nash equilibria {\setcitestyle{semicolon}\citep{Klemperer-Meyer-ECTA-1989, Back+Zender-Review-of-Financial-Studies-1993A, Ausubel+Cramton+Pycia+Rostek+Weretka-The-Review-of-Economic-Studies-2014A, burkett+woodward-2020B}}. The type of multiplicity is starkly different, however. Multiple loss-minimizing bids mean that the minimax-loss best-reply correspondence is multi-valued, while multiple Bayesian Nash equilibria require the coordination of the bidders on one equilibrium. A common assumption in empirical work is that the data is generated by the same equilibrium; this assumption would not lead to the testable prediction that there is more variation in the bids in the uniform-price auction.} Under a natural selection of minimax-loss bids in the uniform-price auction, the optimal bids in the uniform-price auction are both higher and steeper than in the pay-as-bid auction. 
In general, absent the selection we take, the bids in the uniform-price auction may not be uniformly higher than in a pay-as-bid auction. It can be optimal to bid 0 for high quantities, echoing low-revenue ``collusive'' equilibria of the uniform-price auction. On the other hand, there is no optimal bid in the uniform-price auction that is uniformly below the optimal bid in the pay-as-bid auction. In an example (Section~\ref{section: example}), we show that the optimal bid in the pay-as-bid auction may \emph{decrease} in the bidder's value.\footnote{\cite{McAdams-failure-of-monotonicity-JET-2007} provides examples of a uniform-price auction where Bayesian Nash equilibrium bids may decrease in the bidder's value due to risk aversion and affiliated values. We provide an example for the pay-as-bid auction, using a new rationale.}

In the constrained setting, there is a unique minimax-loss bid in both the uniform-price and pay-as-bid auction. Hence, one should not expect more variation in the bids for a given value across the two auction formats. When the marginal values are sufficiently constant, then the bids in two auctions cannot be ranked uniformly: the first bid is higher in the uniform-price auction but the optimal bid drops to zero at a lower quantity in the uniform-price auction. We also provide an example in which the bids can be ranked unambiguously, and bids are higher in the uniform-price auction. 
Using our characterization of the optimal constrained bid function in the case of constant marginal values, a final testable prediction is that the minimax bids are evenly spaced in the quantity space in the pay-as-bid auction but concentrated on intermediate quantities in the uniform-price auction. 
In general, if one knew the bidders' values, then one could test whether they use the minimax-loss bids. Usually, however, the bidders' values are unobserved, and instead observed bids are 
used to estimate the bidders' values. Our uniqueness results and characterizations of the optimal bids in this setting lead to point-identification of the values and a simple estimation procedure.

In the pay-as-bid auction, in any of the three settings, the (multi-dimensional) bid is found by equalizing conditional maximal regret across \emph{all} units; the conditional maximal regret is the higher of the regret from bidding too high and bidding too low. 
In the uniform-price auction, in the constrained and unconstrained setting, the minimax-loss bid is found by considering the iso-loss curves, the curves that trace a certain level of over- and underbidding loss in the bid-quantity space. In the unconstrained case, the upper and the lower iso-loss curves are tangent for the loss-minimizing bid, and look similar to (strictly concave) budget constraints and indifference curves in standard consumer theory. The points of tangency are the only bids that are pinned down; the only requirement for minimax-loss bids on other quantities is that they lie between the two curves. A similar logic applies to the multi-unit case, so that there is also no unique loss-minimizing bid. 
Surprisingly this nonuniqueness does not extend to the constrained case, in which there is a unique minimax-loss bid; this bid minimizes the difference between lower and upper iso-loss curve subject to the number of allowed bid points. The iso-loss curves provide an intuitive, graphical way of understanding the optimal bids.

Ex post payments are not generally comparable between auction formats. For small quantities, the high bids of the uniform-price auction yield higher revenue than the low bids of the pay-as-bid auction, but for large quantities the low bids of the uniform-price auction yield lower revenue than the aggregate payment of both high and low bids in the pay-as-bid auction.\footnote{
A full expected revenue comparison depends on the distribution of opponent strategies, which are unspecified in our model.} Our uniqueness results suggest that a seller interested in certainty over the distribution of revenue may prefer the pay-as-bid auction: in both auction formats the distribution 
of ex post revenue 
depends on the distribution of private information, but in the uniform-price auction it 
also depends 
on the method bidders use to select among optimal bids.\footnote{Revenue volatility in the uniform-price auction with unconstrained bids has been observed experimentally by \citet{sade2006competition} and \citet{morales2013divisible}.} On the other hand, our results also show that selection ambiguity can be disposed of by limiting bidders to a finite number of self-selected bid points, hence in practice the distribution of value-relevant private information is the key determinant of expected revenue.

Some of our results, such as bids tending to be higher and steeper in the uniform-price auction and the ambiguous revenue comparison, are in line with previous theoretical and empirical work {\setcitestyle{semicolon}\citep{Ausubel+Cramton+Pycia+Rostek+Weretka-The-Review-of-Economic-Studies-2014A,burkett+woodward-2020A,pycia+woodward-2020A,Barbosa-De-Silva-Yang-Yoshimoto-AEJ-Micro-2022}}. Hence, central findings in the common prior extreme of Bayesian Nash equilibrium also hold in the opposite extreme of maximal uncertainty. We find it reassuring that the two models lead to the same qualitative conclusions, as it suggests that the standard Bayesian approach and our robust approach are complementary. The Bayesian Nash equilibrium approach convinces with its internal consistency of beliefs, while our chosen robustness approach is more tractable in very complicated (e.g., asymmetric and multi-dimensional) settings. Yet the qualitative insights of the two modeling frameworks largely coincide. A novel comparison of the auction formats shows that minimax loss is lower in the uniform-price than in the pay-as-bid auction, suggesting that it is ``easier to get it right'' in the uniform-price auction.




\cite{savage1951} introduced the minimax loss (regret) decision criterion for statistical decision problems. Since then it has been applied in econometrics \citep{Manski2021}, mechanism design {\setcitestyle{semicolon}\citep{https://doi.org/10.1162/JEEA.2008.6.2-3.560,BERGEMANN20112527,RobustMonopolyRegulation,guo-shmaya-project-choice}}, operations research {\setcitestyle{semicolon}\citep{Perakis-Roels-2008,Besbes-Zeevi-OR-2011}}, and more generally in strategic settings. Our paper belongs to the latter category. A first paper on analyzing games with minimax regret as the players' decision criterion was \cite{LINHART1989152} who study the minimization of worst-case regret in bargaining. \cite{Parakhonyak-Sobolev-EJ} consider Bayesian firms best responding to consumers whose search rules for the lowest price are derived from worst-case regret minimization. \cite{renou-schlag-2010}, \cite{halpern-pass-2012}, \cite{Schlag-Zapechelnyuk-best-compromise}, and \cite{kasberger-moment-equ} propose solution concepts for loss (regret) minimizing players.

Applying the minimax loss (regret) decision criterion to strategic situations requires the specification of the player's perspective. A first possibility takes an ex post perspective and asks what the optimal action would be if the realized opponent actions were known; this is the ex post regret framework as in most of the existing literature \citep{STOYE20112226,BERGEMANN20112527}. An alternative approach takes an interim perspective. Players are uncertain about the distribution of actions (or states), and the loss of an action is the difference between the expected payoff of best responding to the distribution and the expected payoff from the chosen action. The interim perspective is also adopted in the Bayesian approach where players best respond to (their belief of) the distribution of competing actions. As not even the Bayesian approach delivers ex post optimality in games of incomplete information, we prefer the interim perspective. Moreover, note that the interim but not the ex post perspective allows to meaningfully incorporate belief restrictions as in \cite{kasberger-schlag} and \cite{kasberger-moment-equ}. Following \cite{sz-te} and \cite{kasberger-schlag}, we refer to the interim concept as loss and to the ex post equivalent as regret. 


We introduce the model in the next section. Section~\ref{section: example} illustrates our approach and some findings in the simple two-unit case. Section~\ref{section: loss in multi-unit auctions} contains some key theoretical results for the analysis of minimax loss in pay-as-bid and uniform-price auctions, which are applied in Sections~\ref{section: minimax loss multi-unit} and~\ref{section: constrained bidding} to analyze the multi-unit and bidpoint-constrained cases, respectively. 
Section~\ref{section: conclusion} concludes. Proofs, calculations, the analysis of the unconstrained case, and an analysis of the uniform-price auction with a first-rejected-bid pricing rule are provided in the appendix.

%% file: section-model.tex
\section{Model}
\label{section: model}

We consider an auction for quantity $Q > 0$ of a perfectly divisible, homogeneous good. There are $n \geq 2$ bidders participating in the auction. Buyer $i$, $i \in \{ 1, \ldots, n \}$, has marginal value $v^i: [ 0, Q ] \to \mathbb{R}_+$, where $v^i( q )$ is their marginal value for quantity $q$. 
We assume that marginal values are weakly decreasing, so that $v^i( q ) \geq v^i( q^\prime )$ whenever $q \leq q^\prime$. 
For notational simplicity we assume that bidders have a strictly positive value for each unit, hence $v^i( Q ) >0$.\footnote{Our results remain valid when bidders do not strictly demand all units, provided we replace aggregate supply $Q$ with the supremum of all quantities for which marginal value is strictly positive, $\bar Q_i = \sup \{ q\colon v^i( q ) > 0 \}$. 
Additionally, if $\bar Q_i < Q$, all results obtain in the limit with values $v^i( q ) + \varepsilon$, letting $\varepsilon \searrow 0$.}

Bidder $i$ submits a weakly decreasing bid function $b^i: [ 0, Q ] \to \mathbb{R}_+$. After observing the bid profile $( b^j )_{j = 1}^n$ the auctioneer computes a \emph{market-clearing price} $p^\star$, 
\begin{align*}
	p^\star &\in \left\{ p^\LAB, p^\FRB \right\}; \\
	& p^\LAB = \inf \left\{ p\colon \exists q \in \left[ 0, Q \right]^n \text{ s.t. } \sum_{i=1}^n q_i < Q \text{ and } b^i\left( q_i \right) \geq p \; \forall i \right\}, \\
	& p^\FRB = \sup \left\{ p\colon \nexists q \in \left[ 0, Q \right]^n \text{ s.t. } \sum_{i = 1}^n q_i > Q \text{ and } b^i\left( q_i \right) \leq p \; \forall i \right\}.
\end{align*}
The prices $p^\LAB$ and $p^\FRB$ are, respectively, the last bid accepted and the first bid rejected.\footnote{See \citet{burkett+woodward-2020A}. Treasury auctions frequently apply last-accepted-bid pricing (e.g., the United States and Switzerland) while theoretical analyses frequently study first-rejected-bid pricing \citep{Ausubel+Cramton+Pycia+Rostek+Weretka-The-Review-of-Economic-Studies-2014A}.} All bids strictly above the market-clearing price $p^\star$ are awarded, and all bids strictly below the market-clearing price are rejected. 
When there are multiple bids placed at the market-clearing price ties are broken randomly.\footnote{As long as all bids strictly above the market-clearing price are awarded, the precise tiebreaking rule does not affect our results.}

Bidders are risk neutral. If a bidder with value $v^i$ receives $q_i$ units and makes transfer $t_i$, their utility is
\[
	\hat{u}\left( q_i, t_i; v^i \right) = \int_0^{q_i} v^i\left( x \right) dx - t_i.
\]

We consider two common auction formats. In a \emph{pay-as-bid} (or \emph{discriminatory}) auction, transfers are equal to the sum of bids for received units, $t_i^\PAB = \int_0^{q_i} b^i( x ) dx$. 
In a \emph{uniform-price} auction, transfers are equal to the market-clearing price times the number of units received, $t_i^\UPA = p^\star q_i$. If opponent bids $b^{-i}$ are distributed according to the integrable distribution $B^{-i}$, we write the bidder's interim utility as $u( b^i, B^{-i}; v^i ) = \mathbb{E}_{B^{-i}}[ \hat{u}( q^i( b ), t^i( b ); v^i ) ]$, where $q^i$ and $t^i$ are functions that map, according to the auction rules, the bidders' bid functions $b = (b^i,b^{-i})$ to bidder $i$'s quantity $q_i$ and transfer $t_i$, respectively.\footnote{Integrability of $B^{-i}$ is not a constraint on our results, since in all auction formats $\hat u$ is bounded below by $0$ and above by $Q v^i( 0 )$.}

\subsection{Loss and regret}

Given a distribution of opponent bids $B^{-i}$, bidder $i$'s \emph{loss} 
from bidding $b^i$ instead of the interim-optimal bid is
\[
	L\left( b^i; B^{-i}, v^i \right) = \sup_{\tilde b} \mathbb{E}_{B^{-i}}\left[ \hat u\left(q^i( \tilde b, b^{-i}), t^i\left( \tilde b, b^{-i} \right); v^i \right) - \hat u\left( q^i\left( b^i, b^{-i} \right), t^i\left( b^i, b^{-i} \right); v^i \right) \right].
\]
Loss measures the difference between expected utility given bid $b^i$ and the utility obtainable by optimizing the submitted bid with respect to distribution $B^{-i}$. For example, when bid $b^i$ is a best response to distribution $B^{-i}$, loss is zero. Loss is evaluated from an interim perspective; the equivalent ex post concept is \emph{regret},
\[
	R\left( b^i; b^{-i}, v^i \right) = \sup_{\tilde b} \hat u\left( q^i\left( \tilde b, b^{-i} \right), t^i\left( \tilde b, b^{-i} \right); v^i \right) - \hat u\left( q^i\left( b^i, b^{-i} \right), t^i\left( b^i, b^{-i} \right); v^i \right).
\]
Regret measures how much additional utility the bidder could receive if they had known the bids their opponents submitted prior to choosing their own bid. A utility-maximizing bidder with perfect foreknowledge of their opponents' bids will have zero regret.

If bidder $i$ knew the true distribution of opponent bids $B^{-i}$, she would evaluate potential bids by standard expected utility. However, in our model bidders face ambiguity regarding the true distribution $B^{-i}$ and know only that $B^{-i} \in \mathcal{B}$, where $\mathcal B$ is a set of feasible distributions over opponent bids. In the presence of this ambiguity, bidder $i$ evaluates potential bids according to the maximum loss generated by any feasible distribution of opponent bids; the optimal bid $b^{\star}$ minimizes this loss,
\[
	b^\star \in \arginf_{b^i} \sup_{B^{-i} \in \mathcal{B}} L\left( b^i; B^{-i}, v^i \right).\footnotemark
\]\footnotetext{Since loss is bounded below by zero, the infimum of maximum loss always exists; as the $\arginf$ of maximum loss, $b^\star$ is the limit of a sequence of bids approximating minimax loss, and is guaranteed to exist by compactness of the bid space.}

We refer to $b^\star$ as bidder $i$'s minimax-loss or optimal bid. We focus on the case of \emph{maximal uncertainty}, in which $\mathcal{B}$ contains all joint distributions on feasible bid functions; i.e., all distributions over $n - 1$ weakly-decreasing functions mapping $[ 0, Q ]$ to $\mathbb{R}_+$. Note that $\mathcal B$ is rich enough to include uncertainty about the number of bidders and supply.\footnote{There are bid distributions in $\mathcal B$ that put all the mass on bidder $j$ bidding zero, i.e., $b^j(q) = 0$ for all $q$. This effectively reduces the number of bidders so that $n$ is merely an upper bound on the number of bidders. Our model can also be understood as featuring (residual) supply uncertainty: let $Q$ be the upper bound of the support of supply and reduce supply through other bidders that demand units at prohibitively high prices, above $v^i( 0 )$.\label{footnote: supply uncertainty}} 

We offer a descriptive and a prescriptive interpretation of minimax-loss bids. From a prescriptive perspective, a practical advantage of our non-Bayesian approach is that the bids are completely prior-free, i.e., they do not depend on the other bidders' value distributions and strategies. 
All a bidder needs to know is their willingness-to-pay, hence the bids are robust because the bidder need not worry about misspecified beliefs. Indeed, if any bid distribution is deemed possible, then in particular the actual distribution is possible. 
\cite{kasberger-schlag} illustrate empirically that loss-minimizing bids perform well in first-price auctions despite bidders having very coarse beliefs about competitors' behavior. 
On the other hand, group decisionmaking provides a descriptive motivation for minimax loss. Suppose a corporation tasks a team with finding the right bid. Based on information learned after the auction, the executive board or a rival colleague might criticize the bidding team for having missed an opportunity, and the bidding team may want to preemptively defend against such a critique. 
By selecting a minimax-loss bid the bidding team 
can claim, ``Your alternative bid would have been worse than our bid had there been this other bid distribution. This bid distribution was a real possibility.'' 
The minimax bid is then robust to complaints that appeal to the materialized bid distribution.\footnote{\cite{savage1951} also suggests group decision making as a justification for the minimax principle. In his story group members have different subjective probability assessments and the minimax principle seeks to keep the greatest ``violence'' done to anyone's opinion to a minimum. In contrast, we interpret the minimax as a way to defend against ex post complaints.} Minimax bids are a way to justify the choice as an (undisputed) counterfactual case can be presented so that the minimax bid was the compromise between the two cases.\footnote{If the bid was chosen to maximize the payoff guarantee, then many opportunities might indeed be missed. Thus, maxmin expected utility is not robust to complaints about missed opportunities.}

Our subsequent analysis is simplified by the following observation, reducing loss (an interim value taken over beliefs) to a pointwise objective over potential allocations.

\begin{observation}[Reduction to aggregate demand]\label{observation: reduction to residual supply}
	When maximizing loss, it is sufficient to consider the distribution of aggregate opponent demand for each quantity $q \in [ 0, Q ]$.
\end{observation}

Bidder $i$'s ex post utility is unaffected by the specific bids submitted by their opponents, provided that the aggregate demand curve of their opponents remains fixed. Moreover, from a bidder's perspective, it makes no difference whether the aggregate demand of the opponents is considered, or if residual supply is considered (as supply $Q$ is constant). Hence, Observation~\ref{observation: reduction to residual supply} states that maximizing loss over the set of feasible joint distributions of opponent bids can be replaced by maximizing loss over the set of feasible aggregate opponent demand curves (i.e., residual supply curves). Importantly, in our subsequent analysis we do not need to consider the number of opponents bidder $i$ faces: it is sufficient to consider an arbitrary demand curve, independent of its source. For this reason our results depend on bidder $i$ alone.


%% file: section-example.tex
\section{Illustrative example}
\label{section: example}

Before analyzing the general case, we first illustrate our analytical approach in a two-unit example tied to our mult-unit results (Section~\ref{section: minimax loss multi-unit}). Bidder $i$ has value $v_{i1}$ for their first unit and a marginal value of $v_{i2}$ for their second unit. We assume marginal values are decreasing and non-negative, i.e., $v_{i1}\ge v_{i2}\ge 0$. The bidder can submit two bids $(b_{i1},b_{i2})$. Our Lemma~\ref{lemma: maximum loss as maximum regret} reduces the interim loss-minimization problem to an ex post regret-minimization problem; therefore we consider distinct outcomes which might maximize regret.

\subsection*{Pay-as-bid auctions}

In the pay-as-bid auction, there are three relevant outcomes: the bidder wins either zero, one, or two units. We consider these outcomes on a case-by-case basis.

\textit{Case 1: zero units.} Conditional on winning no units, bids are most suboptimal if the bidder could have won as many as they wanted at a price just above their first-unit bid. In this case, loss equals
\[
	\left[ \left( v_{i1} - b_{i1} \right) + \left( v_{i2} - b_{i1} \right)_+ \right] - 0 = \left( v_{i1} - b_{i1} \right) + \left( v_{i2} - b_{i1} \right)_+.
\]
Note that the value $v_{i2}$ needs to be sufficiently high so that bidder $i$ actually wants to win two units at the ``just-lost'' price $b_{i1}$.

\textit{Case 2: one unit.} Conditional on winning one unit, the bidder knows that they have overbid on the first unit and could reduce their bid. The worst-case overpayment occurs when the opponent's bid for their first unit is just above the bidder's bid for their second unit. In this case, the bidder could improve her utility not only by decreasing their bid for the first unit, but also by slightly increasing their bid for the second unit, and loss is
\[
	\left[ \left( v_{i1} - b_{i2} \right) + \left( v_{i2} - b_{i2} \right) \right] - \left[ \left( v_{i1} - b_{i1} \right) \right] = \left( b_{i1} - b_{i2} \right) + \left( v_{i2} - b_{i2} \right).
\]

\textit{Case 3: two units.} Conditional on winning both units, bids are most suboptimal if the bidder could also have won both units by bidding (almost) $0$ for both units. In this case, loss equals \[
	\left[ \left( v_{i1} - 0 \right) + \left( v_{i2} - 0 \right) \right] - \left[ \left( v_{i1} - b_{i1} \right) + \left( v_{i2} - b_{i2} \right) \right] = b_{i1} + b_{i2}.
\]

%

The minimax-loss bid balances the conditional regret of all three outcomes: underbidding regret conditional on losing the auction (winning zero units), (underbidding) regret conditional on winning one unit, and overbidding regret conditional on winning two units. Maximal loss is
\begin{equation}
	\max\left\{ \left( v_{i1} - b_{i1} \right) + \left( v_{i2} - b_{i1} \right)_+, \; \left( b_{i1} - b_{i2} \right) + \left( v_{i2} - b_{i2} \right), \; b_{i1} + b_{i2} \right\}. \label{equation: example pab loss}
\end{equation}
Maximal loss is minimized by equalizing the three expressions.
Thus the minimax-loss bid vector in the pay-as-bid auction is 
\[
	b^\PAB_{i1} = \begin{cases}
		\frac{1}{9} \left( 3 v_{i1} + 2 v_{i2} \right) &\text{if } 7 v_{i2} \geq 3 v_{i1}, \\
		\frac{1}{6} \left( 3 v_{i1} - v_{i2} \right) &\text{if } 7 v_{i2} < 3 v_{i1};
	\end{cases} \hspace{0.35cm} \text{ and } b^\PAB_{i2} = \frac{v_{i2}}{3}.
\]
The case distinction is due to the value for the second good being below or above the bid for the first; i.e., the term $( v_{i2} - b_{i1} )_+$ in equation~\eqref{equation: example pab loss}.

\begin{figure}
	\centering
	\begin{tikzpicture}
		\begin{axis}[
		axis lines = left,
		xlabel = \(v_{i2} / v_{i1}\),
    	ylabel = \(b / v_{i1}\),
		xtick={0, 0.25, 0.4285714286, .5, 1},
		xticklabels={$0$,$\frac{1}{4} $, $\frac{3}{7} $, $\frac{1}{2}$, $1$},
		ytick={0.25, 0.33333, .5},
		yticklabels={$\frac{1}{4} $, $\frac{1}{3}$, $\frac{1}{2}$},
		legend style={at={(0.03,2/3+.25)},anchor=west},
		xmin = 0,
		xmax = 1.05,
		ymin = 0,
		ymax = 0.7
		]
		\addplot [mark=none, color=blue, domain=0:1, thick] { max(.5,(1 + x)/3) };
		\addlegendentry{\(b^\LAB\)}
		
			\fill[blue!15] ( 0.0, 0.0 ) -- ( 0.5, 0.0 ) -- ( 1.0, {1/3} ) -- ( 0.5, 0.25 ) -- ( 0.25, 0.25 ) -- cycle;

			\addplot [mark=none, color=red, domain=0:3/7, thick] { (3 - x)/6 };
			\addplot [mark=none, color=red, domain=3/7:1, thick] { (3 + 2*x)/9 };
			
			\addplot [mark=none, color=red, domain=0:1, thick] { x/3 };
			\addlegendentry{\(b^\PAB\)}

			\addplot [mark=none, color=blue, domain=0:1, thick] { max(.5,(1 + x)/3) };
			


			\addplot [mark=none, color=blue, domain=0:1, thick, dashed] { x/3 };

			\addplot [mark=none, color=blue, domain=.5:1, dashed] { (1 + x)/6 };
			\addplot [mark=none, color=blue, domain=0:1, dashed, name path=B] { max( 0, 1/3 * (-1 + 2 * x)) };

			
			\addplot [mark = none, color = blue, domain = 0:0.5, dashed] { 0.25 };
			\fill[pattern = crosshatch, pattern color = blue!15] ( 0.0, 0.0 ) -- ( 0.25, 0.25 ) -- ( 0.0, 0.25 ) -- cycle;

		\end{axis}
	\end{tikzpicture}
	\caption{\label{figure: demand for two units} First- and second-unit bids in the pay-as-bid and uniform-price auctions, when the bidder demands two units.}
\end{figure}
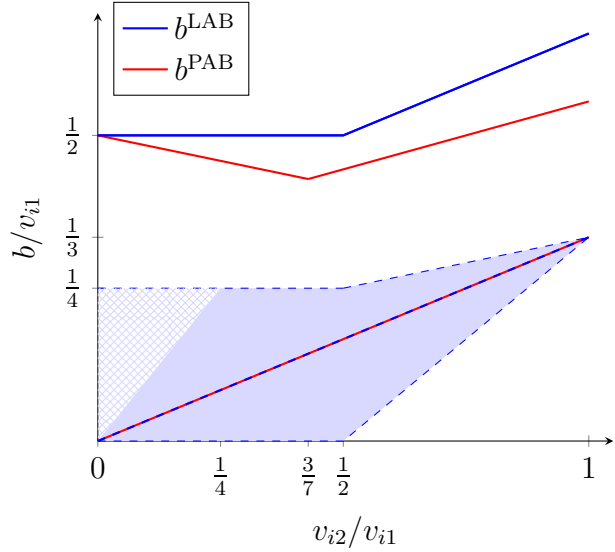

Figure~\ref{figure: demand for two units} illustrates the bidding functions as a function of $v_{i2}$, $v_{i2} \in [0,1]$. If $v_{i2}=0$, then the minimax bid is $b_{i1}^\PAB=1/2$, which is as in the first-price auction for a single good \citep{kasberger-schlag}. The bid $b_{i1}^\PAB$ \emph{decreases} in $v_{i2}$ for $v_{i2}\le 3/7$. This antitonicity arises because increasing $v_{i2}$ in this range increases the loss conditional on receiving a single unit, hence the bid $b_{i1}$ falls so that loss is equalized across outcomes. 
For values above $3/7$, both bids increase in $v_{i2}$, though the second bid $b_{i2}^\PAB$ increases more quickly than $b_{i1}^\PAB$. By corollary, the spread between the two bids uniformly decreases in $v_{i2}$.

\subsection*{Uniform-price auctions}

As in the pay-as-bid auction, three outcomes are focal when evaluating loss in the (last-accepted bid) uniform-price auction: the bidder either receives zero, one, or two units. We consider these outcomes on a case-by-case basis.

\textit{Case 1: zero units.} As in the pay-as-bid auction, if the bidder wins zero units they know that they have underbid two opponent bids. Their bids are most suboptimal if they could marginally increase their bid and win as many units as they desire, in which case loss is
\[
	\left[ \left( v_{i1} - b_{i1} \right) + \left( v_{i2} - b_{i1} \right)_+ \right] - 0 = \left( v_{i1} - b_{i1} \right) + \left( v_{i2} - b_{i1} \right)_+.
\]

\textit{Case 2: one unit.} Conditional on winning one unit, the bidder overbids if $b_{i1}$ sets the market-clearing price and underbids if the market-clearing price is just above $b_{i2}$. In this case, loss is
\[
	\max \left\{ b_{i1}, \left( v_{i2} - b_{i2} \right)_+ \right\}.
\]

\textit{Case 3: two units.} When the bidder wins two units, they set the market-clearing price. In this case, bids are most suboptimal when the bidder could have reduced bids to (almost) zero without losing any units; then loss is
\[
	\left[ \left( v_{i1} - 0 \right) + \left( v_{i2} - 0 \right) \right] - \left[ \left( v_{i1} - b_{i2} \right) + \left( v_{i2} - b_{i2} \right) \right] = 2 b_{i2}.
\]


Maximal loss is then
\[
	\max\left\{ \left( v_{i1} - b_{i1} \right) + \left( v_{i2}-b_{i1} \right)_+, \; b_{i1}, \; 2b_{i2}, \; v_{i2} - b_{i2} \right\}.
\]
Due to the different signs, maximal loss is minimized by equalizing at least \emph{some} of the conditional losses; this contrasts the pay-as-bid auction, in which maximal loss is minimized by equalizing \emph{all} of the conditional losses. 
%
Pairwise equalization of maximum loss gives a minimax-loss bid vector in the uniform-price auction,
\[
	b_{i1}^\LAB = \begin{cases}
		\frac{1}{3} \left( v_{i1} + v_{i2} \right) &\text{if } v_{i1} \leq 2 v_{i2}, \\
		\frac{1}{2} v_{i1} &\text{otherwise;}
	\end{cases}
	\text{ and } b_{i2}^\LAB = \frac{v_{i2}}{3}.
\]
The first bid can be found by equalizing the underbidding regret conditional on losing the auction $v_{i1} - b_{i1} + (v_{i2}-b_{i1})_+$ with the overbidding regret conditional on winning one unit $b_{i1}$. The second bid can be found by equalizing the underbidding regret conditional on winning one unit $v_{i2} - b_{i2}$ and the overbidding regret conditional on winning two units $2b_{i2}$.

While minimax-loss bids must minimize cross-conditional regret for some unit, this will not in general determine the minimax-loss bid for all units. With demand for two units, worst-case loss minimization uniquely determines the bid for the first unit, but the bid for the second unit need only lie within the bounds $v_{i2} - L^\LAB \leq b_{i2} \leq L^\LAB / 2$, where $L^\LAB$ is minimax loss in the uniform-price auction. The range of feasible minimax-loss bids in the uniform-price auction is depicted in Figure~\ref{figure: demand for two units}. The different shades distinguish minimax-loss bids above and below the marginal value.

%% file: section-loss.tex
\section{Loss in auctions for homogeneous goods}
\label{section: loss in multi-unit auctions}

We begin by establishing general properties of the loss-minimization problem in the pay-as-bid and uniform-price auctions. 
In the case of maximal uncertainty, bidders believe every possible distribution of opponent bids is feasible. Following Observation~\ref{observation: reduction to residual supply}, this is equivalent to bidders believing that every distribution of residual supply curves is feasible. In particular, bidders believe that degenerate distributions on specific aggregate demand curves are feasible, which implies that maximum loss is equivalent to maximum regret. 
This is a consequence of linearity of bidder preferences, and is not specific to the analysis of auctions or other features of our model.

\begin{lemma}[Reduction to maximum regret]\label{lemma: maximum loss as maximum regret}
	Under maximal uncertainty, maximizing loss is equivalent to maximizing regret. That is, for all values $v^i$ and bids $b^i$,
	\[
		\sup_{B^{-i}\in\mathcal B} L\left( b^i; B^{-i}, v^i \right)  = \sup_{b^{-i}} R\left( b^i; b^{-i}, v^i \right).
	\]
\end{lemma}

Following Lemma~\ref{lemma: maximum loss as maximum regret}, bidder $i$'s loss maximization problem can be identified with a regret maximization problem. 
As noted in Observation~\ref{observation: reduction to residual supply}, bidder $i$'s utility depends only on the aggregate demand curve submitted by bidders $-i$ and does not depend directly on any other bidder's specific bid. We therefore consider the set of feasible demand functions $\mathcal{S}$,
\[
	\mathcal S = \left\{ S: \left[ 0, Q \right] \to \mathbb R_+\colon S \text{ is decreasing} \right\}.\footnotemark
\]\footnotetext{Each opponent $j \neq i$ submits a decreasing bid $b^j: [ 0, Q ] \to \mathbb{R}_+$, so the aggregate demand of bidder $i$'s opponents is a function mapping $[ 0, ( n - 1 ) Q ]$ to $\mathbb{R}_+$. However, because there are only $Q$ units available demand is only relevant for quantities $q \in [ 0, Q ]$.}Abusing notation, let $q^i( b^i, S )$ be the quantity bidder $i$ receives when they submit bid $b^i$ and face aggregate demand curve $S$, and let $R( b^i; S, v^i )$ be bidder $i$'s regret when submitting bid $b^i$ against opponent aggregate demand $S$.

The regret maximization problem is complicated by the dependence of bidder $i$'s market allocation $q_i$ on both their own bid $b_i$ and opponent demand $S$. To simplify the problem, we decompose the regret maximization problem to the related problem of maximizing \emph{conditional regret}. Given any quantity $q \in [ 0, Q ]$, bidder $i$'s conditional regret from winning $q$ units is
\[
	R_q\left( b^i; v^i \right) = \sup_{S\colon q^i\left( b^i, S \right) = q} R\left( b^i; S, v^i \right).
\]
Given a bid $b^i$ and an opponent demand curve $S$, bidder $i$'s quantity allocation $q^i( b^i, S )$ is deterministic. Since maximum loss is identical to maximum regret, which is derived ex post after opponent demand is realized, it follows that maximum loss is the highest conditional regret from receiving any quantity, 
$\sup_{b^{-i}} R( b^i; b^{-i}, v^i ) = \sup_q R_q( b^i; v^i )$.

Conditional regret forms the basis of our subsequent results on bidding in pay-as-bid and uniform-price auctions under maximal uncertainty.

%
%
\subsection{Pay-as-bid auctions}

To develop intuition for loss minimization in the pay-as-bid auction, consider the potential sources of regret in a canonical single-unit first-price auction. Ex post, bids in single-unit discriminatory auctions are either too high---because the bidder strictly outbid the second-highest bidder---or too low---because the bidder underbid the highest bidder, whose bid was below the bidder's value.\footnote{In the case in which bids are neither too high nor too low, regret is zero. Typically, maximal regret will be nonzero.} This same intution is true pointwise in multi-unit pay-as-bid auctions: the bidder frequently would prefer to increase their bid for large quantities and decrease their bid for small quantities. We use this observation to pin down conditional regret in the pay-as-bid auction.


If bidder $i$ submits bid $b$ and obtains quantity $q$, they know that the market-clearing price is $p^\star \in [ b^i_+( q ), b^i( q ) ]$, where $b^i_+( q ) = \lim_{q^\prime \searrow q} b^i( q^\prime )$.\footnote{For notational simplicity we define $b^i_+( Q ) = 0$.} Their regret is at least
\[
	\int_0^{q} \left( b^i\left( x \right) - p^\star \right) dx + \int_{q}^{Q} \left( v^i\left( x \right) - p^\star \right)_+ dx.
\]
That is, their regret is at least their overpayment for units they received, plus the utility foregone by underbidding for units they value above the market-clearing price. This regret would be realized if, for example, all opponents submitted flat bids at the price $p^\star$. This expression is strictly decreasing in $p^\star$, hence bidder $i$'s conditional regret is at least
\begin{equation}
	\underline R_q^\PAB\left( b^i; v^i \right) = \int_0^q \left( b^i\left( x \right) - b^i_+\left( q \right) \right) dx + \int_q^Q \left( v^i\left( x \right) - b^i_+\left( q \right) \right)_+ dx. \label{equation: pab underbidding regret}
\end{equation}
Because $\underline R_q^\PAB$ is the regret the bidder has in the case in which they wish they had bid slightly more for larger quantities, we refer to $\underline R_q^\PAB$ as \emph{underbidding regret}.

Alternatively, bidder $i$ might be able to obtain the same allocation by bidding just above zero for all units. This will be the case when their opponents, in aggregate, submit extremely high bids 
for $Q - q$ units and zero bids for all remaining units. In this case all nonzero payment is wasted, and regret is
\[
	\overline R_q^\PAB\left( b^i; v^i \right) = \int_0^q b^i\left( x \right) dx.
\]
Because $\overline R_q^\PAB$ is the regret the bidder has in the case in which they wish they had bid nearly zero for all units, we refer to $\overline R_q^\PAB$ as \emph{overbidding regret}. 

Because maximum loss is equal to maximum regret, and ex post regret is obtained at some allocation, maximal loss may be identified with maximizing conditional regret.


\begin{lemma}[Maximum loss in pay-as-bid]\label{lemma: maximal loss in PAB}
	In the pay-as-bid auction, maximal loss given bid $b^i$ is
	\[
		\sup_q \left[ \max\left\{ \overline R_q^\PAB\left( b^i; v^i \right), \underline R_q^\PAB\left( b^i; v^i \right) \right\} \right].
	\]
\end{lemma}
A proof appears in Appendix~\ref{appendix: proofs for loss-minimizing bids}. Since $\underline R_Q^\PAB( b^i; v^i ) = \overline R_Q^\PAB( b^i; v^i )$, and $\overline R_q^\PAB( b^i; v^i )$ is weakly increasing in $q$, Lemmas~\ref{lemma: maximum loss as maximum regret} and~\ref{lemma: maximal loss in PAB} together imply that maximum loss is the supremum of underbidding regret, taken over all quantities $q$.
\begin{corollary}[Maximum loss in pay-as-bid]\label{corollary: maximum loss in pab}
	In the pay-as-bid auction, maximal loss given bid $b^i$ is
	\[
		\sup_q \underline R_q^\PAB\left( b^i; v^i \right).
	\]
\end{corollary}

%
%
\subsection{Uniform-price auctions}

In the uniform-price auction, bids above the market-clearing price are relevant only to the extent that they guarantee a unit is awarded; they do not otherwise affect the bidder's utility. 
This is in contrast to the pay-as-bid auction, where bids above the market-clearing price are paid whenever the unit is awarded. 
We first establish expressions for underbidding and overbidding regret in the last accepted bid uniform price auction, in line with our analysis of overbidding and underbidding regret in pay-as-bid auctions. The market-clearing price is $p^\star = p^{\LAB}$.\footnote{An analysis of the first rejected bid uniform-price auction can be found in Appendix~\ref{appendix: frb}. The analyses differ only in the multi-unit case.}

When bidder $i$ receives quantity $q$, the market-clearing price must be $p^\star \in [ b^i_+( q ), b^i( q ) ]$. The lower is the market-clearing price, the higher is underbidding regret, hence bidder $i$'s underbidding regret is
\[
	\underline R_q^\LAB\left( b^i; v^i \right) = 
	\int_q^Q \left( v^i\left( x \right) - b^i_+\left( q \right) \right)_+ dx.
\]
As in the pay-as-bid auction, underbidding regret accounts not only for the fact that the bidder might regret not bidding just above the market-clearing price, but also for the fact that the bidder might affect their own transfer. In particular, if the bidder sets the market-clearing price at a quantity just to the left of a discontinuity in their bid, they can reduce their bid and also the market-clearing price without affecting their allocation.

Alternatively, bidder $i$ might be able to obtain the same allocation by bidding just above zero for all units. This will be the case when their opponents submit high bids for $Q - q$ units and submit zero bids for all remaining units. In this case all nonzero bids are wasted, and regret is higher the higher is the market-clearing price, hence overbidding regret is
\[
	\overline R_q^\LAB\left( b^i; v^i \right) = q b^i\left( q \right).
\]
This differs from overbidding regret in the pay-as-bid auction, $\overline R_q^\PAB$, since in the uniform-price auction only the marginal bid is relevant.

The conditional regret for any quantity $q$ is
\[
	R^\LAB_q\left( b^i; v^i \right) = \max \left\{ \underline R_q^\LAB\left( b^i; v^i \right), \overline R^\LAB_q\left( b^i; v^i \right) \right\}.
\]
Because maximum loss is equal to maximum regret, and ex post regret is obtained at some allocation, maximal loss may be identified with maximizing conditional regret.
\begin{lemma}[Maximum loss in uniform-price]\label{lemma: maximal loss in UPA}
	In the uniform-price auction, maximal loss given bid $b^i$ is
	\[
		\sup_q R_q^\LAB\left( b^i; v^i \right).
	\]
\end{lemma}

%% file: section-minimax-loss-bids.tex
\section{Minimax-loss bids in multi-unit auctions}
\label{section: minimax loss multi-unit}




In most practical applications the homogeneous commodity up for auction is not perfectly divisible. We first consider the case in which bidder $i$ can bid on $M$ discrete units, and their value for unit $k$ is $v_{ik} = \int_{q_{k-1}}^{q_k} v^i( x ) dx$, where $q_k = k Q / M$. In this multi-unit model each bidder $i$ submits an $M$-dimensional bid vector, $b_i=( b_{ik} )_{k = 1}^M$; this bid vector may be translated to a bid function $\hat b^i$,
\[
	\hat b^i\left( q \right) = \begin{cases}
		b_{ik} &\text{if } q_{k-1} \leq q < q_k, \\
		0 &\text{if } q = Q.
	\end{cases}
\]
For ease of exposition, we define $q_{i0} = 0$ and $b_{iM+1} = 0$.

We now analyze minimax-loss bidding in the pay-as-bid and uniform-price auctions in this context. Most proofs for this section may be found in Appendix~\ref{appendix: proofs for loss-minimizing bids}.

\subsection{Pay-as-bid auctions}

For quantities $q \in ( q_{k-1}, q_k )$, underbidding regret \eqref{equation: pab underbidding regret} is weakly decreasing in $q$: increasing $q$ does not affect the value of the integral $\int_0^q ( \hat b^i( x ) - \hat b^i( q ) ) dx$ since $\hat b^i( x ) = \hat b^i( q )$ for $x$ near $q$; on the other hand, increasing $q$ shrinks the bounds of integration of $\int_q^Q ( v^i( x ) - \hat b^i( q ) )_+ dx$, and since the integrand is weakly positive it follows that underbidding regret $\underline R^\PAB_q( \hat b^i; v^i )$ is weakly decreasing on this range.

An immediate implication is that maximum underbidding regret is obtained at some multi-unit quantity $q_k$. Then Corollary~\ref{corollary: maximum loss in pab} implies 
\[
	L^\PAB\left( \hat b^i; v^i \right) = \max_{k \in \left\{ 0, 1, \ldots, M \right\}} \underline R_{q_k}^\PAB\left( \hat b^i; v^i \right).
\]
Underbidding regret for quantity $q_k$ increases in the bid for quantities $q_{k^\prime} \leq q_k$, decreases in the bid for quantity $q_{k+1}$, and is unaffected by the bid for quantities $q_{k^\prime} > q_{k+1}$. It follows that if $b_i$ is an optimal bid vector then underbidding regret must be constant at all quantities $q_k$.

\begin{theorem}[Equal conditional regret in multi-unit pay-as-bid]\label{theorem:discriminatory auction equal conditional regret}
	If $b_i$ is a minimax-loss bid vector in the multi-unit pay-as-bid auction, then $\underline R^\PAB_{q_k}( \hat b_i; v_i ) = \underline R^\PAB_{q_{k^\prime}}( \hat b_i; v_i )$ for all $k, k^{\prime} \in \{ 0, 1, \ldots, M \}$.
\end{theorem}

Theorem~\ref{theorem:discriminatory auction equal conditional regret} gives a straightforward method for computing minimax-loss bids: minimize conditional regret for any quantity, conditional on equal conditional regret across all quantities. Although computationally straightforward, optimal bids do not 
have a general analytical form. The definition of conditional regret contains a summation over all units which are valued more than a given bid, and the extent of this summation depends not only on the bidder's values but also on the prospective bid, complicating the relationship between bid and loss.


\begin{corollary}\label{theorem:multi unit discriminatory representation}
	The unique minimax-loss bid vector $b^\PAB_i$ is such that $b^\PAB_{iM} = v_{iM} / ( M + 1 )$, and for all $k < M$, $b^\PAB_{ik}$ is defined by the implicit equation
	\begin{equation}
		\left( b^\PAB_{ik} - b^\PAB_{ik+1} \right) k = \left( v_{ik} - b^\PAB_{ik} \right) + \sum_{k^\prime = k + 1}^{M} \left[ \left( v_{ik^\prime} - b^\PAB_{ik} \right)_+ - \left( v_{ik^\prime} - b^\PAB_{ik+1} \right)_+ \right]. \label{equation:multi unit discriminatory implicit}
	\end{equation}
\end{corollary}

The minimax-loss bid is unique for any multi-dimensional valuation. Uniqueness particularly simplifies the estimation of the private values if one believes that the observed bid data is generated by bidders playing the minimax-loss bids under maximal uncertainty. In this case, one can infer bidder $i$'s value $v_{iM}$ from the bid on the $M^{\text{th}}$ unit ($v_{iM}= ( M + 1 ) b_{iM}^\PAB$). Marginal values for units $k < M$ can be inferred by recursively solving 
Equation~\eqref{equation:multi unit discriminatory implicit} for the unknown $v_{ik}$, starting with $k = M - 1$. 
Note that this estimation is straightforward, and involves solving only a sequence of linear equations. In contrast to the approach relying on Bayesian Nash equilibrium as the data-generating model, estimating values from minimax-loss bids 
does not require the difficult estimation of the (opponent) bid distribution. 
From a normative perspective, a unique bid is attractive as it saves one from further assessing the relative merits of all minimax-loss bids.


\begin{observation}\label{remark:multi unit discriminatory bids below value}
	The minimax-loss bid vector is strictly below marginal values wherever $v_{ik} > 0$. Let $\bar k = \max \{ k\colon v_{ik} > 0 \}$. Applying Corollary~\ref{theorem:multi unit discriminatory representation} gives $b^\PAB_{i \bar k} = v_{i \bar k} / ( \bar k + 1 ) < v_{i \bar k}$, and the bid for the last positively-valued unit is below the value for this unit. Then if there is $k$ with $b_{ik} = v_{ik} > 0$, there is a maximal such quantity. Since values are weakly decreasing in quantity, this implies that $( b_{ik} - b_{ik+1} ) k = -\sum_{k^\prime = k + 1}^{M} ( v_{ik^\prime} - b_{ik+1} )_+$; this can only hold if $v_{ik+1} = b_{ik+1}$, contradicting the assumption that $k$ is the maximal quantity with $b_{ik} = v_{ik} > 0$. It follows that $b^\PAB_{ik} < v_{ik}$.
\end{observation}

\begin{observation}\label{observation: pab bids strictly decreasing}
	The minimax-loss bid vector is strictly decreasing in quantity wherever $v_{ik} > 0$. Otherwise, there is $k$ such that $b_{ik} = b_{ik+1}$ and $v_{ik} > b_{ik}$ (see Observation~\ref{remark:multi unit discriminatory bids below value}). In this case, the left-hand side of~\eqref{equation:multi unit discriminatory implicit} is zero and the right-hand side is strictly positive. Increasing $b_{ik}$ increases the left-hand side of~\eqref{equation:multi unit discriminatory implicit} and decreases the right-hand side, and it follows that $b^\PAB_{ik} > b^\PAB_{ik+1}$ whenever $v_{ik} > 0$.
\end{observation}
The tractability of the minimax-loss bid stands in stark contrast to the typical intractability of the Bayesian equilibrium in the pay-as-bid auction. As discussed in the introduction, Bayesian equilibrium characterizations exist only in relatively simple (usually complete information or one-parameter) economic settings. On the other hand it is straightforward to numerically compute minimax-loss bids. 
Equation~\eqref{equation:multi unit discriminatory implicit} provides an implicit definition for minimax-loss bids in the pay-as-bid auction; and although this equation cannot in general be solved in closed form, the 
%
bid representation in Theorem~\ref{theorem:multi unit discriminatory representation} is straightforward to compute numerically: first compute $b^\PAB_{i M}$, then iteratively compute each $b^\PAB_{ik-1}$ from the already-computed $( b^\PAB_{ik^\prime} )_{k^\prime = k}^{M}$.

The following example shows that an analytical solution is available when the marginal values are sufficiently flat.


\begin{example}\label{example: pab closed form}
	Suppose that there are $M$ units available for auction, and that bidder $i$'s value vector is $v_i$. Assume that bidder $i$'s value vector is relatively flat, so that $v_{iM} \not\ll v_{i1}$.\footnote{
	Formally, we require $b_{i1} \leq v_{iM}$, which in light of equation~\eqref{equation: multi-unit pab flat values} is equivalent to $\sum_{k^\prime = 1}^M ( M / [ M + 1 ] )^{k^\prime - 1} v_{ik^\prime} \leq ( M + 1 ) v_{iM}$.} Following Corollary~\ref{theorem:multi unit discriminatory representation}, the minimax-loss bid for unit $M$ is $b^\PAB_{iM} = v_{iM} / ( M + 1 )$. Note that the bid for unit $k$ can be written as
	\[
		\left( M + 1 \right) b^\PAB_{ik} = v_{ik} + M b^\PAB_{ik+1}.
	\]
	Thus for $k < M$, minimax-loss bids in the pay-as-bid auction can be written as
	\begin{equation}
		b^\PAB_{ik} = \frac{1}{M + 1} \sum_{k^\prime = k}^{M} \left( \frac{M}{M+1} \right)^{k^\prime - k} v_{ik^\prime}.
		\label{equation: multi-unit pab flat values}
	\end{equation}
	The minimax-loss bid in this pay-as-bid auction is compared to its uniform-price equivalent in Figure~\ref{figure: multi-unit pab and lab bids with flat values} in Example~\ref{example: multi-unit lab flat values} below.
\end{example}

\subsection{Uniform-price auctions}
\label{subsection: multi-unit lab}

We now analyze the multi-unit uniform-price auction, in which bidder $i$ may bid on $M$ discrete units. 
Following Lemma~\ref{lemma: maximal loss in UPA}, maximum loss is
\[
	L^\LAB\left( \hat b^i;  v^i \right) = 
	\max_{k \in \left\{ 0, 1, \ldots, M \right\}} R_{q_k}^\LAB\left( \hat b^i; v^i \right).
\]
That is, maximum loss is a maximum over conditional regrets, which are defined as the higher of overbidding and underbidding regrets for quantity $q_k$. This can be written equivalently as
\[
	L^\LAB\left ( \hat b^i; v^i \right) = \max_{k \in \left\{ 0, 1, \ldots, M \right\}} \max \left\{ \underline R_{q_{k-1}}^\LAB\left( \hat b^i; v^i \right), \overline R^\LAB_{q_k}\left( \hat b^i; v^i \right) \right\}.\footnotemark
\]\footnotetext{For notational simplicity we define $R^\LAB_{q_{-1}} = 0$.}
The terms $\underline R^\LAB_{q_{k-1}}$ and $\overline R^\LAB_{q_k}$ both depend on $b_{ik}$ and not on $b_{ik^\prime}$ for any $k^\prime \neq k$. Since $\underline R^\LAB_{q_{k-1}}$ is decreasing in $b_{ik}$ and $\overline R^\LAB_{q_k}$ is increasing in $b_{ik}$, if $b_i$ is a minimax-loss bid vector there must be some quantity $q_k$ so that $\underline R^\LAB_{q_{k-1}}( \hat b^i; v^i ) = \overline R^\LAB_{q_k}( \hat b^i; v^i )$. 
The following Theorem is immediate.

\begin{theorem}[No unique optimal bid in uniform-price]\label{theorem: no unique loss-minimizing bid in lab}
	Generically, there is not a unique minimax-loss bid in the multi-unit uniform-price auction unless $M = 1$.
\end{theorem}


If there is a unique minimax-loss bid, then $\overline R^\LAB_{q_k}( \hat b_i; v_i ) = \underline R^\LAB_{q_{k-1}}( \hat b_i; v_i )$ for all units $k$. By definition of overbidding regret $\overline R^\LAB_{q_k}$, this implies that there is a constant $c$ such that $q_k b_{ik} = c$ for all units $k$. This in turn implies $\sum_{k^\prime = k}^{M} ( v_{ik^\prime} - c / q_k )_+ = c$ for all units $k$, and this equation cannot generically be solved simultaneously for all units. On the other hand, if the bidder may submit only a single bid ($M = 1$) then this equation need only be solved for a single unit, and incentives are as in a first-price auction; \citet{kasberger-schlag} show that the unique minimax-loss bid in this case is $b_{i1} = v_{i1} / 2$.

To obtain sharp predictions for optimal strategies, we introduce a selection from the set of minimax-loss bids. We define a \emph{cross-conditional regret minimizing} strategy to be one which minimizes the larger of overbidding regret for unit $q_k$ and underbidding regret for unit $q_{k-1}$, which we term \emph{cross-conditional regret}. By construction cross-conditional regret is independent across bid points; since regret is maximized by cross-conditional regret for some quantity, a cross-conditional regret-minimaxing bid is a minimax-loss bid.

\begin{definition}
	The minimax-loss bid vector $b_i$ is a \emph{cross-conditional regret minimizing} bid if $\underline R_{q_{k-1}}^\LAB( \hat b_i; v_i ) = \overline R_{q_k}^\LAB( \hat b_i; v_i )$ for all $k \in \{ 1, \ldots, M \}$.
\end{definition}

The appeal of cross-conditional regret minimizing bids is that any bid $b_{ik}$ is justifiable ex post. If another minimax bid was chosen so that the bid for unit $k$ was below the respective cross-conditional regret minimizing bid for that unit, then after winning $q_{k-1}$ units, the case can be made that this bid was too low as it would have been profitable to win more units. Only the cross-conditional regret minimizing bid does not allow such complaints as the regret of paying too much for $q_k$ units serves as a defense.



\begin{theorem}[Cross-conditional regret minimizing bids in uniform-price]\label{theorem: conditional regret minimization lab}
	The unique cross-conditional regret minimizing bid $b_i^\LAB$ is such that for all units $k$,
	\begin{equation}
		b_{ik}^\LAB = \frac{1}{k} \sum_{k^\prime = k}^M \left( v_{ik^\prime} - b_{ik}^\LAB \right)_+.\label{equation: conditional regret minimization lab}
	\end{equation}
\end{theorem}

Theorem~\ref{theorem: conditional regret minimization lab} illustrates the nonuniqueness of minimax-loss bids in the uniform-price auction. If $b^\LAB_i$ is a cross-conditional regret minimizing bid, equation~\eqref{equation: conditional regret minimization lab} implies
\[
	b^\LAB_{iM} = \frac{1}{M + 1} v_{iM}, \text{ and } b^\LAB_{i1} \geq \frac{1}{2} v_{i1}.
\]
Overbidding regret at quantity $q = q_M$ is $\overline R^\LAB_{q_M}( \hat b_i; v_i ) = M v_{iM} / ( M + 1 )$ and underbidding regret at quantity $0$ is $\underline R^\LAB_0( b_i; v_i ) \geq v_{i1} / 2$. These are unequal when, for example, $v_{i1} \geq 2 v_{iM}$, in which case there cannot be a unique minimax-loss bid.

As in the pay-as-bid auction (Example~\ref{example: pab closed form}), minimax-loss bids in the uniform-price auction have a convenient analytical expression when a bidder's values are relatively constant.

\begin{example}\label{example: multi-unit lab flat values}
	Suppose that there are $M$ units available for auction, and that bidder $i$'s value vector is $v_i$. Assume that bidder $i$'s value vector is relatively flat, so that $v_{iM} \not\ll v_{i1}$.\footnote{
	Formally, we require $b_{i1} \leq v_{iM}$, which in light of equation~\eqref{equation: multi-unit lab flat values} is equivalent to $\sum_{k^\prime = 1}^M v_{ik^\prime} \leq ( M + 1 ) v_{iM}$.} Following Theorem~\ref{theorem: conditional regret minimization lab}, the cross-conditional regret minimizing bid is
	\begin{equation}
		b^\LAB_{ik} = \frac{1}{M + 1} \sum_{k^\prime = k}^{M} v_{ik^\prime}.
		\label{equation: multi-unit lab flat values}
	\end{equation}
	Figure~\ref{figure: multi-unit pab and lab bids with flat values} illustrates these bids, and compares them to their corresponding pay-as-bid bids.
	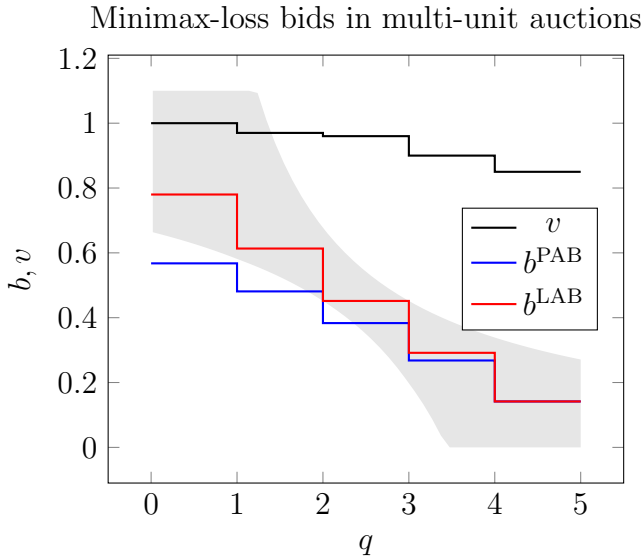
\begin{figure}
		\centering
		\begin{tikzpicture}
			\begin{axis}[
				xlabel = {$q$}, 
				ylabel = {$b, v$},
				title = {Minimax-loss bids in multi-unit auctions}, 
				legend style = {
					at = { ( 0.95, 0.50 ) },
					anchor = east
				}
			]
			\addplot [thick, black] table {
0	1.0
1	1.0
1	0.97
2	0.97
2	0.96
3	0.96
3	0.9
4	0.9
4	0.85
5	0.85
};
\addlegendentry{$v$}
\addplot [thick, blue] table {
0	0.567624742798354
1	0.567624742798354
1	0.4811496913580247
2	0.4811496913580247
2	0.3833796296296296
3	0.3833796296296296
3	0.26805555555555555
4	0.26805555555555555
4	0.14166666666666666
5	0.14166666666666666
};
\addlegendentry{$b^\PAB$}
\addplot [thick, red] table {
0	0.7799999999999999
1	0.7799999999999999
1	0.6133333333333334
2	0.6133333333333334
2	0.45166666666666666
3	0.45166666666666666
3	0.2916666666666667
4	0.2916666666666667
4	0.14166666666666666
5	0.14166666666666666
};
\addlegendentry{$b^\LAB$}

\addplot [draw = none, forget plot, name path = upper] table {
0.02	1.10
0.1216326530612245	1.10
0.223265306122449	1.10
0.32489795918367353	1.10
0.426530612244898	1.10
0.5281632653061226	1.10
0.629795918367347	1.10
0.7314285714285715	1.10
0.833061224489796	1.10
0.9346938775510205	1.10
1.0363265306122451	1.10
1.1379591836734695	1.10
1.239591836734694	1.0931017451432332
1.3412244897959185	1.01027084601339
1.442857142857143	0.9391089108910889
1.5444897959183674	0.8773123678646934
1.646122448979592	0.8231465410364491
1.7477551020408166	0.77528024287716
1.849387755102041	0.7326749062017214
1.9510204081632656	0.6945083682008367
2.05265306122449	0.6601212964804135
2.1542857142857144	0.6289787798408488
2.255918367346939	0.6006423014293468
2.3575510204081636	0.5747489612188365
2.459183673469388	0.5509958506224065
2.5608163265306128	0.5291281479120178
2.662448979591837	0.508929940211559
2.7640816326530615	0.490217070289427
2.865714285714286	0.4728315054835493
2.9673469387755107	0.45663686382393387
3.068979591836735	0.44151482909961426
3.1706122448979595	0.42736225540679706
3.272244897959184	0.41408881127603836
3.3738775510204086	0.4016150496007742
3.4755102040816332	0.389870816206694
3.5771428571428574	0.37879392971246
3.678775510204082	0.36832908021746363
3.7804081632653066	0.3584269056359317
3.882040816326531	0.34904321312164854
3.9836734693877554	0.3401383196721311
4.08530612244898	0.33167649115795783
4.186938775510204	0.32362546305322676
4.288571428571428	0.3159560293137908
4.390204081632653	0.3086416883599851
4.4918367346938775	0.30165833711949114
4.5934693877551025	0.29498400568686683
4.695102040816327	0.2885986264452751
4.796734693877551	0.2824838325391423
4.898367346938776	0.2766227814348804
5.0	0.271
};

\addplot [draw = none, forget plot, name path = lower] table {
0.02	0.6636546184738955
0.1216326530612245	0.656647423025435
0.223265306122449	0.6493420490472528
0.32489795918367353	0.6417190501134974
0.426530612244898	0.6337572512271307
0.5281632653061226	0.6254335523913837
0.629795918367347	0.6167227047725786
0.7314285714285715	0.6075970548862114
0.833061224489796	0.5980262513468508
0.9346938775510205	0.587976907630522
1.0363265306122451	0.5776871588919782
1.1379591836734695	0.5673631367575566
1.239591836734694	0.5564810593726256
1.3412244897959185	0.5449944221329764
1.442857142857143	0.53285140562249
1.5444897959183674	0.5199940940231516
1.646122448979592	0.5063575514177924
1.7477551020408166	0.49186872489959843
1.849387755102041	0.4764451353802306
1.9510204081632656	0.4599933065595716
2.05265306122449	0.4425855144716798
2.1542857142857144	0.42410642570281126
2.255918367346939	0.4042585155436561
2.3575510204081636	0.3828838430645659
2.459183673469388	0.3597991967871485
2.5608163265306128	0.33479082998661297
2.662448979591837	0.30760782259472663
2.7640816326530615	0.2779536327126688
2.865714285714286	0.24547523427041487
2.9673469387755107	0.20974899598393565
3.068979591836735	0.1724054111181568
3.1706122448979595	0.13198348951360983
3.272244897959184	0.08680604772029277
3.3738775510204086	0.03598142570281092
3.4755102040816332	0.0
3.5771428571428574	0.0
3.678775510204082	0.0
3.7804081632653066	0.0
3.882040816326531	0.0
3.9836734693877554	0.0
4.08530612244898	0.0
4.186938775510204	0.0
4.288571428571428	0.0
4.390204081632653	0.0
4.4918367346938775	0.0
4.5934693877551025	0.0
4.695102040816327	0.0
4.796734693877551	0.0
4.898367346938776	0.0	
5.0	0.0
};

\addplot[black!10] fill between [of = lower and upper];
			\end{axis}
		\end{tikzpicture}
		\caption{\label{figure: multi-unit pab and lab bids with flat values}Minimax-loss bids in the pay-as-bid and uniform-price auctions, with $M = 5$ units available. In the pay-as-bid auction the minimax-loss bid is unique, while in the uniform-price auction any bid in the gray region minimizes maximum loss. The red line is the cross-conditional regret minimizing bid in the uniform-price auction.}
	\end{figure}
\end{example}

Although the bidding function of Theorem~\ref{theorem: conditional regret minimization lab} cannot be compared to all Bayesian Nash equilibria of the uniform-price auction, it is apparent that it does not resemble ``collusive'' low-revenue equilibria that are frequently discussed in the literature \citep{Ausubel+Cramton+Pycia+Rostek+Weretka-The-Review-of-Economic-Studies-2014A,marszalec2020epic}.\footnote{The existence of collusive-seeming Bayesian Nash equilibria is linked to the absence of supply uncertainty \citep{Klemperer-Meyer-ECTA-1989,burkett+woodward-2020B}. See footnote~\ref{footnote: supply uncertainty} for an interpretation of supply uncertainty in our model.} Indeed, the bid on the last unit is positive in a cross-conditional regret minimizing strategy under maximal uncertainty, while it is zero in the canonical low-revenue Bayesian Nash equilibrium.

%
%
\subsection{Comparison of auction formats}

Bids in the uniform-price auction may be higher or lower than in the pay-as-bid auction, and the revenue comparison of the two formats is inherently ambiguous. 
To demonstrate revenue ambiguity, we first compare the cross-conditional regret minimizing bids in the uniform-price auction to the regret minimizing bids in the pay-as-bid auction.

\begin{comparison}[Uniform-price bids above pay-as-bid bids]\label{theorem:multi unit bid comparison}
	The unique cross-conditional regret minimizing bid in the multi-unit 
	uniform-price auction is higher 
	than the unique minimax-loss bid in the pay-as-bid auction: $b^\LAB \geq b^\PAB$.
\end{comparison}

Although cross-conditional regret minimizing bids in the uniform-price auction are above the unique minimax-loss bid in the pay-as-bid auction, this is not the case for all selections of minimax-loss bids in the uniform-price auction. 
In the uniform-price auction, underbidding regret for large quantities is necessarily small: uniform pricing implies there is no wedge for overpayment (as there is in the pay-as-bid auction), and there is little utility foregone by not receiving a small number of units. Since conditional regret is the larger of overbidding and underbidding regret, and overbidding regret is increasing in bid, for large quantities there is a conditional regret-minimizing bid which is equal to zero; this zero bid is below the unique minimax-loss bid in the pay-as-bid auction, which is strictly positive for all units the bidder values positively. 
We explore this further in our analysis of iso-loss curves and minimax-loss bids in bidpoint-constrained auctions in Section~\ref{section: constrained bidding}.

\begin{comparison}[Semi-comparability of optimal bids]\label{proposition: semicomparability of bids across auctions}
	Suppose $M \geq 2$ units are available. If $b^\LAB$ is a minimax-loss bid in the multi-unit uniform-price auction, then $b^\LAB \not\leq b^\PAB$. However, there is a minimax-loss bid $b^\LAB$ in the multi-unit uniform-price auction 
	such that $b^\LAB \not\geq b^\PAB$. 
\end{comparison}

While there is a minimax-loss bid in the uniform-price auction which is not everywhere greater than the minimax-loss bid in the pay-as-bid auction, there is no minimax-loss bid in the uniform-price auction which is everywhere below the minimax-loss bid in the pay-as-bid auction.

We now ask whether the minimax-loss bids are 
rankable by elasticity (steepness). Previous theoretical work has identified uniform-price bids as more elastic (i.e., steeper) than pay-as-bid bids {\setcitestyle{semicolon}\citep{malvey1998uniform,Ausubel+Cramton+Pycia+Rostek+Weretka-The-Review-of-Economic-Studies-2014A,pycia+woodward-2020A}} in the Bayesian paradigm. This results from the significant demand-shading incentives for small quantities in the pay-as-bid auction---where bids for small quantities are paid for all larger quantities---and the significant demand-shading incentives for large quantities in the uniform-price auction---where bids are paid times the quantity for which they are offered. This intuition extends to the loss-averse context, provided restriction is made to cross-conditional regret minimizing bids in the uniform-price auction.

\begin{comparison}[Uniform-price bids steeper than pay-as-bid bids]\label{proposition: uniform price bids are steep}
	Define the average slope of the bid $b$ to be $\alpha = ( b_1 - b_M ) / M$. Cross-conditional regret-minimizing bids in the multi-unit 
	uniform price auction are on average steeper 
	than the unique minimax-loss bid in the pay-as-bid auction: $\alpha^\LAB \geq \alpha^\PAB$.
\end{comparison}

Note that the minimax-loss bid for the last unit in the pay-as-bid auction is identical to the cross-conditional regret minimizing bid for the last unit in the uniform-price auction, $b_{iM} = v_{iM} / ( M + 1 )$. The proof of Comparison~\ref{proposition: semicomparability of bids across auctions} 
shows that $b_{i1}^\LAB > b_{i1}^\PAB$, which completes the proof of Comparison~\ref{proposition: uniform price bids are steep}. The comparisons of the bid functions imply that
the auctioneer's revenues cannot be generically compared across the two auction formats.

\begin{comparison}[Ambiguous revenue]\label{proposition: ex post revenue comparison}
	Depending on the joint value distribution, both ex post and expected revenues can be higher in either multi-unit auction format.
\end{comparison}


Minimax-loss bids do not depend on the distribution of opponent values; the joint value distribution is necessary to compute expected revenue. If the distribution places significant probability on each bidder demanding exactly one unit, the uniform-price auction may yield higher revenue; following Comparison~\ref{theorem:multi unit bid comparison}, the cross-conditional regret minimizing bid in the uniform-price auction is higher than the unique minimax-loss bid in the pay-as-bid auction, and therefore the ex post transfer to the auctioneer can be higher in the uniform-price auction than in the pay-as-bid auction. Similarly, although the uniform-price bid for quantity $Q$ may be above the pay-as-bid bid for quantity $Q$, price discrimination in the pay-as-bid auction may yield a higher transfer to the auctioneer than the uniform payment given cross-conditional regret minimizing bids. In total, when the distribution places significant probability on bidders having zero value, with small probability on demanding the entire market, the pay-as-bid auction will yield higher revenue.

While revenue cannot be ranked across auction formats, bidder loss is uniformly lower in the uniform-price auction than in the pay-as-bid auction. 
The existence of multiple minimax-loss bids in the uniform-price auction does not affect this comparison, because even when some bids are not uniquely defined, there remains a quantity $q_k$ for which conditional regret $R^\LAB_{q_k}$ is equal to maximum loss (Lemma~\ref{lemma: maximum loss as maximum regret}).


\begin{comparison}[Minimax loss]\label{proposition: comparison level of minimax loss}
	In the multi-unit case, minimax loss is lower in the uniform-price auction than in the pay-as-bid auction,
	\begin{equation*}
		\sup_{B^{-i} \in \mathcal B} L^\LAB \left( b^\LAB; B^{-i}, v^i \right) \le \sup_{B^{-i} \in \mathcal B} L^\PAB\left( b^\PAB; B^{-i}, v^i \right).
	\end{equation*}
	This comparison is strict whenever $M > 1$.
\end{comparison}

What are the implications of one mechanism having lower minimax loss than another? Suppose a bidder can obtain costly information about the other bidders' behavior
; this information will shrink the set of possible bid distributions $\mathcal B$. 
The bidder will tend to acquire more information when the subsequent auction mechanism yields higher minimax loss. Thus Comparison~\ref{proposition: comparison level of minimax loss} implies that bidders in the pay-as-bid auction may obtain more costly information than bidders in the uniform-price auction.

\section{Minimax-loss bids in bidpoint-constrained auctions}
\label{section: constrained bidding}

As discussed in the introduction, in practice bidders are frequently constrained from submitting a distinct bid for each unit. For example, bidders can submit up to 10 bidpoints in Czech treasury auctions \citep{Kastl_Restud_2011} or 40 steps in the Texas electricity market \citep{10.1257/aer.20172015}. We now consider the case in which bidder $i$ can submit up to $M$ bid points, $\{ ( q_{ik}, b_{ik} ) \}_{k=1}^M$, where $q_{ik} \leq q_{ik+1}$ and $b_{ik} \geq b_{ik+1}$ for all $k$. The implied bid function is as in the multi-unit case; the only distinction is that the quantities at which bids are submitted are now a choice variable for the bidder.

Optimization over quantities in addition to bid levels allows the bidder to reduce loss below what is feasible in the multi-unit context. Nonetheless, the qualitative results of the multi-unit case remain intact.

\subsection{Pay-as-bid auctions}

As in the multi-unit case, the minimax-loss bid in the bidpoint-constrained pay-as-bid auction equates overbidding regret across all units. This leads immediately to an expression for minimax-loss bids.

\begin{theorem}[Constrained minimax-loss bids in pay-as-bid]\label{theorem: constrained bids in pab}
	The unique minimax-loss bid in the constrained pay-as-bid auction solves
	\begin{align*}
		\left( q^\PAB, b^\PAB \right) &\in \argmin_{q^\prime, b^\prime} \int_0^Q \left( v^i\left( x \right) - \hat b^\prime( q_0 ) \right)_+ dx, \\ 
		&\text{ s.t. } \int_0^{q^\prime_k} \left( \hat b^\prime\left( x \right) - \hat b^\prime\left( q^\prime_k \right) \right) dx+ \int_{q^\prime_k}^{Q} \left( v^i\left( x \right) - \hat b^\prime\left( q^\prime_k \right) \right)_+ dx \\
		&\phantom{\text{ s.t. } \int_0^{q^\prime_k} \left( \hat b^\prime\left( x \right) - \hat b^\prime\left( q^\prime_k \right) \right) dx+ \int_{q^\prime_k}^{Q}} = \int_0^Q \left( v^i\left( x \right) - \hat b^\prime( q_0 ) \right)_+ dx 
	\end{align*}
	for all $k=1,2,\dots,M.$
\end{theorem}

Intuitively, the bidder minimizes their maximum payment subject to equal conditional regret across all outcomes. We illustrate Theorem~\ref{theorem: constrained bids in pab} for the case in which the bidder has constant marginal values.
\begin{example}[Pay-as-bid with constant marginal values]\label{example: PAB with constant marginal values}
	Suppose bidder $i$'s marginal value is constant, $v^i( q ) = v$ for all $q$. The constrained loss optimization problem is
	\[
		\min_{q^\prime, b^\prime} \left( v - b_1^\prime \right) Q, \text{ s.t. } \left( Q - q_{k-1}^\prime \right) \left( v - b_{k}^\prime \right) + \sum_{k^\prime = 1}^{k} \left( q^\prime_{k^\prime} - q^\prime_{k^\prime - 1} \right) \left( b_{k^\prime} - b_k^\prime \right) = \left( v - b_1^\prime \right) Q.
	\]
	Equating conditional loss across units requires $R^\PAB_{q_{k+1}} - R_{q_k}^\PAB = 0$, or
	\[
		0 = -Qb_{k+2}-\left(q_{k+1}-q_{k}\right)v+\left(Q+\left(q_{k+1}-q_{k}\right)\right)b_{k+1}.
	\]
	Solving this equation recursively, backwards from $b_{M+1} = 0$, gives a closed-form expression for optimal bids conditional on quantities,
	\[
		b_{k}=\sum_{k^{\prime}=k}^{M}\frac{Q^{k^{\prime}-k} \left( q_{k^\prime} - q_{k^\prime - 1} \right)}{\prod_{j=k}^{k^{\prime}}\left[Q+ \left( q_j - q_{j - 1} \right) \right]}v.
	\]
	Minimizing loss then implies
	\[
		q_k = \frac{k}{M} Q, \text{ and } b_k = \frac{v}{M+1}\sum_{k^{\prime}=k}^{M}\left[\frac{M}{M+1}\right]^{k^{\prime}-k}.
	\]
	Notably, minimax bidpoints are evenly spaced in the quantity space. Figure~\ref{figure: example constrained bidpoints} plots these bids and compares them to minimax-loss bids in the bidpoint-constrained uniform-price auction.
\end{example}

%
%
%
Providing more qualitative insights on the optimal constrained bid, Example~\ref{example: PAB with single bidpoint} in Appendix~\ref{appendix: derivations of examples} shows that the location of a bidstep need not change if the bidder's value for larger quantities changes, provided that preferences are over discrete units. Moreover, similar to the two-unit example in Section~\ref{section: example}, the bid level can decrease in the bidder's value for larger quantities.

Theorem~\ref{theorem: constrained bids in pab} provides a constrained optimization problem for computing minimax-loss bids in the pay-as-bid auction when the bidder may submit at most $M$ bid points. The optimization problem is stated in terms of divisible goods, and in a multi-unit setting with constrained bids it is possible that the minimax-loss bid has bid points which are away from integer quantities. Practically, the bidder may approximate minimax loss in the bidpoint-constrained auction by rounding the quantities at which bids are submitted down to the nearest feasible unit; this approximation is especially tight when the number of units available is large (hence when the rounding has little effect).

\begin{proposition}[Approximate minimax loss in bidpoint-constrained multi-unit pay-as-bid]\label{proposition: approximate minimax loss pab}
	Suppose that $( q_i, b_i )$ is a minimax-loss bid in the constrained pay-as-bid auction with $M_b$ bid points, and $L^\star$ is minimax loss in the constrained multi-unit pay-as-bid auction with $M_q$ units and $M_b$ bid points. Define a bid $( q_i^\prime, b^\prime_i )$ so that $q^\prime_{ik} = \lfloor M_q q_{ik} / Q \rfloor ( Q / M_q )$ and $b^\prime_{ik} = b_{ik}$. Then $( q_i^\prime, b_i^\prime )$ is feasible in the multi-unit auction, and
	\[
		L^\star \leq L^\PAB\left( \hat b_i^\prime; v^i \right) \leq L^\star + \int_{0}^{\frac{Q}{M_q}} v^i\left( x \right) dx.
	\]
\end{proposition}

Importantly, when $M_q$ is relatively large---that is, when the number of discrete units available is large---the right-hand integral will tend to be small relative to $L^\star$, as the integral is bounded above by $Q v^i( 0 ) / M_q$.


\subsection{Uniform-price auctions}
\label{subsection: bidpoint-constrained upa}

In Section~\ref{subsection: multi-unit lab} we showed that there are typically many minimax-loss bids in the multi-unit uniform-price auction. We show below that this is in stark contrast to the bidpoint-constrained uniform-price auction, where there is a unique minimax-loss bid. 
As there is a unique minimax bid when the location of the bid steps can be chosen, nonuniqueness in the multi-unit case derives from the prespecified location of bid points in the multi-unit auction.




\begin{theorem}[Minimax-loss bids in constrained uniform-price auction]\label{theorem: unique loss-minimizing bids in constrained upa}
	In the bidpoint-constrained 
	uniform-price auction with $M$ bid points, the unique minimax-loss bid solves
	\begin{align*}
		\left( q^\UPA, b^\UPA \right) \in & \min_{q^\prime, b^\prime} R, \\
		&\text{ s.t. } q^\prime_k b^\prime_k = R \;\;\; \forall k \in \left\{ 1, \ldots, M \right\}, \\
		& \text{ and } \int_{q^\prime_{k-1}}^{Q} \left( v^i\left( x \right) - b^\prime_k \right)_+ dx = R \;\;\; \forall k \in \left\{ 1, \ldots, M \right\}.
	\end{align*}
\end{theorem}

We provide some intuition for the uniqueness in the constrained case and contrast it with the multiplicity of the multi-unit case. Intuitively, when the bidder receives a small quantity they do not leave a lot of money on the table due to overbidding, because they received a small number of units and their total payment is low; they also do not miss out on significant utility from underbidding, because the market price will tend to be high and they will not desire many units at this price. Thus the main source of loss is bids on intermediate quantities, leaving bids on small (and very large) quantities only partially specified. This stands in contrast to the bidpoint-constrained case where the locations of the bid steps are choice variables. Given the choice, the bidder will submit relatively dense bids for intermediate quantities and relatively sparse bids for extreme quantities; the large gaps between bid points for small units work against the intuition arising from the multi-unit case, where bidpoint gaps are uniform, that bids for small quantities are not uniquely determined.

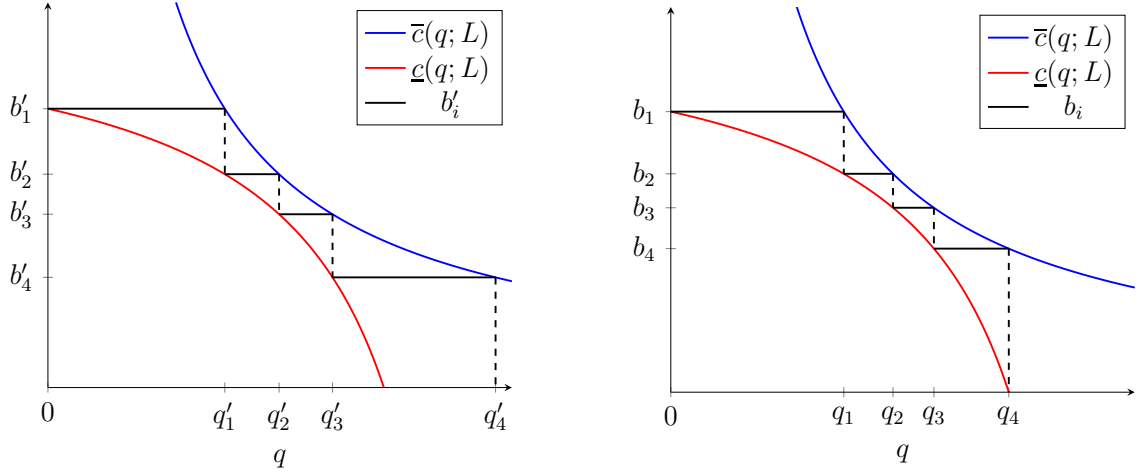
\begin{figure}
\centering
\begin{subfigure}{.5\textwidth}
  \centering
  	\begin{tikzpicture}[scale=.9]
		 \begin{axis}[
		 	samples=200, 
		 	domain=0:1,
		 	axis lines = left,
    		xlabel = \(q\),
    		xtick={0, 0.381215, 0.498227, 0.613401, 0.964758},
    		ytick={0.724, 0.553964, 0.44995, 0.286082},
    		ymax=1,
    		xticklabels={$0$,$q_1'$, $q_2'$, $q_3'$, $q_4'$},
    		yticklabels={$b_1'$, $b_2'$, $b_3'$, $b_4'$}
		 ]
		    \addplot [mark=none, thick, color=blue, domain=0.276:1] { 0.276/x };
		    \addlegendentry{\(\overline c(q;L)\)}
		    
    		\addplot [mark=none, thick, color=red, domain=0:0.724] { (-0.724 + x)/(-1 + x) };
    		\addlegendentry{\(\underline c(q;L)\)}
    		
    		\addplot [mark=none, thick, color=black, domain=0:0.381215] { 0.724 };
    		\addlegendentry{\(b_i'\)}

    		\draw [dashed, thick] (0.381215,.724)--(0.381215, 0.553964);
    		\draw [thick] (0.381215, 0.553964)--(0.498227, 0.553964);
    		\draw [dashed, thick] (0.498227, 0.553964)--(0.498227, 0.44995);
    		\draw [thick] (0.498227, 0.44995)--(0.613401,0.44995);
    		\draw [dashed, thick] (0.613401,0.44995)--(0.613401, 0.286082);
    		\draw [thick] (0.613401, 0.286082)--(0.964758,0.286082);
    		\draw [dashed, thick] (0.964758,0.286082)--(0.964758,0);
		\end{axis}
	\end{tikzpicture}
  \caption{Loss $L$ above constrained minimax loss}
  \label{fig: upa loss above minimax}
\end{subfigure}%
\begin{subfigure}{.5\textwidth}
  \centering
  	\begin{tikzpicture}[scale=.9]
		 \begin{axis}[
		 	samples=200, 
		 	domain=0:1,
		 	axis lines = left,
    		xlabel = \(q\),
    		xtick={0, 0.372786, 0.478891, 0.567047, 0.728446},
    		ytick={0.728446, 0.567047, 0.478891, 0.372786},
    		ymax=1,
    		xticklabels={$0$,$q_1$, $q_2$, $q_3$, $q_4$},
    		yticklabels={$b_1$, $b_2$, $b_3$, $b_4$}
		 ]
		    \addplot [mark=none, thick, color=blue, domain=0.27:1] { 0.271554/x };
		    \addlegendentry{\(\overline c(q;L)\)}
		    
    		\addplot [mark=none, thick, color=red, domain=0:0.728446] { (-0.728446 + x)/(x-1)};
    		\addlegendentry{\(\underline c(q;L)\)}

    		\addplot [mark=none, thick, color=black, domain=0:0.372786] {0.728446};
    		\addlegendentry{\(b_i\)}

    		\draw [dashed, thick] (0.372786,0.728446)-- (0.372786,0.567047);
    		\draw [thick] (0.372786,0.567047)--(0.478891,0.567047);
    		\draw [dashed, thick] (0.478891,0.567047)-- (0.478891,0.478891);
    		\draw [thick] (0.478891,0.478891)--(0.567047,0.478891);
    		\draw [dashed, thick] (0.567047,0.478891)--(0.567047,0.372786);
    		\draw [thick] (0.567047,0.372786)--(0.728446,0.372786);
    		\draw [dashed, thick] (0.728446,0.372786)--(0.728446,0);
		\end{axis}
	\end{tikzpicture}
  \caption{Loss $L$ equal to constrained minimax loss}
  \label{fig: constrained M = 4 upa}
\end{subfigure}
\caption{Iso-loss curves of conditional underbidding and overbidding regret in the uniform-price auction.}
\label{fig:iso loss upa}
\end{figure}

The chief distinction between the multi-unit and constrained-bid cases is that in the constrained-bid case the spacing of bid points is an additional tool for reducing ex post regret. The construction of the minimax-loss bid in the constrained uniform-price auction follows from observing that steps in the implied bid function extend between two \emph{iso-loss curves}. Given loss $L$, the upper iso-loss curve is $\overline c( \cdot; L )$ such that $q \overline c( q; L ) = L$, and the lower iso-loss curve is $\underline c( \cdot; L )$ such that $\int_q^Q ( v^i( x ) - \underline c( q; L ) )_+ dx = L$. The bid $b( q ) = \overline c( q; L )$ induces overbidding loss which is constant in quantity, and the bid $b( q ) = \underline c( q; L )$ induces underbidding loss which is constant in quantity.\footnote{The same logic does not apply to the pay-as-bid auction, since overbidding regret is monotonically increasing in quantity.} Figure \ref{fig:iso loss upa} illustrates the two iso-loss curves. The upper iso-loss curve is always a hyperbola; the lower loss curve depends on marginal values.

Bids above the upper iso-loss curve induce loss above $L$ by inducing overbidding regret above $L$, and bids below the lower iso-loss curve induce loss above $L$ by inducing underbidding regret above $L$. It follows that the minimax-loss bid must lie entirely between the upper and lower iso-loss curves. In particular, the minimax-loss bid in the constrained uniform-price auction extends from the lower iso-loss curve to the upper iso-loss curve, then jumps down to the lower iso-loss curve, and extends again to the upper iso-loss curve; this continues until a bid of zero is reached. Figure \ref{fig: constrained M = 4 upa} illustrates this construction for $M=4$. If the bid did not extend fully between the two iso-loss curves, with a slight perturbation the bid could be made to lie strictly between the two iso-loss curves, which would entail strictly lower loss.

Constructing bidpoint-constrained minimax-loss bids is straightforward. For loss $L$ such that $\overline c( \cdot; L ) \geq \underline c( \cdot; L )$, let $q_0 = 0$ and for all $k \in \{ 1, \ldots, M \}$ let $b_k = \underline c( q_{k-1}; L )$ and let $q_k$ be such that $\overline c( q_k; L ) = b_{k}$.\footnote{In the event that $\overline c( Q; L ) > b_k$, we define $q_k = Q$.} If $\underline c( q_M; L ) > 0$ constrained minimax loss is above $L$, and if $\underline c( q_M; L ) < 0$ constrained minimax loss is below $L$. In either case, a new level of loss $L^\prime$ may be proposed, and the procedure continues until $\underline c( q_M; L ) = 0$ (or is within numerical tolerance). Figure \ref{fig: upa loss above minimax} illustrates the case when the level of loss is above the minimax loss. In the Figure, the final step $q_4'$ is too high, and loss can be decreased.

The construction of minimax-loss bids between the upper and lower iso-loss curves provides an intuitive argument for the uniqueness of minimax-loss bids in the uniform-price auction. Given a level of loss and associated iso-loss curves, either there is no $M$-step step function between them, or there is a single $M$-step step function between them, or there are multiple such step functions between them. If there is no feasible step function between the iso-loss curves, this level of loss is not feasible and minimax loss is above the assumed loss. On the other hand, if there are multiple feasible step functions between the iso-loss curves the iso-loss curves can be brought closer together (by reducing assumed loss) while still allowing for a feasible step function between them. This improvement in loss is infeasible only when there is a unique step function between the iso-loss curves, and at that point maximum loss is minimized.

The following example illustrates Theorem~\ref{theorem: unique loss-minimizing bids in constrained upa} for the case in which the bidder has constant marginal values.\footnote{An additional example is included in Appendix~\ref{appendix: derivations of examples}.}

\begin{example}[Uniform-price with constant marginal values]\label{example: upa with constant marginal values}
	Suppose that bidder $i$'s marginal value $v^i$ is constant, $v^i( q ) = v$ for all $q$. The constrained loss optimization problem is
	\[
		\min_{q^\prime, b^\prime} b^\prime_1 q^\prime_1, \text{ s.t. } b^\prime_k q^\prime_k = \left( v - b^\prime_k \right) \left( Q - q^\prime_k \right) \;\; \forall k.
	\]
	The minimax-loss bid induces loss $C_M Q v$, and solves
	\[
		q_0 = 0, \;\; q_{k} = \left( C_M - \frac{C_M^2}{q_{k-1} - \left( 1 - C_M \right)} \right) Q, \;\; q_M = \left( 1 - C_M \right) Q, \; \text{ and } b_k = \frac{C_M v}{q_k}.
	\]
	The solution to this expression is unique: the recursive equation for $q_k$ increases in $C_M$, while the endpoint condition for $q_M$ decreases in $C_M$.\footnote{In the unconstrained model (Appendix~\ref{appendix: unconstrained upa}) minimax loss is $v Q / 4$. Since loss is higher when bids are constrained than when they are unconstrained, it follows that $q_1 \geq Q / 3$ and $q_M \leq 3 Q / 4$. That is, minimax-loss bid points are all for interior quantities.} Figure~\ref{figure: example constrained bidpoints} illustrates these bids and compares them to the unique minimax-loss bids in the pay-as-bid auction.
	
\begin{figure}
\centering
\begin{tikzpicture}[scale = 0.5]
			\begin{axis}[
				xlabel = {$q$}, 
				ylabel = {$b, v$},
				title = {Minimax-loss bids in bidpoint-constrained auctions},
				legend style = {
					at = { ( 0.95, 0.95 ) },
					anchor =north east
				}
			]
\addplot [draw = none, fill = black!10, forget plot] table {
0.0	1.1
0.34	1.1
0.35000000000000003	1.0913314608790512
0.36	1.0610166980768554
0.37	1.0323405711018052
0.38	1.0051737139675472
0.39	0.9794000289940203
0.4	0.9549150282691699
0.41000000000000003	0.9316244178235803
0.42	0.909442884065876
0.43	0.8882930495527162
0.44	0.8681045711537908
0.45	0.8488133584614843
0.46	0.8303608941471042
0.47000000000000003	0.8126936410801445
0.48	0.7957625235576415
0.49	0.7795224720564652
0.5	0.7639320226153359
0.51	0.7489529633483685
0.52	0.7345500217455152
0.53	0.7206905873729583
0.54	0.7073444653845702
0.55	0.6944836569230326
0.56	0.6820821630494069
0.5700000000000001	0.6701158093116981
0.58	0.6585620884614964
0.59	0.6474000191655389
0.6	0.6366100188461132
0.61	0.6261737890289638
0.62	0.6160742117865612
0.63	0.6062952560439173
0.64	0.5968218926682312
0.65	0.5876400173964123
0.66	0.5787363807691939
0.67	0.5700985243398029
0.68	0.5617147225112763
0.6900000000000001	0.5535739294314027
0.7000000000000001	0.5456657304395256
0.71	0.5379802976164337
0.72	0.5305083490384277
0.73	0.5232411113803671
0.74	0.5161702855509026
0.75	0.5092880150768906
0.76	0.5025868569837736
0.77	0.4960597549450233
0.78	0.48970001449701017
0.79	0.4835012801362885
0.8	0.4774575141345849
0.81	0.47156297692304683
0.8200000000000001	0.46581220891179015
0.8300000000000001	0.4602000136236963
0.84	0.454721442032938
0.85	0.4493717780090211
0.86	0.4441465247763581
0.87	0.4390413923076643
0.88	0.4340522855768954
0.89	0.42917529360412127
0.9	0.42440667923074216
0.91	0.41974286956886586
0.92	0.4151804470735521
0.93	0.4107161411910408
0.9400000000000001	0.40634682054007226
0.9500000000000001	0.4020694855870189
0.96	0.39788126177882077
0.97	0.3937793931006886
0.98	0.3897612360282326
0.99	0.38582425384612923
1.0	0.38196601130766794
1.0	0.0
0.62	0.0
0.61	0.020599971005979656
0.6	0.04508497173083015
0.59	0.0683755821764197
0.58	0.09055711593412408
0.5700000000000001	0.11170695044728374
0.56	0.13189542884620908
0.55	0.15118664153851558
0.54	0.16963910585289577
0.53	0.18730635891985536
0.52	0.20423747644235846
0.51	0.2204775279435348
0.5	0.23606797738466412
0.49	0.2510470366516315
0.48	0.2654499782544848
0.47000000000000003	0.2793094126270417
0.46	0.2926555346154298
0.45	0.3055163430769674
0.44	0.31791783695059306
0.43	0.3298841906883019
0.42	0.34143791153850367
0.41000000000000003	0.3525999808344611
0.4	0.36338998115388677
0.39	0.3738262109710362
0.38	0.3839257882134388
0.37	0.3937047439560827
0.36	0.40317810733176884
0.35000000000000003	0.41235998260358775
0.34	0.42126361923080613
0.33	0.4299014756601971
0.32	0.4382852774887236
0.31	0.4464260705685972
0.3	0.4543342695604743
0.29	0.46201970238356627
0.28	0.4694916509615723
0.27	0.4767588886196329
0.26	0.4838297144490974
0.25	0.4907119849231094
0.24	0.4974131430162264
0.23	0.5039402450549767
0.22	0.5102999855029898
0.21	0.5164987198637114
0.2	0.5225424858654151
0.19	0.5284370230769532
0.18	0.5341877910882098
0.17	0.5397999863763037
0.16	0.545278557967062
0.15	0.5506282219909788
0.14	0.5558534752236419
0.13	0.5609586076923356
0.12	0.5659477144231047
0.11	0.5708247063958787
0.1	0.5755933207692578
0.09	0.5802571304311341
0.08	0.5848195529264479
0.07	0.5892838588089592
0.06	0.5936531794599277
0.05	0.5979305144129812
0.04	0.6021187382211792
0.03	0.6062206068993115
0.02	0.6102387639717675
0.01	0.6141757461538708
0.0	0.6180339886923321
} -- cycle;
			\addplot [ thick, black ] table {
			0.0	1.0
			1.0	1.0
			};
			\addlegendentry{$v$}
			\addplot [thick, blue] table {
0.0	0.5
1.0	0.5
};
\addlegendentry{$b^\PAB$}
\addplot [thick, red] table {
0	0.6180339886923321
0.6180339889005961	0.6180339886923321
0.6180339889005961	-5.452425178020803e-10
};
\addlegendentry{$b^\LAB$}
			\end{axis}
\end{tikzpicture}
\hspace{0.5cm}
\begin{tikzpicture}[scale = 0.5]
			\begin{axis}[
				xlabel = {$q$}, 
				ylabel = {$b, v$},
				title = {Minimax-loss bids in bidpoint-constrained auctions},
				legend style = {
					at = { ( 0.95, 0.95 ) },
					anchor = north east
				}
			]\addplot [draw = none, fill = black!10, forget plot] table {
0.0	1.1
0.27	1.1
0.28	1.0999233156326227
0.29	1.0619949254383945
0.3	1.0265950945904478
0.31	0.9934791237972076
0.32	0.9624329011785449
0.33	0.933268267809498
0.34	0.9058192011092187
0.35000000000000003	0.879938652506098
0.36	0.8554959121587066
0.37	0.8323744010192821
0.38	0.8104698115187746
0.39	0.7896885343003445
0.4	0.7699463209428359
0.41000000000000003	0.7511671423832544
0.42	0.7332822104217485
0.43	0.7162291357607776
0.44	0.6999512008571235
0.45	0.6843967297269652
0.46	0.6695185399502921
0.47000000000000003	0.6552734646322007
0.48	0.6416219341190299
0.49	0.6285276089329273
0.5	0.6159570567542687
0.51	0.6038794674061457
0.52	0.5922664007252584
0.53	0.5810915629757252
0.54	0.5703306081058043
0.55	0.5599609606856988
0.56	0.5499616578163113
0.5700000000000001	0.540313207679183
0.58	0.5309974627191972
0.59	0.5219975057239565
0.6	0.5132975472952239
0.61	0.5048828334051383
0.62	0.4967395618986038
0.63	0.4888548069478323
0.64	0.4812164505892724
0.65	0.4738131205802067
0.66	0.466634133904749
0.67	0.45966944533900644
0.68	0.4529096005546093
0.6900000000000001	0.4463456933001947
0.7000000000000001	0.439969326253049
0.71	0.4337725751790625
0.72	0.4277479560793533
0.73	0.42188839503717035
0.74	0.41618720050964103
0.75	0.41063803783617914
0.76	0.4052349057593873
0.77	0.39997211477549915
0.78	0.39484426715017223
0.79	0.3898462384520688
0.8	0.38497316047141794
0.81	0.38022040540386953
0.8200000000000001	0.3755835711916272
0.8300000000000001	0.37105846792425823
0.84	0.36664110521087423
0.85	0.36232768044368746
0.86	0.3581145678803888
0.87	0.3539983084794648
0.88	0.34997560042856174
0.89	0.3460432903113869
0.9	0.3421983648634826
0.91	0.3384379432715762
0.92	0.33475926997514605
0.93	0.3311597079324025
0.9400000000000001	0.32763673231610035
0.9500000000000001	0.32418792460750984
0.96	0.32081096705951495
0.97	0.3175036375022004
0.98	0.3142638044664636
0.99	0.311089422603166
1.0	0.30797852837713435
1.0	0.0
0.7000000000000001	0.0
0.6900000000000001	0.006520876202792181
0.68	0.037567098821455036
0.67	0.06673173219050188
0.66	0.09418079889078124
0.65	0.12006134749390174
0.64	0.14450408784129343
0.63	0.16762559898071794
0.62	0.18953018848122538
0.61	0.21031146569965553
0.6	0.23005367905716412
0.59	0.24883285761674556
0.58	0.26671778957825165
0.5700000000000001	0.2837708642392224
0.56	0.3000487991428764
0.55	0.3156032702730347
0.54	0.3304814600497079
0.53	0.3447265353677992
0.52	0.3583780658809701
0.51	0.37147239106707275
0.5	0.3840429432457313
0.49	0.39612053259385427
0.48	0.4077335992747416
0.47000000000000003	0.4189084370242748
0.46	0.42966939189419573
0.45	0.4400390393143012
0.44	0.45003834218368866
0.43	0.459686792320817
0.42	0.4690025372808029
0.41000000000000003	0.47800249427604347
0.4	0.4867024527047761
0.39	0.4951171665948617
0.38	0.5032604381013962
0.37	0.5111451930521678
0.36	0.5187835494107276
0.35000000000000003	0.5261868794197933
0.34	0.5333658660952509
0.33	0.5403305546609934
0.32	0.5470903994453906
0.31	0.5536543066998052
0.3	0.5600306737469509
0.29	0.5662274248209376
0.28	0.5722520439206467
0.27	0.5781116049628297
0.26	0.583812799490359
0.25	0.5893619621638209
0.24	0.5947650942406126
0.23	0.6000278852245009
0.22	0.6051557328498278
0.21	0.6101537615479312
0.2	0.6150268395285821
0.19	0.6197795945961304
0.18	0.6244164288083728
0.17	0.6289415320757417
0.16	0.6333588947891258
0.15	0.6376723195563125
0.14	0.6418854321196112
0.13	0.6460016915205352
0.12	0.6500243995714383
0.11	0.6539567096886131
0.1	0.6578016351365175
0.09	0.6615620567284237
0.08	0.665240730024854
0.07	0.6688402920675974
0.06	0.6723632676838995
0.05	0.6758120753924901
0.04	0.679189032940485
0.03	0.6824963624977995
0.02	0.6857361955335364
0.01	0.688910577396834
0.0	0.6920214716228656
} -- cycle;
			\addplot [ thick, black ] table {
			0.0	1.0
			1.0	1.0
			};
			\addlegendentry{$v$}
			\addplot [thick, blue] table {
0.0	0.5555555555555556
0.5	0.5555555555555556
0.5	0.3333333333333333
1.0	0.3333333333333333
};
\addlegendentry{$b^\PAB$};
\addplot [thick, red] table {
0	0.6920214716228656
0.44504186792772654	0.6920214716228656
0.44504186792772654	0.4450418678845026
0.6920214716900771	0.4450418678845026
0.6920214716900771	-2.1823431950451777e-10
};
\addlegendentry{$b^\LAB$};
			\end{axis}
\end{tikzpicture}
\hspace{0.5cm}
\begin{tikzpicture}[scale = 0.5]
			\begin{axis}[
				xlabel = {$q$}, 
				ylabel = {$b, v$},
				title = {Minimax-loss bids in bidpoint-constrained auctions},
				legend style = {
					at = { ( 0.95, 0.95 ) },
					anchor = north east
				}
			]
			
\addplot [draw = none, fill = black!10, forget plot] table {
0.0	1.1
0.24	1.1
0.25	1.0607513368595392
0.26	1.0199532085187877
0.27	0.9821771637588325
0.28	0.9470994079103027
0.29	0.9144408076375338
0.3	0.8839594473829493
0.31	0.8554446264996284
0.32	0.828711981921515
0.33	0.8035994976208629
0.34	0.7799642182790728
0.35000000000000003	0.7576795263282422
0.36	0.7366328728191245
0.37	0.7167238762564454
0.38	0.6978627216181179
0.39	0.6799688056791917
0.4	0.662969585537212
0.41000000000000003	0.6467995956460604
0.42	0.6313996052735352
0.43	0.6167158935229878
0.44	0.6026996232156473
0.45	0.5893062982552996
0.46	0.5764952917714886
0.47000000000000003	0.5642294344997548
0.48	0.5524746546143433
0.49	0.5411996616630302
0.5	0.5303756684297696
0.51	0.519976145519382
0.52	0.5099766042593938
0.53	0.5003544041790279
0.54	0.49108858187941623
0.55	0.4821596985725178
0.56	0.47354970395515134
0.5700000000000001	0.4652418144120785
0.58	0.4572204038187669
0.59	0.4494709054489573
0.6	0.44197972369147465
0.61	0.4347341544506308
0.62	0.4277223132498142
0.63	0.4209330701823568
0.64	0.4143559909607575
0.65	0.40798128340751505
0.66	0.40179974881043146
0.67	0.39580273763415635
0.68	0.3899821091395364
0.6900000000000001	0.38433019451432576
0.7000000000000001	0.3788397631641211
0.71	0.37350399185195043
0.72	0.36831643640956224
0.73	0.36327100577381477
0.74	0.3583619381282227
0.75	0.3535837789531797
0.76	0.34893136080905895
0.77	0.34439978469465554
0.78	0.33998440283959586
0.79	0.3356808028036516
0.8	0.331484792768606
0.81	0.32739238791961084
0.8200000000000001	0.3233997978230302
0.8300000000000001	0.3195034147167286
0.84	0.3156998026367676
0.85	0.3119856873116292
0.86	0.3083579467614939
0.87	0.3048136025458446
0.88	0.30134981160782365
0.89	0.29796385866840985
0.9	0.2946531491276498
0.91	0.2914152024339393
0.92	0.2882476458857443
0.93	0.28514820883320946
0.9400000000000001	0.2821147172498774
0.9500000000000001	0.27914508864724713
0.96	0.27623732730717165
0.97	0.2733895198091596
0.98	0.2705998308315151
0.99	0.2678664992069543
1.0	0.2651878342148848
1.0	0.0
0.74	0.0
0.73	0.01782283624116754
0.72	0.05290059208969733
0.71	0.08555919236246634
0.7000000000000001	0.11604055261705049
0.6900000000000001	0.1445553735003715
0.68	0.17128801807848493
0.67	0.19640050237913687
0.66	0.22003578172092697
0.65	0.24232047367175769
0.64	0.2633671271808755
0.63	0.2832761237435546
0.62	0.3021372783818821
0.61	0.3200311943208083
0.6	0.33703041446278803
0.59	0.3532004043539396
0.58	0.3686003947264649
0.5700000000000001	0.38328410647701205
0.56	0.3973003767843527
0.55	0.41069370174470043
0.54	0.4235047082285113
0.53	0.43577056550024507
0.52	0.4475253453856567
0.51	0.45880033833696976
0.5	0.4696243315702304
0.49	0.480023854480618
0.48	0.49002339574060616
0.47000000000000003	0.4996455958209721
0.46	0.5089114181205838
0.45	0.5178403014274822
0.44	0.5264502960448487
0.43	0.5347581855879215
0.42	0.5427795961812332
0.41000000000000003	0.5505290945510427
0.4	0.5580202763085254
0.39	0.5652658455493692
0.38	0.5722776867501858
0.37	0.5790669298176432
0.36	0.5856440090392425
0.35000000000000003	0.5920187165924848
0.34	0.5982002511895684
0.33	0.6041972623658436
0.32	0.6100178908604634
0.31	0.6156698054856742
0.3	0.6211602368358788
0.29	0.6264960081480495
0.28	0.6316835635904378
0.27	0.6367289942261852
0.26	0.6416380618717773
0.25	0.6464162210468203
0.24	0.651068639190941
0.23	0.6556002153053444
0.22	0.6600155971604041
0.21	0.6643191971963485
0.2	0.668515207231394
0.19	0.6726076120803892
0.18	0.6766002021769698
0.17	0.6804965852832714
0.16	0.6843001973632323
0.15	0.6880143126883709
0.14	0.6916420532385061
0.13	0.6951863974541554
0.12	0.6986501883921763
0.11	0.7020361413315901
0.1	0.7053468508723502
0.09	0.7085847975660606
0.08	0.7117523541142556
0.07	0.7148517911667905
0.06	0.7178852827501225
0.05	0.7208549113527529
0.04	0.7237626726928283
0.03	0.7266104801908404
0.02	0.7294001691684848
0.01	0.7321335007930456
0.0	0.7348121657851152
} -- cycle;
			\addplot [ thick, black ] table {
			0.0	1.0
			1.0	1.0
			};
			\addlegendentry{$v$}
			\addplot [thick, blue] table {
0.0	0.5981224279835392
0.2	0.5981224279835392
0.2	0.517746913580247
0.4	0.517746913580247
0.4	0.4212962962962964
0.6000000000000001	0.4212962962962964
0.6000000000000001	0.30555555555555564
0.8	0.30555555555555564
0.8	0.1666666666666667
1.0	0.1666666666666667
};
\addlegendentry{$b^\PAB$};
\addplot [thick, red] table {
0	0.7348121657851152
0.36089200283114936	0.7348121657851152
0.36089200283114936	0.5850656925126492
0.4532616381521147	0.5850656925126492
0.4532616381521147	0.5149639156129566
0.5149639152856648	0.5149639156129566
0.5149639152856648	0.45326163852104184
0.5850656920364417	0.45326163852104184
0.5850656920364417	0.36089200356463425
0.7348121642916667	0.36089200356463425
0.7348121642916667	5.63166280276306e-09
};
\addlegendentry{$b^\LAB$};
			\end{axis}
\end{tikzpicture}
\caption{\label{figure: example constrained bidpoints}Minimax-loss bids under constant marginal values, when bidders are constrained to $M \in \{ 1, 2, 5 \}$ bidpoints. In bidpojnt-constrained auctions, minimax-loss bids are unique in both the pay-as-bid and uniform-price auctions. As the number of bidpoints increases, the upper and lower iso-loss curves in the uniform-price auction approach tagency.}
\end{figure}
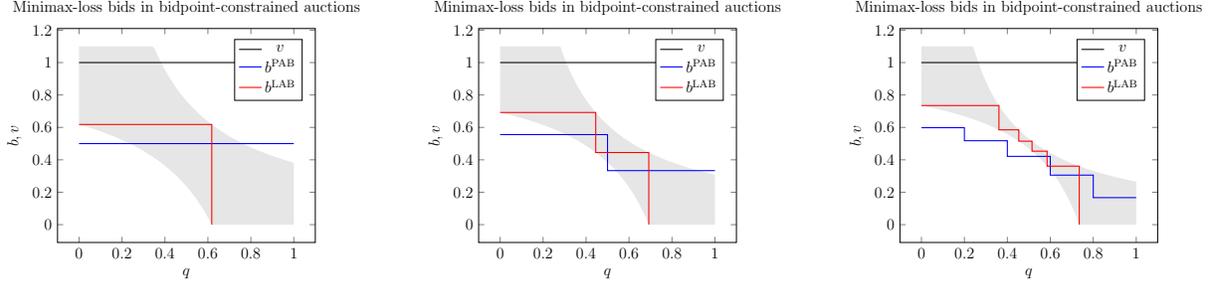

\end{example}

Examples~\ref{example: PAB with constant marginal values} and~\ref{example: upa with constant marginal values} suggest a new testable prediction. With constant marginal values, the bids in the bidpoint-constrained pay-as-bid are evenly spaced, while they are more clustered around intermediate quantities in the bidpoint-constrained uniform-price auction. More generally, the location of the bids in the pay-as-bid auction is more dispersed than in the uniform-price auction.

%
%

As is the case in the bidpoint-constrained pay-as-bid auction, the unique minimax-loss bid in the bidpoint-constrained uniform-price auction may be rounded to an approximate minimax-loss bid in the constrained multi-unit uniform-price auction. In the uniform-price auction this approximation is guaranteed by rounding bid points upward to the nearest feasible quantity, which differs from the pay-as-bid auction in which quantities are rounded down.

\begin{figure}
 	\begin{center}
 		\begin{tikzpicture}
		\begin{axis}[
		axis lines = left,
		xlabel = \( q \),
		ylabel = {$\hat b^\LAB$},
		xtick = { 0.0, {.333333333333}, 0.5, 0.75, 1.0 },
		xticklabels = { $0$, $\frac{1}{3}$, $\frac{1}{2}$, $\frac{3}{4}$, $1$ },
		ytick={ 0.0, 0.5, 1.0 },
		yticklabels={ $0$, $\frac{1}{2}$, $1$ },
		legend style = {at ={ ( 0.03, 0.03 ) }, anchor = south west, legend columns = 2},
		xmin = 0,
		xmax = 1.05,
		ymin = 0,
		ymax = 1.05
		]
	     	
	     	\draw[solid, thick] ( 0.00, 1.00 ) -- ( 1.00, 1.00 );
	     	\draw[dotted, thick] ( 1.00, 1.00 ) -- ( 1.00, 0.00 );

		\tikzset{M1/.style = { color = blue!100!red, thick }};
		\tikzset{M2/.style = { color = blue!067!red, thick }};
		\tikzset{M5/.style = { color = blue!033!red, thick }};
		\tikzset{M10/.style = { color = blue!000!red, thick }};
	     	
	     	%
	     	%
		\draw[solid, M1] ( 0.00000, 0.61803 ) -- ( 0.61803, 0.61803 );
		\draw[dashed, M1] ( 0.61803, 0.00000 ) -- ( 0.61803, 0.61803 );
		\draw[dotted, smooth, samples = 100, domain = 0.38197 : 1.00000, M1, thin] plot ( \x, { 0.381966 / \x } );
		\draw[dotted, smooth, samples = 100, domain = 0 : 0.61803, M1, thin] plot ( \x, {1 - 0.381966 / ( 1 - \x )} );
		
		%
		%
		\draw[solid, M2] ( 0.00000, 0.69202 ) -- ( 0.44504, 0.69202 );
		\draw[solid, M2] ( 0.44504, 0.44504 ) -- ( 0.69202, 0.44504 );
		\draw[dashed, M2] ( 0.69202, 0.00000 ) -- ( 0.69202, 0.44504 );
		\draw[dashed, M2] ( 0.44504, 0.44504 ) -- ( 0.44504, 0.69202 );
		\draw[dotted, smooth, samples = 100, domain = 0.30798 : 1.00000, M2, thin] plot ( \x, { 0.307979 / \x } );
		\draw[dotted, smooth, samples = 100, domain = 0 : 0.69202, M2, thin] plot ( \x, {1 - 0.307979 / ( 1 - \x )} );

		%
		%

		%
		%
		\draw[solid, M5] ( 0.00000, 0.73481 ) -- ( 0.36089, 0.73481 );
		\draw[solid, M5] ( 0.36089, 0.58507 ) -- ( 0.45326, 0.58507 );
		\draw[solid, M5] ( 0.45326, 0.51496 ) -- ( 0.51496, 0.51496 );
		\draw[solid, M5] ( 0.51496, 0.45326 ) -- ( 0.58507, 0.45326 );
		\draw[solid, M5] ( 0.58507, 0.36089 ) -- ( 0.73481, 0.36089 );
		\draw[dashed, M5] ( 0.73481, 0.00000 ) -- ( 0.73481, 0.36089 );
		\draw[dashed, M5] ( 0.58507, 0.36089 ) -- ( 0.58507, 0.45326 );
		\draw[dashed, M5] ( 0.51496, 0.45326 ) -- ( 0.51496, 0.51496 );
		\draw[dashed, M5] ( 0.45326, 0.51496 ) -- ( 0.45326, 0.58507 );
		\draw[dashed, M5] ( 0.36089, 0.58507 ) -- ( 0.36089, 0.73481 );
		\draw[dotted, smooth, samples = 100, domain = 0.26519 : 1.00000, M5, thin] plot ( \x, { 0.265188 / \x } );
		\draw[dotted, smooth, samples = 100, domain = 0 : 0.73481, M5, thin] plot ( \x, {1 - 0.265188 / ( 1 - \x )} );

		%
		%
		\draw[solid, M10] ( 0.00000, 0.74528 ) -- ( 0.34178, 0.74528 );
		\draw[solid, M10] ( 0.34178, 0.61301 ) -- ( 0.41553, 0.61301 );
		\draw[solid, M10] ( 0.41553, 0.56418 ) -- ( 0.45149, 0.56418 );
		\draw[solid, M10] ( 0.45149, 0.53561 ) -- ( 0.47558, 0.53561 );
		\draw[solid, M10] ( 0.47558, 0.51428 ) -- ( 0.49530, 0.51428 );
		\draw[solid, M10] ( 0.49530, 0.49530 ) -- ( 0.51428, 0.49530 );
		\draw[solid, M10] ( 0.51428, 0.47558 ) -- ( 0.53561, 0.47558 );
		\draw[solid, M10] ( 0.53561, 0.45149 ) -- ( 0.56418, 0.45149 );
		\draw[solid, M10] ( 0.56418, 0.41553 ) -- ( 0.61301, 0.41553 );
		\draw[solid, M10] ( 0.61301, 0.34178 ) -- ( 0.74528, 0.34178 );
		\draw[dashed, M10] ( 0.74528, 0.00000 ) -- ( 0.74528, 0.34178 );
		\draw[dashed, M10] ( 0.61301, 0.34178 ) -- ( 0.61301, 0.41553 );
		\draw[dashed, M10] ( 0.56418, 0.41553 ) -- ( 0.56418, 0.45149 );
		\draw[dashed, M10] ( 0.53561, 0.45149 ) -- ( 0.53561, 0.47558 );
		\draw[dashed, M10] ( 0.51428, 0.47558 ) -- ( 0.51428, 0.49530 );
		\draw[dashed, M10] ( 0.49530, 0.49530 ) -- ( 0.49530, 0.51428 );
		\draw[dashed, M10] ( 0.47558, 0.51428 ) -- ( 0.47558, 0.53561 );
		\draw[dashed, M10] ( 0.45149, 0.53561 ) -- ( 0.45149, 0.56418 );
		\draw[dashed, M10] ( 0.41553, 0.56418 ) -- ( 0.41553, 0.61301 );
		\draw[dashed, M10] ( 0.34178, 0.61301 ) -- ( 0.34178, 0.74528 );
		\draw[dotted, smooth, samples = 100, domain = 0.25472 : 1.00000, M10, thin] plot ( \x, { 0.254723 / \x } );
		\draw[dotted, smooth, samples = 100, domain = 0 : 0.74528, M10, thin] plot ( \x, {1 - 0.254723 / ( 1 - \x )} );

		%

	     	\end{axis}
	     \end{tikzpicture}

 	\end{center}
 	\caption[Minimax-loss bids in the constrained uniform-price auction with varying numbers of bid points.]{Minimax-loss bids in the constrained uniform-price auction with $Q = 1$, $v = 1$, and varying numbers of bid points.\footnotemark\ Submitted bids lie on upper and lower iso-loss curves, and more-central iso-loss curves (a lower upper iso-loss curve and a higher lower iso-loss curve) correspond to lower loss.}
 	\label{figure: constrained upa bids}
\end{figure}
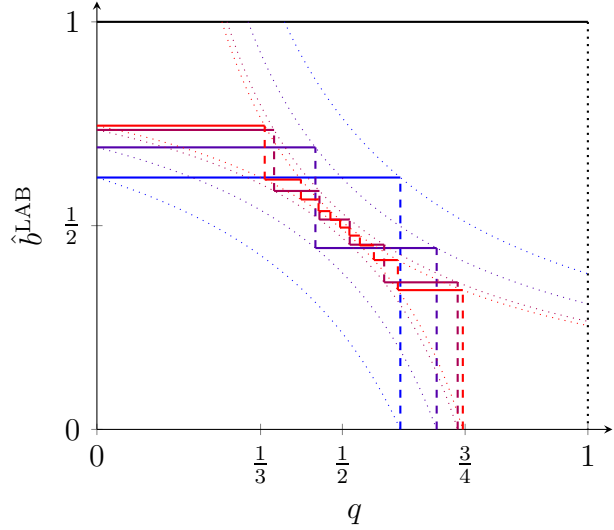\footnotetext{The choice of $Q$ and $v$ is a normalization, as bid levels scale linearly with $v$ and are constant in $Q$, and bid points scale linearly with $Q$ and are constant in $v$.}


\begin{proposition}[Approximate minimax loss in bidpoint-constrained multi-unit uniform-price]\label{proposition: approximate minimax loss lab}
	Suppose that $( q_i, b_i )$ is a minimax-loss bid in the constrained uniform-price auction with $M_b$ bid points, and $L^\star$ is minimax loss in the constrained multi-unit uniform-price auction with $M_q$ units and $M_b$ bid points. Define a constrained bid $( q_i^\prime, b^\prime_i )$ so that $q^\prime_{ik} = \lceil M_q q_{ik} / Q \rceil ( Q / M_q )$ and $b^\prime_{ik} = b_{ik}$. Then $( q_i^\prime, b_i^\prime )$ is feasible in the constrained multi-unit auction, and
	\[
		L^\star \leq L^\LAB\left( \hat b_i^\prime; v^i \right) \leq L^\star + \int_{0}^{\frac{Q}{M_q}} v^i\left( x \right) dx.
	\]
\end{proposition}

As is the case in the pay-as-bid auction, the minimax loss approximation will be close when the number of available units is large.

\subsection{Comparison of auction formats}
We now compare the constrained auction formats. The first comparison relates the two unique minimax-loss bidding functions. We give a condition under which they cannot be uniformly ranked. Note that unlike the multi-unit case it is not enough to compare the levels of the bids $b^\LAB$ and $b^\PAB$; the optimal bid function is a step function and the location of the steps matters for the comparison. However, if $q_M^\LAB < q_M^\PAB$, then the bid in the uniform-price auction is $0$ for smaller quantities than in the pay-as-bid auction.

\begin{comparison}[Sufficient conditions for the semi-comparability of optimal constrained bids]\label{comparison: semicomparability of bids across constrained auctions}
	Let $b^\LAB$ and $b^\PAB$ be the minimax-loss bids in the constrained uniform-price and pay-as-bid auction, respectively. If the marginal values are sufficiently flat, 
	then $b_1^\LAB > b_1^\PAB$ and $q_M^\LAB < q_M^\PAB$. 
\end{comparison}
The proof shows that $q_M^\LAB$ is always below $Q$ and that $q_M^\PAB = Q$ when the marginal values are sufficiently flat. 
Examples~\ref{example: PAB with single bidpoint} and~\ref{example: UPA with single bidpoint} in Appendix~\ref{appendix: derivations of examples} show that the bids in the two auction formats can sometimes be ranked unambiguously. In particular, the bids in the uniform-price auction can be higher than in the pay-as-bid auction.

Comparison~\ref{comparison: semicomparability of bids across constrained auctions} implies that the minimax-loss bid in the constrained uniform-price auction drops to 0 at a quantity at which the minimax-loss bid in the constrained pay-as-bid auction is still positive. The ambiguous revenue comparison is immediate.

\begin{comparison}[Ambiguous revenue]\label{comparison: ex post revenue comparison constrained}
	Depending on the joint value distribution, both ex post and expected revenues can be higher in either constrained auction format.
\end{comparison}

We illustrate the ambiguous revenue comparison by means of the following numerical example.

\begin{example}\label{example: revenue in bidpoint constrained auctions}
We simulate bidpoint-constrained auction outcomes for different choices of the number of allowed bid points $M$. In the simulated auctions the available quantity is $Q = 100$, hence the locations of bidpoints correspond to percentage of aggregate supply. We vary the number of bidders from $n = 2$ to $n = 10$. 
Bidders' marginal values are constant, $v( q ) = v_0$, where $v_0$ follows a truncated lognormal distribution with support $v_0 \in [ 0.5, 2 ]$ and mean $1$. 
For each number of allowed bid points, $M$, we first compute constrained minimax-loss bids in both the pay-as-bid and uniform-price auctions. In the pay-as-bid auction bids are obtained from the expressions in Example~\ref{example: PAB with constant marginal values}; in the uniform-price auction bids are obtained from the simple search procedure outlined in Section~\ref{subsection: bidpoint-constrained upa}. 

Figure \ref{fig: welfare as fun of M} plots average auction revenue as a function of the number of bid points $M$. As expected, increasing the number of bidders increases the seller's expected revenue: the highest value of $n$ independent draws increases in $n$ in expectation. In general, revenue is ambiguous in the auction format and the number of bid points $M$. As observed in Examples~\ref{example: PAB with constant marginal values} and~\ref{example: upa with constant marginal values}, bidders in a pay-as-bid auction with a single bid point will bid half their value for the full market quantity, and bidders in a uniform-price auction with a single bid point will bid more than half their value for less than the full market quantity. Revenue in the pay-as-bid auction is therefore half the highest marginal value, while revenue in the uniform-price auction is more than half the second-highest marginal value. It follows that expected revenue will be higher in the pay-as-bid auction when both the number of bid points and the number of bidders are small. 

\begin{figure}
\centering
\begin{tikzpicture}[scale = 0.85]
	\begin{axis}[
	axis lines = left,
	xlabel = \( M \),
	ylabel = {Revenue},
	xmin = 0,
	xmax = 11,
	ymin = 0,
	ymax = 105,
	legend style = {at ={ ( 0.5, 0.03 ) }, anchor = south, legend columns = 2}
	]

%
%
%
%
		\addplot[red, dashed, mark = square*, mark options = solid] table[x index = 0, y index = 1] {tikz-points-rev-pab-mean-n2.tex};
		\addlegendentry{PAB, $n = 2$};
		\addplot[blue, dashed, mark = square*, mark options = solid] table[x index = 0, y index = 1] {tikz-points-rev-upa-mean-n2.tex};
		\addlegendentry{UPA, $n = 2$};

		\addplot[red, dashed, mark = triangle*, mark options = solid] table[x index = 0, y index = 1] {tikz-points-rev-pab-mean-n5.tex};
		\addlegendentry{PAB, $n = 5$};
		\addplot[blue, dashed, mark = triangle*, mark options = solid] table[x index = 0, y index = 1] {tikz-points-rev-upa-mean-n5.tex};
		\addlegendentry{UPA, $n = 5$};

		\addplot[red, dashed, mark = *, mark options = solid] table[x index = 0, y index = 1] {tikz-points-rev-pab-mean-n10.tex};
		\addlegendentry{PAB, $n = 10$};
		\addplot[blue, dashed, mark = *, mark options = solid] table[x index = 0, y index = 1] {tikz-points-rev-upa-mean-n10.tex};
		\addlegendentry{UPA, $n = 10$};
	\end{axis}
\end{tikzpicture}%
\hspace{1cm}%
\begin{tikzpicture}[scale = 0.85]
	\begin{axis}[
	axis lines = left,
	xlabel = \( M \),
	ylabel = {Revenue higher in UPA (\%)},
	xmin = 0,
	xmax = 11,
	ymin = 0,
	ymax = 105,
	ytick = { 0, 20, 40, 60, 80, 100 },
	yticklabels = { 0\%, 20\%, 40\%, 60\%, 80\%, 100\% },
	legend style = {at ={ ( .5, .9 ) }, anchor = south, legend columns = 3}
	]

%
%
%
%
		\addplot[dashed, mark = square*, mark options = solid] table[x index = 0, y index = 1] {tikz-points-upa-rev-higher-n2.tex};
		\addlegendentry{$n = 2$};

		\addplot[dashed, mark = triangle*, mark options = solid] table[x index = 0, y index = 1] {tikz-points-upa-rev-higher-n5.tex};
		\addlegendentry{$n = 5$};

		\addplot[dashed, mark = *, mark options = solid] table[x index = 0, y index = 1] {tikz-points-upa-rev-higher-n10.tex};
		\addlegendentry{$n = 10$};
	\end{axis}
\end{tikzpicture}

\caption{Average revenue (left) and ex post revenue comparison (right) as a function of number of bid points $M$.}
\label{fig: welfare as fun of M}
\end{figure}
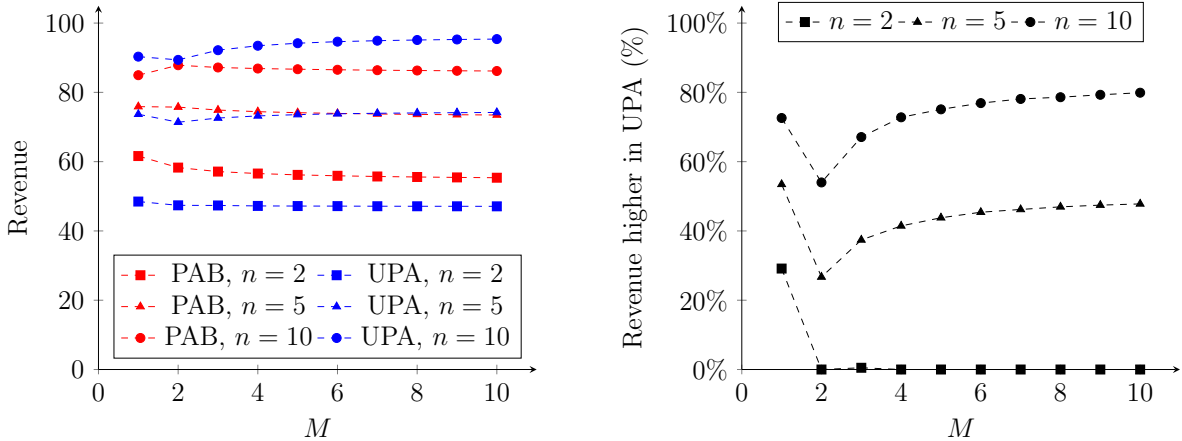

Although average revenues may be ranked, reverse rankings can be observed ex post. Figure \ref{fig: welfare as fun of M} also compares ex post revenues and depicts the share of simulated auctions in which uniform-price revenue is higher than pay-as-bid revenue. As the number of bidders increases, the share of auctions in which revenue is higher in the uniform-price auction increases. Low-revenue outcomes mainly appear in uniform-price auctions with two bidders, and these ``collusive'' outcomes are less likely when there are many bidders. The uniform-price auction dominates the pay-as-bid auction with ten bidders in terms of revenue in expectation and ex post in the majority of auctions. The ambiguous, setting-dependent revenue ranking is in line with empirical results on multi-unit auctions.\footnote{See \citet{pycia+woodward-2020A} for a summary of the ambiguous revenue rankings obtained in the empirical literature.} Nonetheless, it is generically true that increasing the number of bidders increases the performance of the uniform-price auction relative to the pay-as-bid auction. Because initial bids are relatively high in the uniform-price auction and bids are relatively inelastic, increasing the number of bidders has strong upward influence on the market-clearing price, and thus on revenue.
\end{example}

As in multi-unit auctions the minimax loss is lower in the uniform-price auction.
\begin{comparison}[Minimax loss]\label{comparison: comparison level of minimax loss in the constrained case}
	In constrained auctions with $M$ bid points, minimax loss is lower in the uniform-price auction than in the pay-as-bid auction,
	\begin{equation*}
		\sup_{B^{-i} \in \mathcal B} L^\LAB \left( b^\LAB; B^{-i}, v^i \right) <\sup_{B^{-i} \in \mathcal B} L^\PAB\left( b^\PAB; B^{-i}, v^i \right).
	\end{equation*}
\end{comparison}

\begin{figure}
	\centering
	\begin{tikzpicture}[scale = 0.85]
		\begin{axis}[
			axis lines = left,
			xlabel = \( M \),
			ylabel = {Normalized Minimax Loss},
			xmin = 0,
			xmax = 26,
			ymin = 24,
			ymax = 51,
			ytick = { 25, 30, 35, 40, 45, 50 },
			yticklabels = { 25\%, 30\%, 35\%, 40\%, 45\%, 50\% },
			]
			\addplot[red, dashed, mark = square*, mark options = solid] table[x index = 0, y index = 1] {tikz-loss-pab.tex};
			\addlegendentry{PAB};
			\addplot[blue, dashed, mark = square*, mark options = solid] table[x index = 0, y index = 1] {tikz-loss-upa.tex};
			\addlegendentry{UPA};
		\end{axis}
	\end{tikzpicture}
	\caption{Normalized constrained minimax loss as a function of the number of bid points $M$.}
	\label{fig:minimax_loss_function_of_M}
\end{figure}
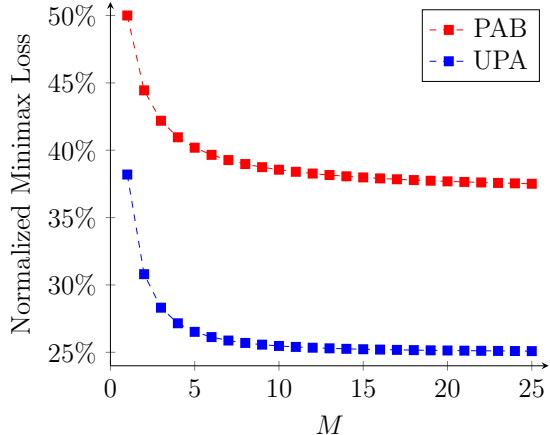

\emph{Example~\ref{example: revenue in bidpoint constrained auctions} (continued)}. In the setting of Example~\ref{example: revenue in bidpoint constrained auctions}, Figure~\ref{fig:minimax_loss_function_of_M} reports the normalized minimax loss, computed as loss divided by the constant marginal value $v_{0}$. As predicted by Comparison~\ref{comparison: comparison level of minimax loss in the constrained case}, the level of minimax loss is higher in the pay-as-bid auction. The figure also shows decreasing gains from adding another bid point and a relatively fast convergence to the unconstrained level of minimax loss. Indeed, minimax loss with four bid points is less than 10\% higher than with 25 bid points in both auction formats.

%% file: section-conclusion.tex
\section{Conclusion}
\label{section: conclusion}

In this paper we have characterized optimal prior-free bids in the pay-as-bid and uniform-price auctions, the two leading auction formats for allocating homogeneous goods such as electricity and government debt. Our analysis considers two natural cases of bid selection: in the 
\textit{multi-unit case} bidders may bid on $M$ discrete units, and in the 
\textit{bidpoint-constrained case} bidders may select up to $M$ bid points at self-selected quantities. In each case the two pricing rules create different incentives for the bidders; our analysis shows that taking a worst-case loss approach to bid optimization enables a tractable analysis of the two formats. Remarkably, our analysis remains tractable even with multi-dimensional private information because we do not require the inversion of strategies as in the canonical Bayesian Nash equilibrium approach. Hence, we believe the worst-case loss approach may also be fruitfully applied to other complex strategic interactions.

%% file: appendix-frb.tex
\section{Uniform-price auctions with a first rejected bid price}
\label{appendix: frb}

%
%

The pricing rule of the uniform-price auction affects bidders' strategic incentives. In the main text, we consider the last accepted bid  pricing rule, in which the market price is the highest possible market-clearing price; this is the uniform-pricing rule commonly used in practice. In this appendix we analyze the multi-unit case of the uniform-price auction with a first rejected bid pricing rule, in which the market price is the lowest possible market-clearing price; this is the uniform-pricing rule commonly analyzed in the literature. 


Our analysis of the first rejected bid pricing rule begins by defining overbidding and underbidding loss, as in our analyses in the main text. In the uniform-price auction with the first rejected bid pricing rule, winning bids below values can never be too high as they do not determine the market-clearing price. In particular, the bid for the first unit, $b_{i1}$, 
can be too high only if it is above value. As such, in this analysis we constrain attention to bids which are below value, and later verify that this assumption is satisfied by the minimax-loss bids we obtain. Conditional on bidder $i$ winning $k$ units, $k\ge 1$, the only bid that may be too high is bidder $i$'s first rejected bid $b_{ik+1}$. For this case, we define \emph{overbidding regret} as
\[
	\overline{R}_{q_k}^\FRB\left( b_i; v_i \right) = k b_{ik+1}.
\]
This is the additional utility the bidder could have received if they reduced their bid $b_{ik+1}$ to zero. The case occurs if the other bidders bid a strictly positive amount for only the $M-k$ units they received. The overbidding regret of winning 0 units is 0. Similar to the other pricing rules, a bid is too low if the bidder wants to win more units given the market-clearing price. Maximal regret arises if the opponents all bid just above $b_{ik+1}$ for the units they received. The resulting \emph{underbidding regret} is
\[
	\underline{R}_{q_k}^\FRB\left( b_i; v_i \right) = \sup_{p \in \left[ b_{ik+1}, b_{ik} \right]} \sum_{k^\prime = k + 1}^{Q} \left( v_{ik^\prime} - p \right)_+
	= 
	\sum_{k^\prime = k + 1}^{M} \left( v_{ik^\prime} - b_{ik+1} \right)_+.
\]
This is the additional utility the bidder could have received if they bid just above $b_{ik+1}$ for all units for which it is profitable to do so. Note that these regret terms correspond to those in the pay-as-bid auction, except that underbidding regret in the uniform-price auction does not consider bids for submarginal quantities.

The conditional regret for unit $k$, $k\in\{0,1,\dots,M-1\}$, is the maximum of overbidding and underbidding regret,
\[
	R^{\FRB}_{q_k}\left( b_i; v_i \right) = \max \left\{ \overline{R}_{q_k}^\FRB\left( b_i; v_i \right), \underline{R}_{q_k}^\FRB\left( b_i; v_i \right) \right\}.
\]

\begin{observation}
	Note that if $b_{ik} > v_{ik}$, then overbidding regret conditional on winning $k - 1$ units is $( k - 1 ) b_{ik} > 0$ and underbidding regret is $\sum_{k^\prime = k}^M ( v_{ik^\prime} - b_{ik} )_+ = 0$. That is, bidding above value equates overbidding and underbidding regret only when $k = 1$. Since the minimax-loss bid must equate overbidding and underbidding regret for some unit (see Lemma~\ref{lemma: maximal loss in frb} below), there is a minimax-loss bid that is weakly below the bidder's value vector.
\end{observation}

\begin{lemma}[Maximal loss in first rejected bid uniform-price auction]\label{lemma: maximal loss in frb}
	In the first rejected bid uniform-price auction, the maximal loss given bid $b_i \leq v_i$ is
	\[
		\max_{k \in \left\{ 1, \ldots, M \right\}} \left[ \max \left\{ \left( k - 1 \right) b_{ik}, \sum_{k^\prime = k}^M \left( v_{ik^\prime} - b_{ik} \right)_+ \right\} \right].
	\]
\end{lemma}

\begin{proof}
	We first consider the augmented problem in which the bidder receives $k$ units at market price $p^\star \in [ b_{ik+1}, b_{ik} ]$. In a uniform-price auction, a bidder facing a known residual supply curve should pick a point on the supply curve to maximize their own utility. The bidder's utility from this optimization increases as the residual supply curve falls, hence the loss-maximizing supply curve must be as low as possible. When receiving $k$ units, the bidder knows that either their opponents demanded $M - k$ units with bids weakly above $p^\star$ and the $M - k + 1^{\text{th}}$ unit at $p^\star$, or their opponents demanded $M - k$ units with bids weakly above $p^\star$ and the market-clearing price is $p^\star = b_{ik+1}$. The loss-maximizing residual supply curve $S$ is given by
	\[
		S\left( k^\prime; p^\star \right) = \begin{cases}
			p^\star &\text{if } k^\prime \in \left\{ 1, \ldots, M - k \right\}, \\
			p^\star &\text{if } k^\prime = M - k + 1 \text{ and } p^\star > b_{ik+1}, \\
			0 &\text{if } k^\prime = M - k + 1 \text{ and } p^\star = b_{ik+1}, \\
			0 &\text{if } k^\prime > M - k + 1.
		\end{cases}
	\]
	The loss-maximization problem is then
	\begin{equation}
		\max_{\tilde k} \sum_{k^\prime = 1}^{\tilde k} \left( v_{ik^\prime} - S\left( M - \tilde k + 1; p^\star \right) \right) - \sum_{k^\prime = 1}^{k} \left( v_{ik^\prime} - p^\star \right).
		\label{eq: uniform frb loss with residual supply}
	\end{equation}
	Note that $S( M - \tilde k + 1; p^\star )$ is increasing and locally constant in $\tilde k$; then loss is obtained at $\tilde k \in \{ k - 1, k, v^{-1}( p^\star ) \}$, where $v^{-1}( p^\star ) = \max \{ k^\prime \colon v_{ik^\prime} > p^\star \}$. It follows that loss, conditional on market-clearing price $p^\star$, is
	\[
		R_k^\FRB\left( b^i; v^i \right) = \max \left\{ \left( k - 1 \right) p^\star - \left( v_{ik} - p^\star \right), k b_{ik+1}, \sum_{k^\prime = k + 1}^M \left( v_{ik^\prime} - p^\star \right)_+ \right\}.
	\]
	The equation is obtained by plugging $\tilde k \in \{ k - 1, k, v^{-1}( p^\star ) \}$ into Equation~\eqref{eq: uniform frb loss with residual supply}.  
	By construction, $p^\star \leq b_{ik}$; then
	\[
		\left( k - 1 \right) p^\star - \left( v_{ik} - p^\star \right) \leq k b_{ik} - v_{ik} \leq \left( k - 1 \right) b_{ik}.
	\]
	The right-hand inequality follows by the assumption that $b_i \leq v_i$. Then the leftmost in the maximization expression for $R_{q_k}^\FRB$ is bounded above by the middle term in $R_{q_{k-1}}^\FRB$, and hence
	\[
		\max_k R_{q_k}^\FRB\left( b_i; v_i \right) = \max_k \left[ \max\left\{ \left( k - 1 \right) b_{ik}, \sum_{k^\prime = k}^M \left( v_{ik^\prime} - b_{ik} \right)_+ \right\} \right].
	\]
\end{proof}

Similar to the analysis of cross-conditional regret minimizing bids in the last accepted bid uniform-price auction, note that $\overline{R}^\FRB_k$ is increasing in $b_{ik+1}$ while $\underline{R}^\FRB_{q_k}$ is decreasing in $b_{ik+1}$, and both terms are independent of $b_{ik^\prime}$ for $k^\prime \neq k+1$. Then if maximum loss is determined by conditional regret for unit $k$, it must be that $\overline{R}^\FRB_{q_k}( b_i; v_i ) = \underline{R}^\FRB_{q_k}( b_i; v_i )$. There is, however, no unique optimal bid.

\begin{theorem}[No unique minimax-loss bid]\label{theorem: no unique loss-minimizing bid in frb}
	If $M > 1$, then there is not a unique minimax-loss bid in the uniform-price auction with the first rejected bid pricing rule.
\end{theorem}

	

\begin{proof}
	It is sufficient to consider $b_{i1}$. When the bidder receives $0$ units, overbidding regret is $0$ and underbidding regret $\underline R_0^\FRB$ is non-negative but arbitrarily close to 0 when $b_{i1}$ is close to $v_{i1}$. As overall minimax loss $L^\FRB$ is strictly positive, any choice of $b_{i1}$ such that $\max \{ b_{i2}, v_{i1} - L^\FRB \} \leq b_{i1} \leq v_{i1}$ minimaxes loss.
\end{proof}

In the specific case of a single unit, $M = 1$, Lemma~\ref{lemma: maximal loss in frb} gives maximum loss as
\[
	\max\left \{ 0, \left( v_{i1} - b_{i1} \right)_+ \right\}.
\]
Then the unique minimax-loss bid is $b_{i1} = v_{i1}$.

In light of Theorem~\ref{theorem: no unique loss-minimizing bid in frb}, minimax-loss bids are not uniquely defined in the uniform-price auction when $M > 1$ indivisible units are available. To obtain sharp predictions for minimax-loss bids, we define a conditional regret minimizing strategy as one which minimizes conditional regret for each unit. Because conditional regret is independent across units, and regret is minimized by conditional regret for some unit, a conditional regret minimizing strategy is a regret minimizing strategy.

\begin{definition}
	The bid vector $b_i$ is \emph{conditionally regret minimizing} if $b_i \in \argmin_{b^\prime} R^{\text{FRB}}_{q_k}( b^\prime; v_i )$ for all units $k \in \{0,1\dots,M-1 \}$.
\end{definition}



The following theorem characterizes the unique conditional regret-minimizing bid vector.

\begin{theorem}[Conditional regret-minimizing bids]\label{theorem: conditional regret minimization frb}
	The unique conditional regret-minimizing bid vector $b_i^\FRB$ is such that $b_{i1}^\FRB = v_{i1}$ and for all $k \in \{1,\dots,M-1\}$,
	\[
		b_{ik + 1}^\FRB = \frac{1}{k} \sum_{k^\prime = k + 1}^{M} \left( v_{ik^\prime} - b_{ik+1}^\FRB \right)_+.
	\]
\end{theorem}

\begin{proof}
	The claim follows immediately from earlier arguments. 
	It remains to be shown that $b_i^\FRB$ is a valid bid (that is, monotone). Suppose to the contrary that there is $k$ such that $b^\FRB_{ik} < b^\FRB_{ik+1}$. Then
	\begin{align*}
		b^\FRB_{ik} &= \frac{1}{k-1} \sum_{k^\prime = k}^{M} \left( v_{ik^\prime} - b_{ik}^\FRB \right)_+ \geq \frac{1}{k-1} \sum_{k^\prime = k+1}^{M} \left( v_{ik^\prime} - b_{ik}^\FRB \right)_+ \\
		&\hspace{1cm}\geq \frac{1}{k} \sum_{k^\prime = k+1}^{M} \left( v_{ik^\prime} - b_{ik}^\FRB \right)_+ 
		\geq \frac{1}{k} \sum_{k^\prime = k+1}^{M} \left( v_{ik^\prime} - b_{ik+1}^\FRB \right)_+ = b^\FRB_{ik+1}.
	\end{align*}
	This is a contradiction, so it cannot be that $b^\FRB_{ik} < b^\FRB_{ik+1}$.
\end{proof}

Similar to minimax-loss bids in the auction formats analyzed in the main text, conditional regret minimizing strategies in the first rejected bid uniform-price auction are straightforward to compute but potentially infeasible to represent in closed form. In particular, determination of $b_{ik}^\FRB$ still faces issues of potential nonlinearities in $\underline{R}^{\FRB}_{q_k}( \cdot; v_i )$. We consider two examples.


\begin{example}[Two-unit demand in first rejected bid uniform-price]
	In the first rejected bid price auction with demand for two units, the conditional regret minimizing bid vector $b_i^\star$ is such that
	\[
		b_{i1}^\FRB = v_{i1}; \;\; \text{ and } b_{i2}^\FRB = \frac{v_{i2}}{2}.
	\]
	This follows immediately from Theorem~\ref{theorem: conditional regret minimization frb}. The first bid $b_{i1}$ cannot be too high, provided that it is below value. Thus, the overbidding regret conditional on losing is 0. The bid is too low if one could win more units by marginally raising it, leading to a worst-case regret conditional on losing the auction of $v_{i1} - b_{i1} + (v_{i2} - b_{i1})_+$. These two types of conditional regret are equalized by bidding value on the first unit. The bid $b_{i2}$ is found by equalizing the overbidding regret conditional on winning one unit $b_{i2}$ with the underbidding regret conditional on winning one unit $v_{i2}-b_{i2}$.
\end{example}

\begin{example}[Multi-unit FRB with flat marginal values]\label{proposition:uniform price strategies flat values frb}
	Suppose that marginal values are relatively flat, so that $( M - 1 ) v_{iM} \geq ( M - 2 ) v_{i2}$. The conditionally regret-minimizing bid vector $b_i^\FRB$ is such that
	\[
		b_{i1}^\FRB = v_{i1}, \text{ and } b_{ik}^\FRB = \frac{1}{M} \sum_{k^\prime = k}^{M} v_{ik^\prime}.
	\]

	Bids are decreasing in quantity. Then following Theorem~\ref{theorem: conditional regret minimization frb}, potential nonlinearities are irrelevant when $b_{i2}^\FRB \leq v_{iM}$. When this is true, bids are as given as stated. We then check
	\begin{align*}
		b_{i2}^\FRB \leq v_{iM} \;\; \iff \;\; \left( M - 1 \right) v_{iM} \geq \sum_{k^\prime = 2}^{M-1} v_{ik^\prime}.
	\end{align*}
	Since $v_{ik^\prime} \leq v_{i2}$ for all $k^\prime \geq 2$, the condition in the proposition is immediate.
\end{example}

%% file: appendix-divisible-goods.tex
\section{Unconstrained bidding}
\label{appendix: unconstrained bidding}

The minimax-loss bids derived in the main text are computationally tractable, but are not necessarily expressable in closed form. The apparent analytical complexity of optimal bids arises from the recursive structure to the loss-minimization problem (in pay-as-bid), and from the simultaneous optimization over bid levels and bid points in the constrained bidpoint model. In this appendix we show that minimax-loss bids may be tractable in an unconstrained divisible-good context, where there is no need to optimize over the location of bid points, and the recursive structure of minimax-loss bids can be expressed as a differential equation. We also show that loss in this unconstrained case is approximated by loss in the multi-unit and constrained bidpoint cases as the number of bid points becomes large.

In this appendix, we assume that the marginal value function $v^i$ is Lipschitz continuous.

\subsection{Pay as bid}
\label{appendix: unconstrained pab}

With unconstrained bids and divisible goods, the equal conditional regret condition from the multi-unit and constrained cases requires that the derivative of conditional regret is equal to zero,
\[
	v^i\left( q \right) - b^i\left( q \right) = -v^{-1}\left( b^i\left( q \right) \right) \frac{db^i \left(q\right)}{dq}.
\]
Regret conditional on receiving the maximum possible allocation is $\int_0^Q b^i( x ) dx$, so long as $b^i( q ) > 0$ everywhere $v^i( q ) > 0$. The fundamental theorem of differential equations implies that solutions to the system cannot cross, hence the bid for quantity $Q$ must be minimal, and optimal unconstrained bids may be computed as the solution to a differential equation.

\begin{proposition}[Unconstrained pay-as-bid bids]\label{proposition: unconstrained divisible-good PAB}
	The unique minimax-loss bid in the unconstrained divisible-good pay-as-bid auction solves
	\[
		v^i\left( q \right) - b^i\left( q \right) = -v^{-1}\left( b^i\left( q \right) \right) \frac{db^i}{dq}\left( q \right), \text{ s.t. } b^i\left( Q \right) = 0.
	\]
\end{proposition}

\begin{proof}
	Arguments in the main text (preceding the statement of Proposition~\ref{proposition: unconstrained divisible-good PAB}) establish the basic differentiable form. It remains to establish the initial condition. Because $b^i( Q ) \geq 0$ by constraint, it is sufficient to show that $b^i( Q )$ cannot be strictly positive. By the fundamental theorem of differential equations (the Picard--Lindel\"of theorem), if there are solutions $b^i$ and $\tilde b^i$ with $b^i( Q ) = 0 < \tilde b^i ( Q )$, then $b^i \leq \tilde b^i$. The differential form ensures equal conditional regret for all units, and conditional regret for unit $q = Q$ under bid $\tilde b^i$ is $\int_0^Q \tilde b^i( x ) dx > \int_0^Q b^i( x ) dx$. Then maximum loss is lower under bid $b^i$ than under bid $\tilde b^i$, and $\tilde b^i$ is not a minimax-loss bid. Then $b^i( Q ) = 0$ for any minimax-loss bid, and uniqueness follows from the fundamental theorem of differential equations.
\end{proof}

The differential equation defining minimax-loss bids in the pay-as-bid auction is similar to the first-order condition defining best responses in a standard Bayesian Nash equilibrium; see, e.g., \citet{Hortacsu+McAdams-Journal-of-Political-Economy-2010A}, \citet{pycia+woodward-2020A}, and \citet{woodward-2021A}. The distinction is that in Bayesian Nash equilibrium the first-order condition contains probabilistic effects---increasing the bid for a particular quantity increases the probability that this quantity is received---while the differential equation in Proposition~\ref{proposition: unconstrained divisible-good PAB} does not. Intuitively, this is because loss is maximized conditional on receiving any particular quantity, and hence 
the loss-maximizing probability a quantity is won is constant across all quantities.

Because any bid which is feasible when $M$ bid points are allowed is also feasible when $M^\prime > M$ bid points are allowed, minimax loss decreases as the constraint on the number of bid points is increased. Since the unconstrained-optimal bid $b^i$ may be arbitrarily approximated by step functions with small step widths, it follows that minimax loss in the multi-unit and constrained pay-as-bid auctions converges to minimax loss in the unconstrained pay-as-bid auction. Because the minimax-loss bid is unique in the pay-as-bid auction, minimax-loss bids in the multi-unit and constrained pay-as-bid auctions converge to the minimax-loss bid in the unconstrained pay-as-bid auction.

\begin{proposition}[Convergence to unconstrained minimax-loss bid]\label{proposition: convergence in pab}
	Let $L^{M_q}$ and $b^{M_q}$ be minimax loss and the minimax-loss bid (respectively) in the multi-unit pay-as-bid auction with $M_q$ units, and let $L^{M_b}$ and $b^{M_b}$ be minimax loss and the minimax-loss bid (respectively) in the constrained pay-as-bid auction with $M_b$ bid points. Let $L^\star$ and $b^\star$ be minimax loss and the minimax-loss bid (respectively) in the unconstrained pay-as-bid auction. Then
	\[
		\lim_{M_q \nearrow \infty} L^{M_q} = L^\star, \; \lim_{M_q \nearrow \infty} \left\| \hat b^{M_q} - b^\star \right\|_1 = 0, \; \lim_{M_b \nearrow \infty} L^{M_b} = L^\star, \text{ and } \lim_{M_b \nearrow \infty} \left\| \hat b^{M_b} - b^\star \right\|_1 = 0,
	\]
	where $\| \cdot \|_1$ represents the $L_1$ norm.
\end{proposition}

\begin{proof}
	Because $L^\star$ is minimax loss when bids are unconstrained, $L^\star \leq L^{M_q}$ and $L^\star \leq L^{M_b}$ for all $M_q$ and $M_b$. Since maximum loss is continuous in bid and the minimax-loss bid $b^\star$ can be arbitrarily approximated by a step function (when the number of steps grows large), it follows that $\lim_{M_q \nearrow \infty} L^{M_q} = L^\star$ and $\lim_{M_b \nearrow \infty} L^{M_b} = L^\star$.
	
	Now suppose that $| \hat b^{M_q} - b^\star |$ does not converge to $0$ as $M_q$ grows large. Then there is a $\varepsilon > 0$ such that for all $\bar M$, there is $M_q > \bar M$ with $| b^{M_q} - b^\star | > \varepsilon$. Let $\langle b^{M_qk} \rangle_{k = 1}^\infty$ be a sequence of such minimax-loss bids, where $M_{qk} M_{qk^\prime}$ whenever $k < k^\prime$. Bids are decreasing in quantity, hence by Helly's selection theorem it is without loss of generality to assume that $\hat b^{M_{qk}} \to \tilde b^\star$ in the $L_1$ norm, and since minimax loss converges the maximum loss associated with bid $\tilde b^\star$ is $L^\star$, the maximum loss associated with bid $b^\star$. It follows that $\tilde b^\star$ is a minimax-loss bid in the unconstrained pay-as-bid auction. Since there is a unique minimax-loss bid in the pay-as-bid auction (Proposition~\ref{proposition: unconstrained divisible-good PAB}), this contradicts the assumption that $\lim_{k \nearrow \infty} b^{M_{qk}} \neq b^\star$.
	
	Showing that $| \hat b^{M_b} - b^\star |$ converges to $0$ is essentially identical to the argument above, and is omitted.
\end{proof}

%
%

\subsection{Uniform price}
\label{appendix: unconstrained upa}

When bids are completely unconstrained, cross-conditional regret minimization requires
\[
	q b^\LAB\left( q \right) = \int_{q}^{Q} \left( v^i\left( x \right) - b^\LAB\left( q \right) \right)_+ dx
\]
for all $q\in [0,Q]$. The cross-conditional regret minimizing bid is unique because overbidding regret increases in bid while underbidding regret decreases in bid. Note that the divisibility of the auctioned good turns cross-conditional regret into conditional regret.

\begin{proposition}[Cross-conditional regret minimizing bid in unconstrained uniform-price auction]\label{proposition: cross conditional regret minimizing bid in unconstrained upa}
	In the unconstrained uniform-price auction there is a unique cross-conditional regret minimizing bid, $b^\LAB$, and this bid solves
	\[
		q b^\LAB\left( q \right) = \int_q^Q \left( v^i\left( x \right) - b^\LAB\left( q \right) \right)_+ dx, \; \forall q.
	\]
\end{proposition}


The proposition implies that $b^\LAB(0)=v^i(0)$, i.e., it is optimal to bid value for the ``first unit.'' Moreover, it is optimal to bid 0 for the $Q$, $b^\LAB(Q)=0$.

\begin{figure}
	\centering
	\begin{tikzpicture}
		 \begin{axis}[
		 	samples=200, 
		 	domain=0.0:0.52,
		 	axis lines = left,
    		xlabel = \(q\),
    		ylabel = \(b\),
    		xtick={0, .33},
    		ytick={2},
    		ymax=1,
    		xticklabels={$0$, $\hat q$},
		 ]

		 	\addplot [mark=none, thick, color=blue, domain=0:0.5, name path=A] { 0.12873/x };
		 	\addlegendentry{\(\overline c(q;L^\star)\)}

		 	\addplot [name path=B, thick, mark=none, color=red, domain=0:0.5] { .66 - 0.5240762698787824 *x - 0.890406129553854 *x*x };
    		\addlegendentry{\(\underline c(q;L^\star)\)}

    		\addplot[gray!15] fill between[of=A and B];

		\end{axis}
				 \begin{axis}[
		 	samples=200, 
		 	domain=0.0:0.52,
		 	axis lines = left,
    		xlabel = \(q\),
    		ylabel = \(b\),
    		xtick={0, .33},
    		ytick={2},
    		ymax=1,
    		xticklabels={$0$, $\hat q$},
		 ]

		 	\addplot [mark=none, thick, color=blue, domain=0:0.5, name path=A] { 0.12873/x };
		 	\addlegendentry{\(\overline c(q;L^\star)\)}

		 	\addplot [name path=B, thick, mark=none, color=red, domain=0:0.5] { .66 - 0.5240762698787824 *x - 0.890406129553854 *x*x };
    		\addlegendentry{\(\underline c(q;L^\star)\)}
		\end{axis}

	\end{tikzpicture}
	\caption{Iso-loss curves of unconstrained underbidding and overbidding regret in the uniform-price auction.}
	\label{fig: upa loss minimax}
\end{figure}
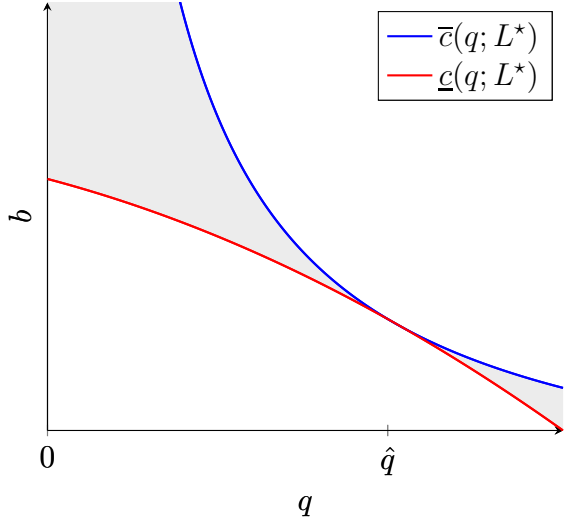

Figure \ref{fig: upa loss minimax} illustrates the upper and lower iso-loss curves for a loss-level equal to minimax loss. In the unconstrained case the upper and lower iso-loss curves are tangent to each other. The bids at the points of tangency are uniquely determined and equal to the cross-conditional regret minimizing bids. In the example depicted in the figure, there is a single point of tangency $\hat q$. Other bids are only partially determined; any bid must be below the upper iso-loss curve and above the lower iso-loss curve. In the figure any decreasing bidding function in the shaded area is a minimax bid. All minimax bidding functions agree on $\hat q$.

The figure illustrates the nonuniqueness of the minimax bid. Indeed, as in the multi-unit uniform-price auction, there is not a unique minimax-loss bid in the unconstrained uniform-price auction. If there is, overbidding regret must be equal across all quantities, giving $q b( q ) = L$ for all quantities $q$. This would imply that high bids for small quantities give zero underbidding regret; these bids can be reduced without affecting maximum loss, and the minimax-loss bid is nonunique.

The following theorem formally states that any weakly decreasing bid below marginal values and between the upper and lower iso-loss curves minimizes worst-case loss.

\begin{theorem}[Minimax-loss bid in unconstrained uniform-price auction]\label{theorem: minimax loss in unconstrained upa}
	Let $L^\star$ be such that $\overline c( \cdot; L^\star ) \geq \underline c( \cdot; L^\star )$ and there exists $q$ with $\underline c( q; L^\star ) = \overline c( q; L^\star )$. The bid $b^\LAB$ minimizes maximal loss if and only if $\underline c( \cdot; L^\star ) \leq b^\LAB \leq \overline c( \cdot; L^\star )$.
\end{theorem}

\begin{proof}
	Suppose that $\underline c( \cdot; L ) \leq b \leq \overline c( \cdot; L )$. At any quantity $q$, overbidding regret is $q b( q ) \leq q \overline c( \cdot; L ) = L$; and at any quantity $q$, underbidding regret is
	\[
		\int_q^Q \left( v^i\left( x \right) - b_+\left( q \right) \right)_+ dx \leq \int_q^Q \left( v^i\left( x \right) - \underline c\left( q; L \right) \right)_+ dx = L.
	\]
	The left-hand inequality follows from the fact that $\underline c( \cdot; L )$ is continuous. Then conditional regret at quantity $q$ is such that $R^\LAB_q( b; v^i ) \leq L$, and it follows that the loss of bid $b$ is weakly below $L$.
\end{proof}

Finally, as the number of available bid points becomes large---either because the commodity becomes divisible, or because the limited-bid-step constraint is weakend---constrained bids can arbitrarily approximate an unconstrained minimax-loss bid. Since loss is continuous in bid, minimax loss will converge to unconstrained minimax loss; and, moreover, the limit of a sequence of constrained bids will be a minimax-loss bid in the unconstrained model.

\begin{proposition}[Convergence to unconstrained minimax-loss bid]\label{proposition: convergence of minimax loss upa}
	Let $L^{M_q}$ be minimax loss and a minimax-loss bid (respectively) in the multi-unit uniform-price auction with $M_q$ units, and let $L^{M_b}$ and $b^{M_b}$ be minimax loss and the minimax-loss bid (respectively) in the constrained uniform-price auction with $M_b$ bid points. Let $L^\star$ be minimax loss in the unconstrained uniform-price auction. Then along any convergent sequence $\langle b^{M_q} \rangle \to b^{q\star}$ and any convergent sequence $\langle b^{M_b} \rangle \to b^{b\star}$,
	\begin{align*}
		\lim_{M_q \nearrow \infty} L^{M_q} &= L^\star, &\; L^\LAB\left( \lim_{M_q \nearrow \infty} \hat b^{M_q}; v^i \right) &= L^\star, \\
		\lim_{M_b \nearrow \infty} L^{M_b} &= L^\star, &\; \text{ and } L^\LAB\left( \lim_{M_b \nearrow \infty} \hat b^{M_b}; v^i \right) &= L^\star.
	\end{align*}
\end{proposition}

\begin{proof}
	The convergence of optimal loss follows from the fact that, as $M_q$ and $M_b$ tend toward infinity, multi-unit and constrained-bid bids can arbitrarily approximate the unconstrained cross-conditional regret minimizing bid $b^\LAB$. Since loss is converging, in the limit bids must lie between the limiting upper and lower iso-loss curves, which are continuous in loss. Theorem~\ref{theorem: minimax loss in unconstrained upa} implies the desired result.
\end{proof}

%% file: appendix-loss-minimizing-bids.tex
\section{Omitted Proofs}
\label{appendix: proofs for loss-minimizing bids}

\begin{proof}[Proof of Lemma~\ref{lemma: maximum loss as maximum regret}]
	Consider the maximization of loss 
	\[\sup_{\tilde b} \sup_{B^{-i}\in\mathcal B}\mathbb{E}_{B^{-i}}\left[ \hat u\left(q^i( \tilde b, b^{-i}), t^i\left( \tilde b, b^{-i} \right); v^i \right) - \hat u\left( q^i\left( b^i, b^{-i} \right), t^i\left( b^i, b^{-i} \right); v^i \right) \right],\]
	where we have swapped the order of the suprema. Observe that the inner maximization problem is linear in the choice variable $B^{-i}$. \cite{Winkler_1988} proves that the extreme points of $\mathcal B$ are distributions with a single point in the support. Since loss is linear in $B^{-i}$, maximum loss is attained at an extreme point. 
\end{proof}

%
%
\subsection{Analysis of pay-as-bid auctions}


\begin{proof}[Proof of Lemma~\ref{lemma: maximal loss in PAB}]
	In a pay-as-bid auction, a bidder facing a known residual supply curve should bid a constant amount for all units they desire: because bids are paid, a bid above the resulting market-clearing price can be reduced to save payment without affecting allocation. Since maximizing loss is equivalent to finding an ex post residual supply curve that maximizes regret, the loss-maximization problem is equivalent to solving
	\[
		R_q\left( b^i; v^i \right) = \sup_{S\colon q^i\left( b^i, S \right) = q} \sup_{\tilde q} \int_0^{\tilde q} v^i(x) - S\left( Q - \tilde q \right) dx= \sup_{S\colon q^i\left( b^i, S \right) = q} U^\star\left( S; v^i \right)
		\footnote{To beat the opponent bid for unit $Q - q$ with certainty, bidder $i$ must bid strictly above $S( Q - q )$, or $b_{i}(q) = S( Q - q ) + \varepsilon$ for any $\varepsilon > 0$. Since regret is defined by a supremum, we let $\varepsilon = 0$ while retaining the assumption that bidder $i$ wins unit $q$ for sure.}
	\]
	for some $q$. Note that $U^\star$ is decreasing in $S$. Let $\tilde S < S$, and consider
	\[
		q^\star \in \arg \sup_{\tilde q} \int_0^{\tilde q} v^i(x) - S\left( Q - \tilde q \right), \text{ and } \tilde q^\star \in \arg \sup_{\tilde q} \int_0^{\tilde q} v^i(x) - \tilde S\left( Q - \tilde q \right).
	\]
	If $q^\star = \tilde q ^\star$, then $U^\star( S; v^i ) \leq U^\star( \tilde S; v^i )$ since the required bid under $\tilde S$ is lower than the required bid under $S$. If $q^\star \neq \tilde q^\star$, then $U^\star( S; v^i ) \leq U^\star( \tilde S; v^i )$ since loss is higher under $\tilde S$ with selected quantity $\tilde q^\star$ than with selected quantity $q^\star$, and the required bid is lower even with selected quantity $q^\star$.
	
	Then when considering maximum loss, it is sufficient to consider residual supply curves which are as low as possible. Conditional on bidder $i$ receiving share $q$, the only constraint on the residual supply curve is $S( Q - q ) \geq b^i_+(q)$; 
	that is, bidder $i$'s opponents bid more for their aggregate $Q-q$ unit than bidder $i$ bid for their ``next'' unit. Because bids are monotone, the lowest residual supply curve satisfying this constraint is
	\[
		S_q\left( \tilde q; b^ii \right) = \begin{cases}
			0 &\text{if }\tilde q < Q - q, \\
			b^i_+(q) &\text{if } \tilde q \geq Q - q.
		\end{cases}
	\]
	
	Given this residual supply curve, bidder $i$'s optimal bid will either win $q$ units at a price of $0$, or will win as many units as desired at a price of $b^i_+(q)$. In light of Lemma~\ref{lemma: maximum loss as maximum regret}, which shows that maximum loss is equivalent to maximum regret, the result follows from evaluating the ex post utility of this decision.
\end{proof}

%
%
\subsubsection{The multi-unit case}

\begin{proof}[Proof of Theorem~\ref{theorem:discriminatory auction equal conditional regret}]
	We show that $\underline R_{q_k}^\PAB( b_i; v_i ) = \sum_{k^\prime = 1}^M b_{ik^\prime}$ for all $k$. First, since $\sum_{k^\prime = 1}^k b_{ik^\prime}$ is increasing in $k$, Lemma~\ref{lemma: maximal loss in PAB} implies that maximum loss is
	\[
		\max\left\{ \max_k \underline R^\PAB_{q_k}\left( b_i; v_i \right), \sum_{k = 1}^{M} b_{ik} \right\}.
	\]
	Importantly, loss is continuous in bid. Note that increasing all bids by $\varepsilon > 0$ will weakly decrease $R_{q_k}( b_i; v_i )$ for all $k$ and strictly increase $\sum_{k = 1}^{M} b_{ik}$; then if $b_i$ is loss-minimizing, it must be that $\sum_{k = 1}^{M} b_{ik} \geq \max_k \underline R_{q_k}^\PAB( b_i; v_i )$. Similarly, decreasing all bids by $\varepsilon > 0$ strictly decreases $\sum_{k^\prime = 1}^{M} b_{ik^\prime}$ and continuously affects $\underline R_{q_k}^\PAB( b_i; v_i )$, thus $\sum_{k = 1}^{M} b_{ik} = \max_k \underline R_{q_k}^\PAB( b_i; v_i )$.\footnote{A bid vector which is not strictly positive---i.e., for which there exists $k$ with $b_{ik} = 0$---cannot be uniformly decreased by $\varepsilon$. Nonetheless, decreasing the bid by $\varepsilon$ where possible will decrease $\sum_{k = 1}^M b_{ik}$ and will continuously affect $\max_k \underline R_{q_k}^\PAB( b_i; v_i )$.}
	
	Now, suppose that there is $k$ with $\underline R_{q_k}^\PAB( b_i; v_i ) < \sum_{k^\prime = 1}^{M} b_{ik^\prime}$. If $b_{ik+1} = 0$, then
	\[
		\underline R_{q_k}^\PAB\left( b_i; v_i \right) = \sum_{k^\prime = 1}^{k} \left( b_{ik^\prime} - b_{ik+1} \right) + \sum_{k^\prime = k+1}^{M} \left( v_{ik^\prime} - b_{ik+1} \right)_+ = \sum_{k^\prime = 1}^{k} b_{ik^\prime} + \sum_{k^\prime = k+1}^{M} v_{ik^\prime} \geq \sum_{k^\prime = 1}^{M} b_{ik^\prime}.
	\]
	This is a contradiction, and it must be that $b_{ik+1} > 0$. In this case, reducing $b_{ik+1}$ will weakly increase $\underline R_{q_k}^\PAB( b_i; v_i )$, strictly decrease $\underline R_{q_{k^\prime}}^\PAB( b_i; v_i )$ for all $k^\prime > k$, and will not affect $\underline R_{q_{k^\prime}}^\PAB( b_i; v_i )$ for $k^\prime < k$; reducing $b_{ik+1}$ also reduces $\sum_{k^\prime = 1}^{M} b_{ik^\prime}$, and the arguments above show that increasing all bids by some small amount will strictly reduce loss.
	
	It follows that $\underline R_{q_k}^\PAB( b_i; v_i ) = \sum_{k^\prime = 1}^{M} b_{ik^\prime}$ for all $k$, and the result is immediate.
\end{proof}

\begin{proof}[Proof of Corollary~\ref{theorem:multi unit discriminatory representation}]
	Following Theorem~\ref{theorem:discriminatory auction equal conditional regret}, conditional regret is equalized across all units. Then for all units $k$, $1\le k\le M$,
	\begin{align}
			& R_{q_{k-1}}^\PAB\left( b_i; v_i \right) - R_{q_k}^\PAB\left( b_i; v_i \right) = 0 \notag \\
			& \iff \;\; \left[ k b_{ik+1} - k b_{ik} \right] + \left( v_{ik} - b_{ik} \right) + \sum_{k^\prime = k+1}^{M}\left[ \left( v_{ik^\prime} - b_{ik} \right)_+ - \left( v_{ik^\prime} - b_{ik+1} \right)_+ \right] = 0. \label{equation:multi unit discriminatory implicit proof}
	\end{align}
	From this, it immediately follows that $b^\PAB_{iM} = v_{iM} / ( M + 1 )$. Fixing $b^\PAB_{ik+1}$, the left-hand side of~\eqref{equation:multi unit discriminatory implicit proof} is strictly positive when $b_{ik} = b^\PAB_{ik+1}$, strictly negative when $b_{ik} = v_{ik}$, and continuous and monotone in $b_{ik}$. Then there is a unique $b_{ik}$ that solves equation~\eqref{equation:multi unit discriminatory implicit proof} conditional on $b^\PAB_{ik+1}$.
\end{proof}

%
%
\subsubsection{The bidpoint-constrained case}

\begin{proof}[Proof of Theorem~\ref{theorem: constrained bids in pab}]
	This proof is substantially similar to proof of the equivalent result for the multi-unit pay-as-bid auction (Theorem~\ref{theorem:discriminatory auction equal conditional regret}). As in the proof of Theorem~\ref{theorem:discriminatory auction equal conditional regret}, Lemma~\ref{lemma: maximal loss in PAB} implies that the loss minimization problem is
	\[
		\left( q^\star, b^\star \right) \in \argmin_{\left( q^\prime, b^\prime \right)} \left[ \max_{k \in \left\{ 0, 1, \ldots, M \right\}} \left[ \max\left\{ \overline R_{q^\prime_k}\left( b^\prime; v_i \right), \underline R_{q^\prime_k}\left( b^\prime; v_i \right) \right\} \right] \right].
	\]
	By definition, $\underline R^\PAB_{q_M}( b; v_i ) \geq \overline R^\PAB_{q_k}( b; v_i )$ for all $k$. Then the loss optimization problem in the pay-as-bid auction can be written
	\[
		\left( q^\star, b^\star \right) \in \argmin_{\left( q^\prime, b^\prime \right)} \left[ \max_{k \in \left\{ 0, 1, \ldots, M \right\}} \underline R^\PAB_{q_k}\left( b^\prime; v_i \right) \right].
	\]
	Recall that
	\[
		\underline R^\PAB_{q_k}\left( b^\prime; v_i \right) = \int_{0}^{q_k} \left( \hat b^\prime\left( x \right) - \hat b^\prime\left( q_k \right) \right) dx + \int_{q_k}^{Q} \left( v^i\left( x \right) - \hat b^\prime\left( q_k \right) \right)_+ dx.
	\]
	Note that $\underline R_{q_k}^\PAB$ decreases as $q_k$ increases while, for all $k^\prime > k$, $\underline R_{q_{k^\prime}}^\PAB$ increases as $q_k$ increases. It follows that if $( q^\star, b^\star )$ is optimal, then $\underline R_{q_k}^\PAB( b^\star; v_i ) = \underline R_{q_{k^\prime}}^\PAB( b^\star; v_i )$ for all $k, k^\prime$.
\end{proof}

\begin{proof}[Proof of Proposition~\ref{proposition: approximate minimax loss pab}]
	Let $( q, b )$ minimax loss in the constrained pay-as-bid auction. Loss is
	\[
		\max_k \left[ \max\left\{ \int_0^{q_k} b\left( x \right) dx, \int_0^{q_k} \left( b\left( x \right) - b_+\left( q_k \right) \right) + \int_{q_k}^{Q} \left( v^i\left( x \right) - b_+\left( q_k \right) \right)_+ dx \right\} \right].
	\]
	Now consider the set of bid points $q^\prime$, where
	\[
		q^\prime_k = \left\lfloor \frac{q_k}{Q} M_q \right\rfloor \frac{Q}{M_q}.
	\]
	That is, $q^\prime_k$ is the feasible bid point nearest to (but below) $q_k$. Define the bid vector $b^\prime$ so that $b^\prime_k = b( q_k )$. By construction, $( q^\prime, b^\prime )$ is feasible in the multi-unit auction. Since $( q, b )$ is optimal, loss is higher under $( q^\prime, b^\prime )$, and since $b^\prime \leq b$ the loss is bounded above by
	\[
		\max_k \int_{q^\prime_k}^{q_k} \left( v^i\left( x \right) - \hat b^\prime_+\left( q_k \right) \right)_+ dx \leq \int_{q^\prime_k}^{q_k} v^i\left( x \right) dx \leq \int_0^{Q/M_q} v^i\left( x \right) dx.
	\]
	Then there is a feasible bid in the multi-unit auction with loss no more than $\int_0^{Q/M_q} v^i\left( x \right) dx$ higher than the optimal bid in the constrained auction.
\end{proof}

%
%
\subsection{Analysis of uniform-price auctions}



\begin{proof}[Proof of Lemma~\ref{lemma: maximal loss in UPA}]
	The proof of this claim is substantially similar to the proof of the equivalent result for the pay-as-bid auction (Lemma~\ref{lemma: maximal loss in PAB}) and is omitted.
%
%
%
\end{proof}

%
%
\subsubsection{The multi-unit case}

\begin{proof}[Proof of Theorem~\ref{theorem: no unique loss-minimizing bid in lab}]
	If there is a unique minimax-loss bid $b_i$, then $\underline{R}^\LAB_{q_k}( b_i; v_i ) = \overline{R}^\LAB_{q_{k+1}}( b_i; v_i )$ for all $k \in \{ 1, \ldots, M - 1 \}$. Otherwise, increasing $b_{ik+1}$ weakly decreases $\underline{R}^\LAB_{q_k}( b_i; v_i )$ and weakly increases $\overline{R}^\LAB_{q_{k+1}}( b_i; v_i )$, and if these terms are nonequal $b_{ik+1}$ can be adjusted without affecting loss, since $b_i$ is optimal. The same argument is sufficient to show that $\underline R^\LAB_{q_k}( b_i; v_i ) = \overline R^\LAB_{q_k}( b_i; v_i )$ for all $k \in \{ 1, \ldots, M - 1 \}$.
	
	Then if there is a unique minimax-loss bid $b_i$, $\overline R^\LAB_{q_k}( b_i; v_i ) = k b_{ik+1} = c$ is constant. By corollary, for all $k \in \{ 1, \ldots, M - 1 \}$,
	\[
		\underline R^\LAB_{q_k}\left( b_i; v_i \right) = 
		\sum_{k^\prime = k + 1}^{M} \left( v_{ik^\prime} - b_{ik+1} \right)_+ = 
		\sum_{k^\prime = k + 1}^{M} \left( v_{ik^\prime} - \frac{c}{k} \right)_+ = c.
	\]
	Except in special cases, this equality cannot simultaneously hold for all $k$.
\end{proof}

\begin{proof}[Proof of Theorem~\ref{theorem: conditional regret minimization lab}]
	The claim follows immediately from earlier arguments. Bids are weakly below values, since $\underline R_{q_k}^\LAB$ is zero and $\overline R_{q_k}^\LAB$ is strictly positive when $b_{ik} > v_{ik}$. We now show that the proposed bid is monotone. Suppose to the contrary that there is $k$ such that $b^\LAB_{ik} < b^\LAB_{ik+1}$. Then
	\begin{align*}
		b^\LAB_{ik} &= \frac{1}{k} \sum_{k^\prime = k}^{M} \left( v_{ik^\prime} - b_{ik}^\LAB \right)_+ \geq \frac{1}{k} \sum_{k^\prime = k+1}^{M} \left( v_{ik^\prime} - b_{ik}^\LAB \right)_+ \\
		&\hspace{1cm}\geq \frac{1}{k+1} \sum_{k^\prime = k+1}^{M} \left( v_{ik^\prime} - b_{ik}^\LAB \right)_+ 
		\geq \frac{1}{k+1} \sum_{k^\prime = k+1}^{M} \left( v_{ik^\prime} - b_{ik+1}^\LAB \right)_+ = b^\LAB_{ik+1}.
	\end{align*}
	The penultimate inequality follows since we have assumed, by way of contradiction, that $b^\LAB_{ik} < b^\LAB_{ik+1}$; but this assumption implies $b^\LAB_{ik} \geq b^\LAB_{ik+1}$, a contradiction, thus 
	it cannot be that $b^\LAB_{ik} < b^\LAB_{ik+1}$.
\end{proof}

%
%
\subsubsection{The bidpoint-constrained case}


\begin{proof}[Proof of Theorem~\ref{theorem: unique loss-minimizing bids in constrained upa}]
	We first prove that the minimax bid $(b_i,q_i)$ must solve
	\begin{align*}
		b_1q_1 &= b_kq_k &\text{for } k \in \left\{1,2,\dots,M \right\}, \text{ and } \\
		b_1q_1 &= \int_{q_{k-1}}^{Q} \left( v^i\left( x \right) - b_k \right)_+ dx  &\text{for } k \in \left\{ 1,2,\dots,M+1 \right\}.
	\end{align*}
	Let $k$ denote the largest index for which maximal loss is attained, i.e, either $k=M+1$ if $\sup_{B^{-i} \in \mathcal B} L^\LAB(b_i; B^{-i}, v^i) = \int_{q_M}^Qv(x)\,dx$ or
	\begin{equation*}
		k = \max\left\{k'\colon \sup_{B^{-i} \in \mathcal B} L^\LAB\left( b_i; B^{-i}, v^i \right) = \max \left\{\underline R_{q_{{k'}-1}}^\LAB, \overline R_{q_{k'}}^\LAB \right\} \right\}.
	\end{equation*}
	Let $k<M+1$. We show that $\underline R_{q_{k-1}}^\LAB = \overline R_{q_k}^\LAB$. Suppose $\underline R_{q_{k-1}}^\LAB > \overline R_{q_k}^\LAB$. As $b_k$ appears in only these two expressions, raising $b_k$ decreases only $\underline R_{q_{k-1}}$ and increases only $\overline R_{q_k}^\LAB$. Suppose $\underline R_{q_{k-1}}^\LAB < \overline R_{q_k}$. Decreasing $b_k$ decreases $\overline R_{q_k}^\LAB$ and increases $\underline R_{q_{k-1}}^\LAB$. We do not have to worry about the effect on $\underline R_{q_k}^\LAB$ as $\underline R_{q_k}^\LAB < \overline R_{q_k}^\LAB$.

	Let $k=M+1$. Observe that $\int_{q_M}^Q v^i( x )\,dx = \underline R_{q_{M}}^\LAB \le \underline R_{q_{M-1}}^\LAB$ as underbidding regret decreases in $b_k$ and $q_{k-1}$. As regret is maximized by $M+1$, the inequality must hold with equality. The argument of the previous paragraph implies $\underline R_{q_{M-1}}^\LAB = \overline R_{q_M}^\LAB$. The result follows.

	We now prove that a unique solution exists. 
	To do so, note that we can express $b_k$ as a function of $q_{k-1}$ and $q_k$ by solving
	\begin{equation*}
		b_{k} q_k = \int_{q_{k-1}}^{Q} \left( v^i\left( x \right) - b_{k} \right)_+ dx
	\end{equation*}
	for $b_k$. The left-hand side increases in $b_k$ and is 0 at $b_k=0$. The right-hand side decreases in $b_k$, is positive for $b_k=0$, and tends to 0 as $b_k$ increases. Thus, there is a unique $b_k(q_{k-1},q_k)$ that solves the equation. The bid $b_k(q_{k-1},q_k)$ decreases in $q_{k-1}$ and $q_k$.

	We then proceed by expressing $q_{k'}$ as a function of $q_1$ by solving $b_1(q_0,q_1)q_1 = b_{k'}(q_{k'-1},q_{k'})q_{k'}$ iteratively for $q_{k'}$, $k' \in \{ 2,3,\dots,M \}$. There is a unique $q_{k'}$ for each $q_1$. Finally, the condition $b_M(q_{M-1}(q_1),q_M(q_1))q_M(q_1) = \int_{q_M(q_1)}^{Q}v(x) \,dx$ pins down $q_1$.
\end{proof}


\begin{proof}[Proof of Proposition~\ref{proposition: approximate minimax loss lab}]
	Let $( q, b )$ be optimal in the constrained uniform-price auction. Loss is
	\[
		\max_k \left[ \max\left\{ q_k b\left( q_k \right), \int_{q_k}^Q \left( v^i\left( x \right) - b_+\left( q \right) \right)_+ dx \right\} \right].
	\]
	Here, $b_+( q ) = \lim_{\varepsilon \searrow 0} b( q + \varepsilon )$. Now consider the set of bid points $q^\prime$, where
	\[
		q^\prime_k = \left\lceil \frac{q_k}{Q} M_q \right\rceil \frac{Q}{M_q}.
	\]
	That is, $q^\prime_k$ is the feasible bid point nearest to (but above) $q_k$. Define the bid vector $b^\prime$ so that $b^\prime_k = b( q_k )$. By construction, $( q^\prime, b^\prime )$ is feasible in the multi-unit auction. Since $( q, b )$ is optimal, loss is higher under $( q^\prime, b^\prime )$, and the difference for a given quantity $q_k$ is
	\begin{align*}
		& \max\left\{ q^\prime_k b^\prime_k, \int_{q^\prime_k}^Q \left( v^i\left( x \right) - \hat b^\prime_+\left( q^\prime_k \right) \right)_+ dx \right\} - \max\left\{ q_k b_k, \int_{q_k}^Q \left( v^i\left( x \right) - \hat b_+\left( q \right) \right)_+ dx \right\} \\
		&\phantom{\leq} \leq \left( q^\prime_k - q_k \right) \hat b\left( q_k \right) = \int_{q_k}^{q^\prime_k} \hat b\left( q_k \right) dx \leq \int_{0}^{Q/M_q} v^i\left( x \right) dx.
	\end{align*}
	Then there is a feasible bid in the multi-unit auction with loss no more than $\int_{0}^{Q/M_q} v^i\left( x \right) dx$ higher than the optimal constrained bid in the divisible-good auction.
\end{proof}

%
%
\subsection{Comparison of auction formats}

\begin{proof}[Proof of Comparison~\ref{theorem:multi unit bid comparison}]	
	We show that when $b^\LAB_i$ is cross-conditionally regret minimizing and $b^\PAB_i$ minimizes loss in the pay-as-bid auction, $b^\LAB_{ik} \geq b^\PAB_{ik}$ for all $k$. Note first that $b^\PAB_{iM} = v_{iM} / ( M + 1 ) = b^\LAB_{iM}$. 
	Additionally, $b^\PAB_{ik+1} \leq b^\LAB_{ik+1}$ implies $b^\PAB_{ik} < b^\LAB_{ik}$. To see this, observe that loss minimization in the pay-as-bid auction requires
	\begin{equation}
		k b_{ik}^\PAB - \sum_{k^\prime = k}^{M} \left( v_{ik^\prime} - b^\PAB_{ik} \right)_+ = k b_{ik+1}^\PAB - \sum_{k^\prime = k+1}^{M} \left( v_{ik^\prime} - b^\PAB_{ik+1} \right)_+. \label{equation:rearranged pab minimization}
	\end{equation}
	Cross-conditional regret minimization in the uniform-price auction requires $( k + 1 ) b^\LAB_{ik+1} = \sum_{k^\prime = k+1}^{M} ( v_{ik^\prime} - b^\LAB_{ik+1} )_+$; then under the assumption that $b^\PAB_{ik+1} \leq b^\LAB_{ik+1}$, it must be that
	\begin{equation}
		k b^\PAB_{ik+1} - \sum_{k^\prime = k+1}^{M} \left( v_{ik^\prime} - b^\PAB_{ik+1} \right)_+ \leq 0. \label{equation:inequality bpabk+1}
	\end{equation}
	This inequality will be strict whenever $b^\PAB_{ik+1} > 0$, which is true whenever $v_{ik+1} > 0$. Substituting inequality~\eqref{equation:inequality bpabk+1} into equation~\eqref{equation:rearranged pab minimization} gives
	\[
		k b_{ik}^\PAB - \sum_{k^\prime = k}^{M} \left( v_{ik^\prime} - b^\PAB_{ik} \right)_+ \leq 0 \;\; \iff \;\; k b_{ik}^\PAB \leq \sum_{k^\prime = k}^{M} \left( v_{ik^\prime} - b^\PAB_{ik} \right)_+.
	\]
Since the left-hand side of the above inequality is increasing in $b^\PAB_{ik}$ and the right-hand side is decreasing in $b^\PAB_{ik}$, the fact that $k b^\LAB_{ik} = \sum_{k^\prime = k}^{M} ( v_{ik^\prime} - b_{ik}^\LAB )_+$ implies that $b^\LAB_{ik} \geq b^\PAB_{ik}$.
\end{proof}

\begin{proof}[Proof of Comparison~\ref{proposition: semicomparability of bids across auctions}]
	In light of Comparison~\ref{proposition: comparison level of minimax loss}, we show that the initial bid in the uniform-price auction must lie above the initial bid in the pay-as-bid auction, and that the lower iso-loss curve (see Section~\ref{subsection: bidpoint-constrained upa}) in the uniform-price auction reaches zero at some $q < Q$, implying the existence of a minimax-loss bid which is zero at the maximum quantity. In the multi-unit case this is below the pay-as-bid bid at the maximum quantity, which must be strictly positive (Corollary~\ref{theorem:multi unit discriminatory representation} and the assumption $v^i(Q)>0$).
	
	First, conditional regret in the pay-as-bid auction at quantity $q = 0$ is $\int_0^Q ( v^i( x ) - b^\PAB( 0 ) )_+ dx = L^\PAB$. Conditional regret in the uniform-price auction at quantity $q = 0$ is $\int_0^Q ( v^i( x ) - b^\LAB( 0 ) )_+ dx \leq L^\LAB \leq L^\PAB$. It follows that $b^\LAB( 0 ) \geq b^\PAB( 0 )$, and thus it cannot be that $b^\LAB < b^\PAB$. Note that Comparison~\ref{proposition: comparison level of minimax loss} implies $b^\LAB( 0 ) > b^\PAB( 0 )$ except in the multi-unit case with a single bid point, $M = 1$.
	
	Second, observe that for $q$ close to $Q$ underbidding loss becomes arbitrarily close to 0. Thus, the lower iso-loss curve must intersect the horizontal axis at some $q<Q$. Then since, in the multi-unit case, the minimax-loss bid in the pay-as-bid auction is always positive (Corollary~\ref{theorem:multi unit discriminatory representation}), there exists a minimax-loss bid in the multi-unit uniform-price auction which is not everywhere above the unique minimax-loss bid in the multi-unit pay-as-bid auction.
	%
\end{proof}

	%

\begin{proof}[Proof of Comparison~\ref{proposition: ex post revenue comparison}]
	When bidder $i$ has unit demand, the ex post transfer to the auctioneer is identical in the last accepted bid uniform-price auction and the pay-as-bid auction. We therefore assume the bidder demands at least two units, $M \geq 2$.
		
	We now show that when bidder $i$ is awarded a small quantity, the ex post transfer to the auctioneer can be larger in the last accepted bid auction than in the pay-as-bid auction; and, when bidder $i$ is awarded a large quantity, the ex post transfer to the auctioneer can be smaller in the last accepted bid auction than in the pay-as-bid auction. The former claim follows from Comparison~\ref{theorem:multi unit bid comparison}, which implies that $b_{i1}^\LAB > b_{i1}^\PAB$ whenever $v_{i2} > 0$. Then the transfer is higher in the last accepted bid auction when the market-clearing price (which is bounded above by $b_{i1}^\LAB$) is relatively close to $b_{i1}^\LAB$. The latter claim is immediate: since $b_{iM}^\LAB = b_{iM}^\PAB$ and bids are strictly decreasing in the pay-as-bid auction, $b_{i2}^\PAB > 0$ implies that $\sum_{k = 1}^{M} b_{ik}^\PAB > M b_{iM}^\PAB$.
	
	Because the comparison of ex post transfers is ambiguous and depends on the quantity allocated, expected transfers are also ambiguous: quantity distributions which place significant weight on quantities under which uniform-price revenue is higher will have higher expected revenue in the uniform-price auction, and quantity distributions which place significant weight on quantities under which pay-as-bid revenue is higher will have higher expected revenue in the pay-as-bid auction.
\end{proof}

\begin{proof}[Proof of Comparison~\ref{proposition: comparison level of minimax loss}]
	Given a bid function $\hat b$, for any quantity $q$ conditional underbidding regret is higher in the pay-as-bid and uniform-price auctions, $\underline R^\PAB_q( \hat b; v^i ) \ge \underline R^\LAB_q( \hat b; v^i )$. Moreover, overbidding regret is weakly higher in the pay-as-bid auction, $\overline R^\PAB_q( \hat b; v^i ) \geq \overline R^\LAB_q( \hat b; v^i )$. Since loss is the supremum of the higher of conditional overbidding and underbidding regrets, taken over all units, it follows that loss is weakly lower in the uniform-price auction.
	
	In the multi-unit case with quantity $M > 1$ the comparison is strict. The proof of Comparison~\ref{theorem:multi unit bid comparison} shows that $b^\LAB_{ik} > b^\PAB_{ik}$ except at $k = M$. 
Let $q$ be the quantity for which worst-case loss equals conditional regret in the uniform-price auction, and let $b^\LAB$ denote the cross-conditional regret minimizing bids of the uniform-price auction. Then we have that
	\begin{align*}
		\sup_{B^{-i} \in \mathcal B} L^\LAB \left( b^\LAB; B^{-i}, v^i \right) &= \int_q^Q \left( v^i\left( x \right) - b^\LAB_+\left( q \right) \right)_+ dx \\
																	&\le \int_q^Q \left( v^i\left( x \right) - b^\PAB_+\left( q \right) \right)_+ dx \\
																	&\le \int_q^Q \left( v^i\left( x \right) - b^\PAB_+\left( q \right) \right)_+ dx + \int_0^q b^\PAB(x) dx\\
																	&= \sup_{B^{-i} \in \mathcal B} L^\PAB \left( b^\PAB; B^{-i}, v^i \right),
	\end{align*}
	where we use that $b^\PAB \le b^\LAB$ (Comparison \ref{theorem:multi unit bid comparison}) and the fact that underbidding regret involves lowering the bids on $[0,q]$.
\end{proof}

\begin{proof}[Proof of Comparison~\ref{comparison: semicomparability of bids across constrained auctions}]
	The statement $b_1^\LAB > b_1^\PAB$ follows essentially from the proof of Comparison~\ref{comparison: comparison level of minimax loss in the constrained case}. See also the proof of Comparison~\ref{proposition: semicomparability of bids across auctions}. The proof that $q_M^\LAB < Q$ also follows the lines of the proof of Comparison~\ref{proposition: semicomparability of bids across auctions}.

	To show that $q_M^\LAB < q_M^\PAB$, recall that the unique minimax-loss bid in the pay-as-bid auction equates underbidding regret across all bidpoints (Theorem~\ref{theorem: constrained bids in pab}). Underbidding regret for quantity $q^\PAB_M$ is
	\[
		\underline R^\PAB_{q^\PAB_M}\left( b^\PAB; v^i \right) = \int_{0}^{q^\PAB_M} \hat b^\PAB\left( x \right) dx + \int_{q^\PAB_M}^Q v^i\left( x \right) dx.
	\]
	This expression is strictly decreasing in $q_M$ so long as $b^\PAB_M < v^i( q^\PAB_M )$, thus the unique minimax-loss bid in the bidpoint-constrained pay-as-bid auction is either such that $q^\PAB_M = Q$, or such that $b^\PAB_M = v^i( q^\PAB_M )$. In the former case, as in the multi-unit case, we have $b^\PAB_M > \hat b^\LAB( Q )$, and the minimax-loss bid in the pay-as-bid auction is above the minimax-loss bid in the uniform-price auction for quantities $q$ near $Q$. Note that this case applies when the marginal values are sufficiently flat. When $v^i(Q) \approx v^i(0)$, then $b^\PAB_M$ cannot be equal to $v^i(Q)$ due to the bid-shading incentives.
%
%
%
\end{proof}

\begin{proof}[Proof of Comparison~\ref{comparison: comparison level of minimax loss in the constrained case}]
	We show that minimax loss in the constrained uniform-price auction with $M$ bid points is lower than in the constrained pay-as-bid auction with $M$ bid points. Consider the minimax loss $L^\PAB$ in the constrained pay-as-bid auction, generated by bid $( q^\PAB, b^\PAB )$. This bid is feasible in the constrained uniform-price auction. Given the same bid, conditional regret is identical in the two auction formats for any quantity $q \in [ 0, q_1^\PAB ]$, and is strictly lower in the uniform-price auction for all quantities $q > q_1^\PAB$ since conditional regret in the pay-as-bid auction includes payments for lower units but conditional regret in the uniform-price auction does not. Since conditional regret is continuous in bid, it follows that a small deviation from $( q^\PAB, b^\PAB )$---namely, increasing $b_{i1}^\PAB$ slightly to $b_{i1}^\LAB > b_{i1}^\PAB$ and decreasing $q_1^\PAB$ slightly to $q_1^\LAB < q_1^\PAB$---will strictly lower conditional regret for quantities $q \in [ 0, \tilde q_1^\PAB ]$ while keeping conditional regret for higher quantities below $L^\PAB$. Since minimax loss is the maximum of conditional regret, taken over all bid points, it follows that minimax loss in the constrained uniform-price auction with $M$ bid points is strictly below optimal loss in the constrained pay-as-bid auction with $M$ bid points.
\end{proof}

%% file: appendix-applications.tex
\section{Examples}
\label{appendix: derivations of examples}

\subsection{Constrained bids}
\subsubsection{Constant marginal values}

\begin{proof}[Calculations for Example~\ref{example: PAB with constant marginal values}]
Equating conditional loss across units requires $R_{k+1}-R_{k}=0$ for all k. This is
	\begin{align*}
		0 &= \left[\sum_{k^{\prime}=0}^{k+1}\left(b_{k^{\prime}}-b_{k+2}\right)\left(q_{k^{\prime}}-q_{k^{\prime}-1}\right)+\left(Q-q_{k+1}\right)\left(v-b_{k+2}\right)\right] - \\
		&\phantom{=\sum} \left[\sum_{k^{\prime}=0}^{k}\left(b_{k^{\prime}}-b_{k+1}\right)\left(q_{k^{\prime}}-q_{k^{\prime}-1}\right)+\left(Q-q_{k}\right)\left(v-b_{k+1}\right)\right] \\
	&=\left(b_{k+1}-b_{k+2}\right)\left(q_{k+1}-q_{k}\right)+\left(Q-q_{k+1}\right)\left(v-b_{k+2}\right) \\
	&\phantom{=\sum} +\sum_{k^{\prime}=0}^{k}\left(b_{k+1}-b_{k+2}\right)\left(q_{k^{\prime}}-q_{k^{\prime}-1}\right)-\left(Q-q_{k}\right)\left(v-b_{k+1}\right) \\
	&=\left(b_{k+1}-b_{k+2}\right)q_{k+1}+\left(Q-q_{k+1}\right)\left(v-b_{k+2}\right)-\left(Q-q_{k}\right)\left(v-b_{k+1}\right) \\
	&=-Qb_{k+2}-\left(q_{k+1}-q_{k}\right)v+\left(Q+\left(q_{k+1}-q_{k}\right)\right)b_{k+1}.
\end{align*}
Let $g_{k}\equiv q_{k}-q_{k-1}$ be the gap between the $k^{\text{th}}$ and $k+1^{\text{th}}$ bid points. Then we have
\begin{align*}
	\left(Q+g_{k+1}\right)b_{k+1}=g_{k+1}v+Qb_{k+2}&\;\;\iff\;\;b_{k+1}=\frac{g_{k+1}}{Q+g_{k+1}}v+\frac{Q}{Q+g_{k+1}}b_{k+2}\\
	&\;\;\iff\;\;b_{k}=\frac{g_{k}}{Q+g_{k}}v+\frac{Q}{Q+g_{k}}b_{k+1}.
\end{align*}
We now solve recursively for optimal bids, conditional on bid points. When $k=M$, we have $b_{k+1}=0$ by assumption, and $b_{M}=\frac{g_{M}}{Q+g_{M}}v$. For $k<M$, we have
\[
	b_{k}=\sum_{k^{\prime}=k}^{M}\frac{Q^{k^{\prime}-k}g_{k^{\prime}}}{\prod_{j=k}^{k^{\prime}}\left[Q+g_{j}\right]}v.
\]
Since $R_{0}=(v-b_{1})Q$, the loss-minimization problem is (dropping the irrelevant constants v and Q)
\begin{align*}
	&\min_{g}1-\sum_{k=1}^{M}\frac{Q^{k-1}g_{k}}{\prod_{k^{\prime}=1}^{k}\left[Q+g_{k^{\prime}}\right]} \\
	&=\min_{g}1-\frac{\sum_{k=1}^{M}\frac{1}{Q+g_{k}}\prod_{k^{\prime}=k}^{M}\left[Q+g_{k^{\prime}}\right]Q^{k-1}g_{k}}{\prod_{k^{\prime}=1}^{M}\left[Q+g_{k^{\prime}}\right]} \\
	&=\min_{g}\frac{\prod_{k^{\prime}=1}^{M}\left[Q+g_{k^{\prime}}\right]-\sum_{k=1}^{M}\frac{1}{Q+g_{k}}\prod_{k^{\prime}=k}^{M}\left[Q+g_{k^{\prime}}\right]Q^{k-1}g_{k}}{\prod_{k^{\prime}=1}^{M}\left[Q+g_{k^{\prime}}\right]}.
\end{align*}
Denote the numerator by $A_{M}$. We show that $A_{M}=Q^{M}$. First, $A_1 = Q$:
\[
	A_1 = \left[ Q + g_1 \right] - \frac{1}{Q + g_1} \left[ Q + g_1 \right] g_1 = Q.
\]
The result follows by induction on $M$; assuming $A_{M} = Q^{M}$, we have
\begin{align*}
	&\prod_{k^{\prime}=1}^{M+1}\left[Q+g_{k^{\prime}}\right]-\sum_{k=1}^{M+1}\frac{1}{Q+g_{k}}\prod_{k^{\prime}=k}^{M+1}\left[Q+g_{k^{\prime}}\right]Q^{k-1}g_{k} \\
	&=\left[Q+g_{M+1}\right]\left[Q^{M}+\sum_{k=1}^{M}\frac{1}{Q+g_{k}}\prod_{k^{\prime}=k}^{M}\left[Q+g_{k^{\prime}}\right]Q^{k-1}g_{k}\right]-\sum_{k=1}^{M+1}\frac{1}{Q+g_{k}}\prod_{k^{\prime}=k}^{M+1}\left[Q+g_{k^{\prime}}\right]Q^{k-1}g_{k} \\
	&=\left[Q+g_{M+1}\right]Q^{M}+\sum_{k=1}^{M}\frac{1}{Q+g_{k}}\prod_{k^{\prime}=k}^{M+1}\left[Q+g_{k^{\prime}}\right]Q^{k-1}g_{k}-\sum_{k=1}^{M+1}\frac{1}{Q+g_{k}}\prod_{k^{\prime}=k}^{M+1}\left[Q+g_{k^{\prime}}\right]Q^{k-1}g_{k} \\
	&=\left[Q+g_{M+1}\right]Q^{M}-Q^{M}g_{M+1}=Q^{M+1}.
\end{align*}

Then the loss minimization problem is
\[
	\min_{g}\frac{Q^{M}}{\prod_{k=1}^{K}\left[Q+g_{k}\right]}, \text{ s.t. } g_k \geq 0 \text{ and } \sum_{k = 1}^{M} g_k \leq Q.
\]
This is solved by $g_{k}=Q/M$. The resulting bids are
\begin{align*}
	b_{k|M} =\sum_{k^{\prime}=k}^{M}\frac{Q^{k^{\prime}-k}g_{k^{\prime}}}{\prod_{j=k}^{k^{\prime}}\left[Q+g_{j}\right]}v 
	&=\sum_{k^{\prime}=k}^{M}\frac{\frac{1}{M}Q^{k^{\prime}-k+1}}{\prod_{j=k}^{k^{\prime}}\left[\frac{M+1}{M}Q\right]}v \\
	&=\sum_{k^{\prime}=k}^{M}\frac{\frac{1}{M}Q^{k^{\prime}-k+1}}{\left[\frac{M+1}{M}Q\right]^{k^{\prime}-k+1}}v 
	=\frac{v}{M}\sum_{k^{\prime}=k}^{M}\left[\frac{M}{M+1}\right]^{k^{\prime}-k+1}.
\end{align*}
\end{proof}

%
%


\subsection{Example: bidpoint-constrained auctions}

We illustrate some perhaps unexpected behavior of minimax-loss bids in bidpoint-constrained auctions in an example where the bidder may submit a single bid point. The bidder has strictly positive marginal value for up to $Q = 5$ units of the good, where
\[
	v\left( q \right)
	= \begin{cases}
		1 &\text{if } q \leq 1, \\
		\varepsilon &\text{if } 1 < q \leq 5.
	\end{cases}
\]
Assume that $\varepsilon \in [ 0, 1 / 6 ]$; we consider the effect of $\varepsilon$ on bids.

\begin{example}[Pay-as-bid with a single bidpoint]\label{example: PAB with single bidpoint}
	If $q\le 1$ and $\varepsilon \le b \le 1=v\left(0\right)$, then the minimax-loss bid solves
	\[
		\min_{b,q}1-b\;\text{s.t. }1 - b = b q + 4 \left( 1 - q \right)\varepsilon.
	\]
	Since the bidder wants maximum loss $1-b$ to be as low as possible, subject to the constraint, and $1-b$ is smaller the larger is $q$ (based on the constraint), it follows that $q \geq 1$. 
	
	If, on the other hand, $q \geq 1$, the right-hand side of the constraint is $bq+( 5 - q ) \varepsilon$, which is increasing in $q$. Then to minimize the right-hand side it must be that $q \leq 1$. It follows that the minimax-loss bid point is $q = 1$.
	
	The minimax-loss bid level follows by solving
	\[
		\min_{b}1-b\text{ s.t. } 1 - b = b + 4 \varepsilon\;\;\iff\;\;b=\frac{1}{2}-2 \varepsilon.
	\]
	Note that this bid is \emph{decreasing} in $\varepsilon$, akin to the example in Section~\ref{section: example}. We have that $b\geq\varepsilon$, and thus that our initial assumption $\varepsilon \leq b$ is satisfied, whenever
	\[
		\frac{1}{2} - 2 \varepsilon \geq \varepsilon \;\; \iff \;\; 1 \geq 6 \varepsilon.
	\]
	Thus the above solution remains valid for all $\varepsilon \in [ 0, 1 / 6 ]$, as assumed.
\end{example}

\begin{example}[Uniform-price with a single bidpoint]\label{example: UPA with single bidpoint}
	In the uniform-price auction the minimax-loss bid solves
	\begin{equation}
		b q = \left( 1 - b \right) = \left( 5 - \max\left\{ 1, q\right\} \right) \varepsilon+\min\left\{ 0, 1 - q \right\}. \label{equation: upa with a single bid point}
	\end{equation}
	The first term is overbidding regret when the bidder receives $q$ units (note that this is independent of whether bids are above or below value), the second term is underbidding regret when the bidder receives $0$ units and $b>\varepsilon$ (we will constrain our analysis to ensure this assumption is valid), and the third term is underbidding regret when the bidder receives just above $q$ units. Together, the first two terms imply
	\[
		b=\frac{1}{1+q}.
	\]
	When $\varepsilon \leq 1/6$, as assumed, Equation~\eqref{equation: upa with a single bid point} has no solution with $q < 1$. Then unique solution is such that
	\[
		1 - \frac{1}{1+q} = \left( 5 - q \right) \varepsilon.
	\]
\end{example}

Consider the solution to Examples~\ref{example: PAB with single bidpoint} and~\ref{example: UPA with single bidpoint} when $\varepsilon = 1 / 6$. In this case, we have
\[
	\left( b^\PAB, q^\PAB \right) = \left( \frac{1}{6}, 1 \right), \text{ and } \left( b^\LAB, q^\LAB \right) = \left( \sqrt{6} - 1, \frac{1}{\sqrt{6}} \right) \approx \left( 0.41, 1.45 \right) \gg \left( b^\PAB, q^\PAB \right).
\]
In particular, the bid is higher in the uniform-price auction and drops to 0 later, and the bids in the pay-as-bid auction are never uniformly higher than in the uniform-price auction.